\documentclass[reprint, aps, twocolumn, twoside,notitlepage, superscriptaddress, nofootinbib, prx]{revtex4-2}

\usepackage{comment}
\usepackage[english]{babel} 
\usepackage{microtype} 
\usepackage[svgnames]{xcolor} 
\usepackage{graphicx} 
\usepackage{enumitem} 
\setlist{noitemsep} 
\usepackage{lastpage}
\usepackage{fancyhdr}
\usepackage{lipsum}
\usepackage[caption=false]{subfig}
\usepackage{graphicx}
\usepackage[T1]{fontenc} 
\usepackage[utf8]{inputenc} 
\usepackage{XCharter} 

\renewcommand{\Re}[0]{\text{Re}}
\renewcommand{\Im}[0]{\text{Im}}
\newcommand{\7}[1]{\pmb{#1}}
\newcommand{\matr}[1]{\ensuremath{\underline{\pmb{#1}}\hskip1pt}}

\usepackage[autostyle=true]{csquotes} 

\usepackage{amsmath}
\usepackage{amsfonts}
\usepackage{amssymb}
\usepackage{textgreek}
\usepackage{cleveref}
\usepackage{braket}
\usepackage{bbold}
\usepackage{dsfont}
\usepackage{amsthm}
\usepackage{listings}
\usepackage[compat=1.0.0]{tikz-feynman}
\usepackage{feynmf}
\usepackage{tikz}
\usetikzlibrary{decorations.markings}

\tikzset{
photon/.style={decorate, decoration={snake}, draw=black},
particle/.style={thin,draw=black, postaction={decorate},
decoration={markings,mark=at position .5 with {\arrow[black]{stealth}}}},
density/.style={decorate, draw=black, decoration={snake=coil}},
propagator/.style={line width=0.75mm, draw=black, postaction={decorate}, decoration={markings,mark=at position .5 with {\arrow[black]{stealth}}}},
}
\tikzset{
photon/.style={decorate, decoration={snake}, draw=black},
particle/.style={thin,draw=black, postaction={decorate},
decoration={markings,mark=at position .5 with {\arrow[black]{stealth}}}},
density/.style={decorate, draw=black, decoration={snake=coil}},
propagator/.style={line width=0.75mm, draw=black, postaction={decorate}, decoration={markings,mark=at position .5 with {\arrow[black]{stealth}}}},
}

\newcommand{\MF}[1]{\textrm{{\textcolor{black}{#1}}}}

\newcommand{\PS}[1]{\textrm{{\textcolor{cyan}{#1}}}}

\newcommand{\bsq}[1]{\begin{subequations}\label{#1}}
	\newcommand{\esq}{\end{subequations}}
\newcommand{\beq}[1]{\begin{equation}\label{#1}}
	\newcommand{\eeq}{\end{equation}}
\newcommand{\beqa}[1]{\begin{eqnarray}\label{#1}}
	\newcommand{\eeqa}{\end{eqnarray}}

\newcommand{\mV}{\matr{V}}
\newcommand{\Tr}{\text{Tr}}

\addto\captionsenglish{}

\begin{document}
	
	\title{A self-consistent current response theory of jamming and vibrational modes in low-temperature amorphous solids}
	\author{Florian Vogel}
	\affiliation{University of Konstanz - D-78457 Konstanz, Germany} 
	\author{Philipp Baumg\"artel }
	\affiliation{University of Konstanz - D-78457 Konstanz, Germany} 
	\author{Matthias Fuchs} \affiliation{University of Konstanz - D-78457 Konstanz, Germany} 
	\date{\today}

	\begin{abstract}
		\MF{We study amorphous solids  with strong elastic disorder and find an un-jamming instability that exists, inter alia, in an harmonic model built using Euclidean random matrices (ERM). }
		Employing the Zwanzig-Mori projection operator formalism \MF{and Gaussian factorization approximations}, we develop a first-principles, \MF{self-consistent theory of transverse momentum correlations} in athermal disordered materials, extending beyond the standard  Born approximation. The vibrational anomalies in glass at low temperatures are recovered in the stable solid limit, and floppy modes lacking restoring forces are predicted in unstable states below the jamming transition. Near the un-jamming transition, the speed of sound $v_0^\perp$ vanishes with $ \propto \sqrt{\epsilon}$, where $\epsilon$ denotes the distance from the critical point. Additionally,  the density of states develops a plateau, independent of $\epsilon$ above a frequency $\omega_*$ which vanishes at the transition, $\omega_*\propto |\epsilon|$.  We identify a characteristic length scale in the un-jammed phase, \MF{$\lambda_-^\perp\propto1/\sqrt{\epsilon}$,} indicating the distance over which injected momentum remains correlated. We confirm the theoretical predictions with numerical solutions of a scalar ERM model, demonstrating overall good qualitative and partly quantitative agreement.		
	\end{abstract}
	\begin{widetext}
		
		\maketitle 
		
	\end{widetext}
	\section{Introduction }
	
	The low-temperature excitations in amorphous solids differ characteristically from the ones in ordered states of the same chemical substances  \cite{Ramos2022}. This does not only hold at very low temperatures, where quantum mechanical tunnelling phenomena dominate in glass. One also observes quite universal anomalies in the vibrational properties at temperatures beyond the quantum regime, where the classical equations of motion hold. Sound waves propagating through glass are damped even in the absence of thermal fluctuations and anharmonicities, and the vibrational density of states (vDOS) deviates from Debye’s law valid in crystals. Structural disorder is held responsible for these distinctive vibrational features in glass.

 In the amorphous solid state, a non-equilibrium phase transition, the  jamming transition, can occur \cite{Ness2022,Charbonneau_Glass_and_jamming, Ikeda_Berthier_Sollich, Mean_field_Paris_2018, DeGiuli2014}. Generally, the glass state exhibits  \textit{entropic} elasticity \cite{Kauzmann1948TheNO,Charbonneau_Glass_and_jamming}. The particles get stuck in so-called \textit{cages}, which  are themselves part of collective motion \cite{g2001supercooled}, and elasticity originates from the structural arrest. The strength of the (shear) resilience to deformation is set by thermal energy times particle density.   On the other hand,  jamming occurs with the emergence of mechanical stability \cite{DeGiuli2014, Mean_field_Paris_2018, Thorpe_1985, Franz_2016}. 
	In jammed structures, particles cannot be moved  without paying an energy cost owing to the interaction potential. For general potentials and with finite thermal fluctuations, it is difficult to separate entropic and enthalpic contributions to the elasticity of glass. Yet, approaching zero temperature the entropic elasticity vanishes and mechanical instability of the glass state below the un-jamming transition becomes visible \cite{Ikeda_Berthier_Sollich}.
	This enables one to perform an harmonic expansion around the force free stable structure.  Then, the Hessian, which gives the dynamical matrix describing the jammed phase, has positive eigenvalues in general.   Only six  eigenvalues are zero in three dimension due to global symmetries.  When approaching the un-jamming transition, more and more  eigenvalues become additionally zero. But,  on average an extensive number of zero eigenvalues is supposed to only exist in the un-jammed state \cite{Mean_field_Paris_2018, ALEXANDER199865}.  
	
	Even though detailed simulations have provided qualitative and quantitative understanding of the low-temperature properties of glass \cite{Ikeda_Phonon_transport,Ikeda_2013,Horbach2001HighFS,Wang_Stable_glasses}, developing an microscopic approach is worthwhile. It should encompass the salient features of the un-jammed and the jammed states in a single first-principles theory by correctly describing the transition and the vibrational features of each phase. It would open the possibility to study amorphous solids at finite temperatures, to include quantum effects, and to develop chemically realistic models for quantitative comparisons to experimental investigations.  There have been a number of approaches  rationalizing aspects of the jammed state and of the (un-) jamming transition. In mesoscopic theories \cite{Schirmacher_Heterogeneous_Elasticity, Schirmacher2007, DeGiuli2014} the peculiarity of the spectrum of the jammed state (the Boson-peak) was rationalized as precursor of the instability. The transition itself was marked as the point when a sufficiently large number of eigenvalues of the Hessian became negative.  While this addresses the glass transition when the structure starts to flow, it would be interesting to study the un-jammed glass state also, where eigenvalues should be non-negative and floppy modes exist but viscous flow is absent.  This state is addressed by mean-field theories or effective medium theories  \cite{Franz2015, Mean_field_Paris_2018, Thorpe_1985}, which often concentrate on the local vibrational modes and do not consider sound waves.  For example, the Debye-law $D_D(\omega)\propto \omega^{(d-1)}$, valid in the vDOS  for $\omega\to0$, becomes irrelevant in the limit of infinite dimension, which is amenable to mean field theory. Yet, in three dimensions the Debye-law and the excess beyond it, called boson peak, are well established in the vDOS and should be described in a first principles approach. The question of the damping of sound also should be addressed, with Rayleigh-damping being observed in a majority of studies \cite{Ikeda_Phonon_transport, BaldiEmergence, PhysRevLett.112.125502, PhysRevLett.104.195501}. Euclidean random matrix models \cite{ciliberti2003brillouin, goetschy2013euclidean,grigera2011high, Vogel_ERM, Ganter_Schirmacher}
	have been suggested to describe these phenomena. They are based on the idea that the particles perform harmonic motion around random positions. The randomness leads to a distribution of local elastic constants \cite{Szamel2022}. These models have proven very promising for capturing the vibrations in stable glass  \cite{schirmacher2019self,Vogel_ERM, Szamel_2023_Elastic_Constant, Baumgaertel2023properties}. 
		Their field theoretic treatments has revealed that multiple local interaction events \cite{grigera2011high,Ganter_Schirmacher, Vogel_ERM} play an important role in e.g. capturing the damping of long wavelength vibrations, viz.~sound. This goes beyond mean-field or effective medium theories, where interaction events on a local level are taken into account. These approaches  miss this class of correlated local events. These multiple local scattering events and the associated occurrence of quasi-localised modes are subject of ongoing research right now \cite{Schober_Localised_modes,SCHOBER2011, Lerner_2021}.
	
	In this work, we aim to analytically investigate the non-equilibrium jamming transition and the vibrational features of both phases on either side. We  develop a self-consistent transverse current response theory taking techniques from the mode-coupling theory (MCT) of the glass transition and its recent generalization \cite{Szamel2003,Janssen2015,Janssen2018}. By doing so, we are able to construct a non-linear self-consistent model which successfully describes  the salient features of the jammed  and un-jammed states at $T=0$  and the critical dynamics. Even though the details of the jammed state highly depend on the preparation protocol (since it is a non-equilibrium state) the transition is supposed to exhibit quite universal features. In the athermal limit $T=0$ the main feature of the transition is the emergence of a finite yield stress, \textit{i.e.} that stresses do not decay to zero but have a finite long-time limit. Such a theory, as will be developed hereinafter, can arguably be generalized to finite temperatures  \cite{T_Franosch_1994} as required for addressing the phase diagram presented in Ref.~\cite{Ikeda_Berthier_Sollich}.  In the athermal limit $T=0$ the system is completely at rest, both in the jammed and un-jammed state.   Hence, we will adopt the concept of a  \textit{fluidity} introduced for yield stress fluids \cite{Goyon_Spatial_cooperativity_2008}, which encodes the potential of the system to yield to applied forces and hence to flow. While it has a finite low-frequency contribution in the un-jammed state, it vanishes with the frequency above the jamming transition. 
	We also include Langevin damping in our approach in order to address the question whether details of the short time motion affect the transition. In order to test the theory, we apply it to two ERM models.

	This work is organized as follows:. In Section  \ref{sec_the_model}, the investigated system is specified and solved by utilising the Zwanzig-Mori projection-operator-formalism \cite{HansenMcDonald}. Eventually, a self-consistent model is constructed, which is shown to predict the correct sound attenuation in the hydrodynamic limit. This model allows the description of two distinct phases. 
	In Section  \ref{sec.jammed_State}  the jammed state is investigated, where stresses do not decay. Additionally, the connection to the ERM model is shown.  The un-jammed state is  analyzed in Section \ref{sec.Unjammed_state}, and the un-jamming transition is analyzed next in Section \ref{sect:crit}. Lastly, we give a conclusion and an outlook.
	
	\section{Theory \label{sec_the_model}}
	
	We construct a self-consistent field theory for force, respectively, stress  fluctuations in athermal disordered materials.  Stress fluctuations are determined from two fields: Firstly,  the  local density $\rho(\7q)= \sum_{i=1}^N e^{-i \7q \cdot \7r_i}$ or rather the fluctuating density $\delta \rho(\7q)=\rho(\7q)-N\delta_{\7q,0}$ since both encode the structure of the disordered system. Here, $\delta$ denotes the Kronecker-delta. 
	A plane wave decomposition of fluctuations is used, with $\7q$ the wavevector, in order to exploit the homogeneity of the system.  	
	Secondly, the displacement field is chosen, or rather its time derivative, viz.~the velocity field  $ \7v(\7q)= \sum_{i=1}^N  \7v_i e^{-i \7q \cdot \7r_i}$. 
	Generally, $\7r_i$ is the position of the $i^\text{th}$ particle and $m \7v_i$ is its momentum. 	\MF{The momentum field $m\7v(\7q)$ will encode the dynamics and indicate forces acting among the particles.}
	\MF{We aim for} an enthalpic theory for the stability of amorphous systems at the bottom of the potential energy landscape. Structural  changes are excluded, leading, as one consequence, to an  incompressible system. Thus, we assume the density correlation function to be time-independent. We ask if the system can support (transverse) sound waves, indicating a stable or jammed system. The decay  of restoring forces on transverse particle displacements serves as an indicator of the un-jammed phase \cite{Voigtmann2011YieldSA}. Longitudinal velocity modes are neglected since they determine the changes of the  local density. Furthermore, it has been shown that the transverse component of the velocity autocorrelation influences the relaxation process much more than the longitudinal component since the peaks are $6-8$ times stronger \cite{Horbach2001HighFS}. Additionally, the transverse modes and not the longitudinal ones change their character at solidification.  Lastly, one of the prominent vibrational anomalies of  disordered materials is the Boson peak. This excess over the Debye-vDOS has been shown to coincide approximately with the transverse Ioffe-Regel limit \cite{tanguy2023vibrations}. The  Ioffe-Regel limit, where the mean free path length and the wavelength become equal, marks the transition from propagation waves to the diffusive regime.

	We proceed by deriving our equations for small temperatures  $T$ and take  the athermal limit at the very end. More concretely, we develop a response functions formalism. The responses are evaluated with the choice of an unperturbed probability distribution function that varies like $\psi_0  \propto  e^{-H/(k_B T)} $ , where $H$ is Hamilton’s energy. Then all manipulations can be performed at small temperatures $T$ and finally the limit $T \to 0$  can be taken. Throughout, we use that the pair correlation function and the structure factor become $T$-independent in the athermal limit as found within the Ornstein-Zernike approach using the hyper-netted-chain approximation \cite{Yang_Schweizer_Glassy_dynamics_2011, Jacquin_Berthier_2010, Mohan_Short_ranged_pair_distribution_2012}. It is noteworthy, that we consider an equilibrium distribution function $\psi_0$, even though  jamming is an non-equilibrium phenomenon \cite{DeGiuli2014, Charbonneau_Glass_and_jamming}.  We do this for two reasons. First, the critical dynamics is supposed to be universal \cite{Wyart_2010, Mean_field_Paris_2018, Charbonneau_Glass_and_jamming},  \textit{i.e} independent of the concrete realisation of the non-equilibrium state. Second, we will re-introduce the dependence on the preparation protocol via the pair correlation function $g(r)$, which is not unique in nonequilibrium systems.

	To keep this text concise, all technical details on the derivation have been moved to the appendix section \ref{app_Derivation_Model}.

	\subsection{Microscopic starting system}

	We consider $N$ particles having masses $m$ contained in a volume $V$ in the thermodynamic limit $N,V \to \infty$ with the density $n= \frac{N}{V}$ staying constant.   The particles are subject to Langevin-dynamics 
	\begin{align} \notag
		m \dot{\7v}_i(t) = \7F_i(t) - \zeta_0 \7v_i(t) + \sqrt{2\zeta_0 k_BT}\; \7w_i(t) \;,
	\end{align}
	where $\zeta_0$ is the Langevin friction coefficient, and $\7w_i(t)$ is uncorrelated Gaussian white noise of unit strength. 
	The interaction forces among the particles are taken from a pair potential $U(r)$ and, because of Newton's third law, the force field  $\7F(\7q, t) = \sum_{i=1}^N  \7F_i(t) e^{-i \7q \cdot \7r_i(t)}$
	arises from the divergence of a stress tensor, $\7F(\7q, t) = i\7q \cdot  \matr{\sigma}(\7q, t)$  \cite{HessKlein, HansenMcDonald, Uniqueness_Stress_Tensor_1995}; here and throughout a  tensor of second rank is indicated by an underbar. Spatial indices are denoted by Greek letters.   The state of the system is assumed to possess rotational and translational invariance, which simplifies the structure of the considered correlation functions \cite{Gotze,Lemaitre2018}.

	\subsection{Response formalism }
	To test the dynamics at zero temperature, we study the temporal evolution of the momentum field, whose initial amplitude $m\7v(\7q, 0)$ was generated by an external velocity perturbation. A linear response calculation gives the response function $\braket{\7v(\7q, t)}^{lin \; resp}|/\7v(\7q, 0) = \7K_q( t)$. The system’s isotropy can be used to decompose the current response function into a longitudinal and a transverse one. With $\hat{\7q}=\7q/q$, the direction of the wavevector,  the transverse response  shall be denoted $\7K^\perp_q (t)=( \matr{1}-\hat{\7q} \hat{\7q}) K_q^\perp(t)$, where the (scalar)  function $K_q^\perp(t)$ depends on time and wavenumber.
	
	\subsection{ Zwanzig-Mori reformulations} 
	
	Using the Zwanzig-Mori (ZM) formalism, the transverse response function $K^\perp_q$ can be related to a memory kernel $G^\perp_q(t)$, the generalized shear modulus. It describes the shear stress response to the internal velocity field \cite{HansenMcDonald, Martin}. Note, that we divided the conventional shear modulus by the mass density for notational simplicity.  The relation simplifies after considering the
	Laplace transformation (see Eq.~\eqref{app_eq_Conventions_Troafos} in the appendix for the used conventions)  to the  expression \cite{HessKlein}
	\begin{align}
		\label{eq_Transverse_velocity_Autocorrelation}
		\hat{K}^\perp_q(s)= \frac{1}{s + \xi + q^2 \hat{G}_q^\perp(s)}\;.
	\end{align}
	Here, $s=-i\omega+0^+$ is the Laplace variable for frequency $\omega$, the $\xi = \zeta_0/m$ is the Langevin damping rate, and the factor $q^2$ in front of the shear modulus arises from the conservation of momentum obeyed by the interparticle forces, viz. Newton’s third law. In the hydrodynamic limit of a fluid, the shear modulus becomes the kinematic shear viscosity, $\lim_{s,q \to 0} \hat{G}^\perp_q (s) \to    \eta /(mn)$, with mass density $(mn)$.

	The phenomenology of the transition from a fluid to a solid can be captured in a generalized Maxwell-model \cite{Maieretall, Vogel_2019}, with the replacement
	\begin{align}
		\hat{G}^\perp_q (s) \to \hat{G}^{gM}(s) =
		\frac{G_\infty/(nm)}{s+ \frac{1}{\tau}} . \notag
	\end{align} An increase of the relaxation time $\tau$ models 
	the increase of the viscosity and the approach to solidification, while $\tau = \infty$ holds in the solid state. When considering plastic flow in driven soft glassy dispersions, the concept of a spatially varying fluidity was introduced \cite{Goyon_Spatial_cooperativity_2008} and tested under shear and in confinement \cite{Microscale_Rheology_2012}. This corresponds to generalizing $1/\tau$ to a response kernel, which we denote: $\hat{W}_q^\perp (s)$. The ZM formalism then expresses the shear modulus by its instantaneous value $G^\perp_q (t = 0) = (c^\perp_q )^2$ and the retarded  kernel:
	\begin{align}
		\label{eq_Transverse_Shear}
		\hat{G}_q^\perp(s)= \frac{(c^\perp_q)^2}{s+ (c_q^\perp)^2\hat{W}_q^\perp(s)}
	\end{align}
	Equation \eqref{eq_Transverse_velocity_Autocorrelation} and the vanishing of the response kernel at high frequencies, $\hat{W}_q^\perp(s) \to0 $ for $s\to\infty$, already indicate (high frequency) vibrations with the  (later on called `bare') dispersion relation $ (q c^\perp_q)^2$.   As  Eq.~\eqref{ap10} in the SM shows, we find the same  expression for the $c^\perp_q$ as given by Zwanzig and Mountain \cite{Zwanzig_High_Frequency_elastic_constants}. 
	
	Note, that all the quantities introduced above have a finite $T \to 0$ limit. All non-zeroth-order terms of the temperature have been set to zero. \MF{Density fluctuations are neglected because they are taken to be frozen-in in the limit of vanishing thermal fluctuations \cite{HansenMcDonald}.}

	\subsection{Self-consistent closure}
	
	\MF{We now search for a closure of the equations, leading to a self-consistent kinetic approach. This requires to approximate the fluidity kernel which contains  the fluctuating time derivative of the  stress. 	We start from the observation that its  potential part can be expressed with the pair state of local density  and velocity field  \cite{Schmid_PHD_Theses}.	The kinetic part of the stress tensor only contributes at finite temperature and can  be neglected. The time-derivative of the stress in the athermal case thus reads
		\begin{align} \notag
			\dot{ \sigma}_{\alpha \beta}(\7q)  = \frac{-2i }{V} \sum_{\7k} \Re\{ X_{\alpha\beta} (\7k,\7q)\} \rho(\7q-\7k) ( \7k \cdot \7v (\7k)),
		\end{align}
		where the weights $X_{\alpha \beta}$ are derived from the pair potential $U(r)$ and  are independent of any phase space variable; see appendix \ref{app_Derivation_Model} for further information.  Inserting this relation into the definition of the fluidity kernel (see Eq.~\eqref{ap12}), and factorizing the appearing four point correlation function leads to the random phase approximation (RPA) \cite{Zaccarelli2001}. 
		Here, the RPA would only give a coupling to the longitudinal momentum field.  Yet, a proper projection of the fluctuating stress variation onto density and momentum following the mode coupling approach \cite{Ohta1976,Gotze} identifies the searched-for coupling to the transverse momentum; see Eq.~\eqref{eq:translong} in appendix \ref{app_Derivation_Model}. This shows that the projection operator formalism is well attuned to  picking up the fluctuations in the system. 
	} 
	
	Therefore, in order to build a theory for an incompressible system at the bottom of the energy landscape ($T=0$) we utilize the projection operator that contains the pair fluctuations of density and transverse momentum: 
	\begin{align} \notag
	P_2 =   \sum_{\7k}  \frac{ \ket{ \7v^\perp(-\7k) \delta\varrho(\7k-\7q)}  \cdot   \bra{ \7v^{\perp}(\7k) \delta\varrho(\7q-\7k)} }{NS_{|\7q-\7k|}   \braket{|\7v^\perp(\7k)|^2}/(d-1)   }  \; .
	\end{align}
	\MF{With it, we express the fluctuating time derivative of the stress in the fluidity kernel, see  Eqs.~\eqref{app_eq_Fluidity_mit_RM} and 
		\eqref{app_eq_Def_Rernormalisef_Vertex}
		in the Appendix. After this projection, we follow the ideas of the generalized mode coupling theory (gMCT) of the glass transition expounded by Szamel,  Janssen, Reichman, and others \cite{Szamel2003,Janssen2015,Janssen2018}, and do not immediately perform the decoupling approximation of the four point correlation function into the product of two point correlation functions.
		This approximation would lead to a self-consistent Born theory, whose properties and failures we describe below.   Rather, we perform additional ZM projections. }
	
		While in gMCT the number of ZM projections is pushed higher as the quality of the approximation is seen to improve, we only perform two more ZM projections in order to obtain the fluctuation kernel that contains the acceleration (second time-derivative) of the transverse momentum field. The reason is threefold. On the one hand, this is the first level where time-reversal symmetry (of the Newtonian system at $\xi=0$)   allows a coupling back to the momentum field. On the other hand, on this level a specific symmetry of the perturbation expansion for transport in random media can be implemented, which was discovered by Leutheusser when studying the Lorentz gas \cite{Leutheusser}. He showed that the collision sequence of a tracer moving through fixed random obstacles has to be invariant under the exchange of the scatterers. This leads to two equally important sequences of binary collisions, while Born theory only keeps one of them.  We recently found the importance of this symmetry for the vibrational modes in random solids \cite{Vogel_ERM}, where it corresponds to a relation between {\em planar} and {\em non-planar} scattering diagrams arising in the perturbative analysis of the ERM model \cite{grigera2011high}.
	The diagrammatic interpretation of this symmetry in the present approach will be given in Sect.~\ref{sec_diagrams}.  There also a third reason, as only on this ZM level the bare Green's function arises naturally in  the high-frequency or high-density expansion.

	After the projections of the fluctuating variables onto the structural and dynamical fields of interest, viz.~the density and transverse momentum, we perform a decoupling/ factorization approximation as for Gaussian statistics; for the details of the calculation see  appendix section \ref{app_Derivation_Model}. As already announced, the structure is taken as frozen at $T=0$, viz.~$S_q(t)=S_q$.	The described steps give the following equation for the fluidity which can sensibly be expressed with a renormalized frequency-dependent vertex:
	\begin{align}
		\label{eq_fluidity_transverse_def}
		\hat{W}_q^\perp(s) =\frac{1}{ N} \sum_{\7k}  \matr{V}_{\7k,\7q} :  \matr{\hat{\mathcal{V}}}_{\7k,\7q}(s)\;,
	\end{align}
where we defined the abbreviation $\matr{A}:\matr{B} =  \Tr \left\{ \matr{A} \cdot \matr{B} \right\}/(d-1)$ that contracts two tensors; all appearing tensors are transverse, but, in order to simplify the notation, we forbear the index $^\perp$ in all quantities that are not measurable directly.
	The  fluidity $W^\perp_q(t)$ captures the scattering of transverse momentum excitations by the frozen-in amorphous structure. The vertex $\matr{V}$ is a  tensorial quantity  of the wavevectors $ \7q,\7k$, which determines the amplitude of the scattering events (see Eq.~\ref{app_eq_Exact_Expression_Vertex}).  It depends on the pair potential and, via density correlation functions, on the structure.  The renormalised vertex $\matr{\hat{\mathcal{V}}}_{\7k,\7q}(s)$ is a functional of the unrenormalised vertex and the transverse current autocorrelation $K^\perp_q(t)$. It contains all topological information about the  scattering events taking into account the disorder.  The described steps lead to its constituting equation 
\begin{align}
&	 \sum_{\7p} \left[	\Big(s(s+\xi) +  (p c_p^\perp)^2  \Big) \matr{\delta}_{\7k,\7p}^\perp + s \;	\hat{\matr{\Xi}} (\7q,\7k,\7p,s) \right] \cdot \hat{\matr{\mathcal{V}}}_{\7q,\7p}(s) \notag  \\ & = \sum_{\7p} \left[	s  \matr{\delta}_{\7k,\7p}^\perp + 	\hat{\matr{\Xi}} (\7q,\7k,\7p,s) \right] \cdot \matr{V}^\dagger_{\7p,\7q}    S_{|\7q-\7p|}  
	   \;,
  	\label{eq_Renormalised_Transverse_Vertex_FE}
\end{align}
where $\matr{\delta}_{\7k,\7p}^\perp = (\matr{1}- \hat{\7p} \hat{\7p} ) \delta_{\7k,\7p}$. 
This equation is derived in detail  in the appendix Sect.~\ref{app_Derivation_Model}.
It results from the combination of two generalized Langevin equations that naturally arise in the ZM formalism \cite{HansenMcDonald}. The memory kernel $\hat{\matr{\Xi}}$ captures non-Markovian dissipation.  For large frequencies, where  it vanishes, the renormalized vertex $\matr{\hat{\mathcal{V}}}_{\7k,\7q}(s)$ is determined by the bare dispersion relation. For small frequencies, a Markovian approximation, where $\hat{\matr{\Xi}} (\7q,\7k,\7p,s)$ is replaced by a transport coefficient, could be a reasonable approximation. Yet, we did not follow this route and  instead aim for a self-consistent closure that leads to a first principles approach.  Performing the decoupling approximation inherent to the mode coupling approach, the  memory kernel $\hat{\matr{\Xi}}$  consists of two parts: 
	\begin{align}
		\label{Main_text_eq_Decomposition_Transverse_zweite_Sigma_memory_function}
		\frac{	\hat{\matr{\Xi}} (\7q,\7k,\7p,s) }{(c_k^\perp)^2 } \approx \hat{ {M}}_k(s)\,  \matr{\delta}_{\7k,\7p}^\perp +   \frac 1N  \Tilde{\matr{V}}(\7q,\7k, \7p) \; \hat{K}^\perp_{|\7q-\7k-\7p|}(s)\;.
	\end{align}
	The first term results from a factorization of a six-point memory kernel diagonal in the contained density fluctuations, while the second term contains their off-diagonal  correlations.  Leutheusser's symmetry dictates to keep both terms.  We will call the first planar and the second non-planar as explained in Sect.~\ref{sec_diagrams} below. Latter it will be seen that the interference of both terms is required in order to correctly capture the elastic disorder in the system. Their amplitudes read
	\beq{Main_text_mct2}
	\begin{split}
		{M}_k(t) &=    \frac 1N \sum_{\7b}  \; \matr{V}_{\7k,\7b}  : \matr{V}_{\7b,\7k}^\dagger  \;   S_{|\7k-\7b|} \; K^\perp_b(t) \;. \\
		\tilde{\matr{V}}(\7q,\7k, \7p)  &  =  \frac{k}{p} \matr{V}_{\7k,\7p+\7k-\7q} \cdot  \matr{V}_{\7p+\7k-\7q,\7p}^\dagger S_{|\7q-\7k|}\;.
	\end{split}
	\eeq
	Neglecting the non-diagonal contribution in Eq.~\eqref{Main_text_eq_Decomposition_Transverse_zweite_Sigma_memory_function}  is equivalent to setting 
	\beq{Main_text_mct_planar}
	\begin{split}
		\;	\hat{\matr{\Xi}} (\7q,\7k,\7p,s) \approx (c_k^\perp)^2\,  \hat{{M}}_k(s) \; \matr{\delta}_{\7k,\7p}^\perp \;.
	\end{split}
	\eeq
	We refer to this diagonal approximation as a self-consistent Born approximation \cite{Kamenev_2011, Altland_Simons_2010}. 
	The fluidity in this approximation becomes 
	\begin{align}
		\label{eq_SCB}
		W_q^{SCB,\perp}(t)= \frac{1}{N} \sum_{\7k}  \matr{V}_{\7q,\7k} :  \matr{V}_{\7k,\7q}^\dagger  S_{|\7q-\7k|}\;   K_k^\perp(t) \;.
	\end{align}
	This approximation is in accordance with classic mode coupling approximations in other contexts \cite{pihlajamaa2023unveiling, Gotze, Hertz_2016,Leutheusser1983}, which involves a diagonalization and factorization of the occurring higher-point correlators. Nevertheless, this approximation would be qualitatively wrong applied here to the vibrations in amorphous solids \cite{ciliberti2003brillouin, goetschy2013euclidean, grigera2011high,Schirmacher_Heterogeneous_Elasticity, Vogel_ERM}. The wavevector dependence of the damping of sound in glass would be $\propto  q^2$ instead of $\propto q^4$ (in $d=3$). Additionally, Eq.~\eqref{eq_SCB} predicts the wrong scaling of the observables at the un-jamming transition compared to simulations \cite{Ikeda_Phonon_transport}.
	We will discuss the sound attenuation in Section \ref{sec_Rayleigh_damping} and the critical dynamics in Section \ref{sect:beta}.  Equations \eqref{eq_Transverse_velocity_Autocorrelation}  to \eqref{Main_text_mct2}  constitute our self-consistent theory of transverse momentum fluctuations in athermal disordered systems. In the thermodynamic limit, $V,N \to \infty$, the sums over the wavevectors become integrals $\sum_{\7k} =  \frac{V}{(2 \pi)^d} \int d^d \7k$, and the equations depend on $n$ and density correlation functions, which encode the arrested structure via the vertex $\matr{V}_{\7q,\7k}$, see  Eq.~\eqref{app_eq_Exact_Expression_Vertex}  derived in Appendix Sect.~\ref{app_sec_Vertex_Evaluation}.

	\subsection{Phase transition}
	
	The set of self-consistent equations allows two different scenarios
	\begin{align}\label{eq:pt}
		\lim_{t \to \infty}  G_q^\perp(t) =		\lim_{s \to 0} s \hat{G}_q^\perp(s ) = \begin{cases}
			=0  \hspace{0.3cm} \hat{=} \hspace{0.15cm} \text{un-jammed} \\
			>0  \hspace{0.3cm} \hat{=} \hspace{0.15cm}  \text{jammed} 
		\end{cases}
	\end{align} In the jammed state, forces and stresses do not completely decay for $t \to \infty$. Hence, the system is an elastic solid and allows the presence of low-frequency transverse vibrational modes on all length scales. On the other hand, a decaying shear modulus implies a mechanically floppy systems with a vanishing macroscopic shear elasticity. This hence characterizes the un-jammed state.  Reassuringly, this characterization does not depend on the Langevin rate $\xi$. In this sense, the two phases and the transition between them are independent on the dynamical details.
	
	We proceed by discussing the two phases  individually, starting with the jammed phase before turning to the critical dynamics close to the transition.
	
	\section{The jammed state \label{sec.jammed_State}}
	
	The system allows the propagation of sound modes, if it is so densely packed, that it is elastic for small frequencies and mechanically stable. Elastic contributions dominate in the shear modulus at small frequencies and the transverse response function vanishes with the frequency.  Hence,  a susceptibility  $\hat{\chi}_q(s)=   \frac 1s \hat{K}^\perp_q(s)$ arises in this phase, along with a self-energy $\hat{\sigma}_q(s)$. The latter is the energy a mode with frequency $s-0^+=-i \omega$ and wavenumber $q$ has because of the changes it inflicts in its environment, \textit{i.e} structural rearrangements or  density fluctuations. The susceptibility quantifies how the system reacts to a perturbation.  It is the displacements' Green's function in the jammed state. The constituting equations follow from Eqs.~\eqref{eq_Transverse_velocity_Autocorrelation} and \eqref{eq_Transverse_Shear}:
	\begin{subequations}\label{eq16}
		\begin{align} \label{susceptibility}
			\hat{\chi}_q(s)&= \frac{1}{s(s + \xi) +q^2(c_q^\perp)^2 +\hat{\sigma}_q(s)}\;,
		\end{align}
		\begin{align}
			\label{Main_text_eq_self_energy}
			\hat{\sigma}_q(s) &= -\frac{q^2 (c_q^\perp)^4 \hat{w}_q(s)}{1+ (c_q^\perp) ^2 \hat{w}_q(s) }    
		\end{align}
	\end{subequations}
	with $\hat{w}_q(s)= \frac{1}{s} \hat{W}_q^\perp(s)$. The self-energy $\hat{\sigma}_q(s) $ is  evaluated via Eqs.~\eqref{eq_fluidity_transverse_def} to \eqref{Main_text_mct2}  with the time-integrated vertex $\matr{\hat{\mathcal{U}}}_{\7q,\7k}(s) = \frac 1s \matr{\hat{\mathcal{V}}}_{\7q,\7k}(s) $. All of these quantities  stay finite for $s \to 0$; see Eq. \eqref{jammed} in the section \ref{app_sec_Jammed_Phase} of the appendix.

	\subsection{Dispersion relation \label{sec_dispersion_relation}}  The non-decaying transverse stresses lead to finite restoring forces in the long time limit. We introduce the transverse renormalized dispersion relation as $(v_q^\perp)^2 =G_q^\perp(t \to  \infty) 
	=  [(c_q^\perp)^{-2} + \hat{w}_q(0)]^{-1}$. \MF{In general, this elastic constant  is a $q$-dependent generalization of Lamé's shear constant and encodes the zero-frequency shear elasticity of the amorphous solid.}	In the small wavevctor limit, it defines the speed of sound, which is given by $v_0^\perp$. The susceptibility features a sound pole   $\hat{\chi}_q(s=-i\omega+0^+)= 1/[-\omega^2 + (q v_0^\perp)^2- i\omega\, \Gamma_q^\perp ]$ for small wavenumber $q$ and small frequency  $s=-i\omega +0^+$ (the sound damping $\Gamma_q^\perp$ will be discussed in Sect.~\ref{sec_Rayleigh_damping}). Overall, the system can  be regarded as a stable, disordered solid. Thus, $(v_q^\perp)^2$ is the most important observable in the jammed phase, and the vanishing of this elastic constant indicates that the system is loosing its stability. 
	
At $s=0$,  Eq.~\eqref{eq_Renormalised_Transverse_Vertex_FE} strongly simplifies. The renormalized vertex can be obtained explicitly, and the fluidity can be summed. 
Because of isotropy,  Eqs.~(\ref{eq_fluidity_transverse_def}-\ref{Main_text_mct2}) and (\ref{eq16}) lead to a concise linear, one-dimensional integral equation in the compliance, viz.~the inverse of the shear elasticity $1/(v_p^\perp)^2$: 
	\begin{align}
		\label{Main_text_eq_dispersion_relation}
		\begin{split}
			\sum_p  \Big( \delta_{q,p}-  p^{d-3}C_{q,p} \Big) \frac{1}{(v_p^\perp)^2} = \Delta_q+ \frac{1}{(c_p^\perp)^2} \equiv \frac{1}{(\Tilde{c}_q^\perp)^2}\;.
		\end{split}
	\end{align} The matrix $ p^{d-3} C_{q,p}$ is an important quantity, called stability matrix. It and the zero-frequency  contribution $\Delta_q$ can be calculated from the pair potential and the radial pair-correlation function. Their concrete expressions are given in the Eqs.~\eqref{eq_app_one_loop_cotribution} and \eqref{eq_app_Def_stability_matrix} in the appendix. 
	The correction term $\Delta_q= \hat{\Delta}_q(s=0)$ arises from the additional terms in the short-time ZM expansion and gives a renormalization of the bare dispersion relation. More importantly, this equation for the shear elasticity has only a unique, non-zero and positive solution for $(v_q^\perp)^2$, if the maximal eigenvalue of the  matrix $p^{d-3}C_{q,p}$ is smaller than unity. As announced in Sect.~\ref{sec.jammed_State},  the stability of the system is independent of $\xi$. 
	
	\begin{figure}
		\centering
		\includegraphics[width=0.5\textwidth]{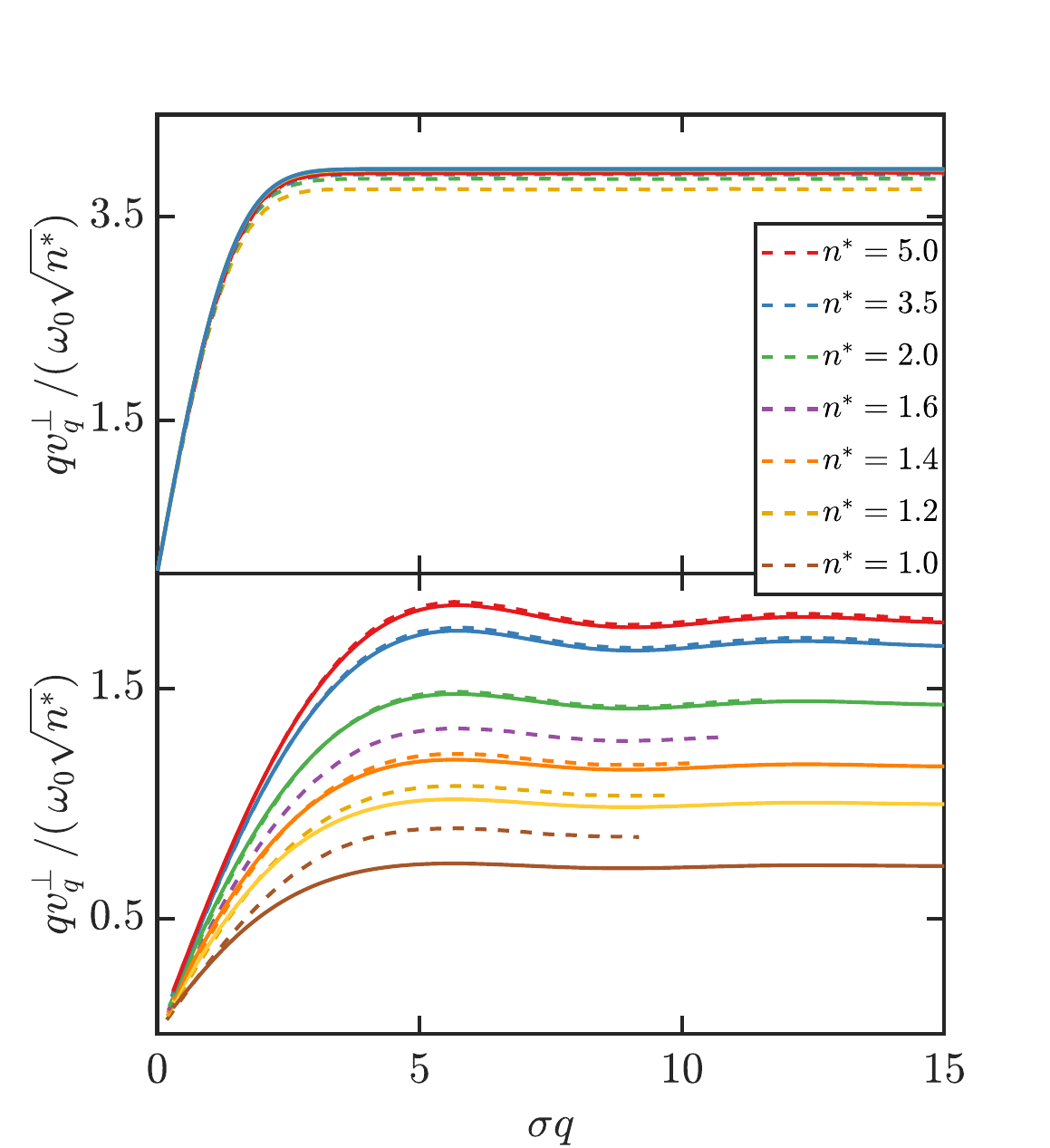}
		\caption{Comparison of the dispersion relations calculated with Eq.~\eqref{Main_text_eq_dispersion_relation} (solid lines) with the numerical solution of the ERM-model calculated with Eq.~\eqref{eq:DispersionNumerics}(dashed lines). The upper panel shows the data for the Gaussian spring function $f^G$ and the lower panel for the theta spring function $f^\Theta$. The colour code for the rescaled density $n^*$ is the same in all panels.}
		\label{fig_Dis_Vergleich}
	\end{figure}

We show two examples for the dispersion relation in Fig.~\ref{fig_Dis_Vergleich} within  the scalar ERM approximation (Eq.~\ref{springf}) for different spring functions $f(r)$ that model the distance dependence of  harmonic pair interactions. A Gaussian $f^G$ gives an infinite coordination number $z$ \cite{DeGiuli2014}, while a Heaviside step function $f^\Theta$ gives a density dependent one, $z=(4\pi/3) n^*$. The models will be introduced in Sect.~\ref{sec_Connection_to_ERM}; $\omega_0^2$  sets the frequency, and $n^*$ is a rescaled density. The results for both ERM models lie remarkably close to the numerically calculated dispersion relations. \MF{There is a regime of sound modes for small wavevectors, $q\sigma\ll1$, where $\sigma$ denotes the interaction range. Then, the sound propagation speed softens, and for large $q$ the vibration frequency saturates, while an overall scaling $(v_q^\perp)^2\propto n$ originates in the use of pair interactions. }
	
	Let us note in passing that a presumably convenient approximate model  can be invented by reintroducing the frequency in Eq.~\eqref{Main_text_eq_dispersion_relation}, as described in more detail in Section \ref{sec_app_F1_model} in the Appendix:
	\begin{align}
		\label{eq_F1_model_Main_text}
		\hat{w}_q(s)
		&= \hat{\Delta}_q(s) +   \sum_{p}   p^{d-1}C_{q,p}    \hat{\chi}_p(s)\;. 
	\end{align}	
This equation can also directly be  obtained from  Eq.~\eqref{eq_Renormalised_Transverse_Vertex_FE} by neglecting the memory kernel $\hat{\matr{\Xi}}$ in the square bracket on the left hand side.  The approximation becomes exact for $s \to 0 $.
	It resembles the well-studied F1-model \cite{Gotze} and  can potentially be solved efficiently for non-zero frequencies  in order to be compared to experiments. 
	
	\subsection{Rayleigh-damping \label{sec_Rayleigh_damping}}  
	Due to the disorder, sound waves are damped even at zero temperature ($T=0$). We set $\xi=0$ in this subsection to investigate the disorder-induced attenuation. The sound modes lose their momentum by exciting other modes via the coupling induced by the disordered structure.   \MF{This is captured in the imaginary part of the self-energy, which is finite even though, for $\xi=0$, the equations of motion are reversible, suggesting an expansion of $\hat{\sigma}_q(s)$ symmetric in $s$ and thus real (as $s^2=-\omega^2$). This expectation of a real self-energy is not correct because $\hat{\sigma}_q(s)$ is non-analytic in $s$.} \MF{Let us show that the low-frequency expansion of  $\hat{\sigma}_q(s)$ is $ \propto q^4 s^{d-2}$ in accordance with Rayleigh-damping. Note that this proof requires excluding hydrodynamic damping $ q^2 s$ in 3D  \cite{grigera2011high,Schirmacher_Heterogeneous_Elasticity}.} We utilize the   Sokhotski–Plemelj theorem
	\begin{align}
		\label{eq_main_text_Sokhotski_Plemelj_theorem}
		\frac{1}{x \pm i0^+}= \mathcal{P} \frac{1}{x} \mp i \pi \delta (x)\;,
	\end{align}
	to calculate the damping for $s \to 0$.   Here $\mathcal{P}$ is  the Cauchy principal value. We define the dissipation via the imaginary part of the self-energy: $\tilde{\Gamma}_q(\omega) = \Im\{ \sigma_q(s=-i \omega+0^+)\}$, and  find for $ \omega \to 0$  
		\begin{align}
	\label{eq_Main_text_Rayleigh_damping_general_equation}
		\begin{split}
			\tilde{\Gamma}_q&(\omega)   \underset{\omega \to 0}{= }   q^2 (v_q^\perp)^4 \frac{\pi}{2}   \Bigg[ \frac{C^{(0)}_{q,\omega/c_0^\perp}}{(c_0^\perp)^d} +\frac{C_{q,\omega/v_0^\perp}}{(v_0^\perp)^d} \Bigg] \omega^{d-2}\;. 
		\end{split}
	\end{align}
	If the square bracket was a constant for $\omega,q$ going to zero, the incorrect hydrodynamic damping would be predicted. Yet,
	the planar and non-planar contributions, which encode Leutheusser's symmetry, cancel each other exactly in the long wavelength limit.  This is trivial in matrix $C^{(0)}$ given in Eq.~\eqref{app_eq_First_order_Matrix_C0} in App.~\ref{App_Paragraph_Proof_Rayleigh}, and can be checked explicitly for the stability matrix 	
\begin{align} \label{eq_Cancelation_Planar_Non_Planar}
		\begin{split}
			&	C_{0,0}=  \frac{1}{N^2(d-1)}  \int d^{d-1} \hat{p}\;     \text{Tr} \Bigg\{\sum_{\7k  } \frac{1}{k^2} \matr{V}_{0,\7k}    \\&   \times   \matr{V}_{\7k,0} S_{k}    \cdot  \Big[ \matr{V}_{-\7k,0}    \cdot   	S_{k}  \matr{V}_{0-\7k}  +   \matr{V}_{-\7k,0}\cdot 	   S_{k}   \matr{V}^\dagger_{0,\7k}  \Big] \bigg\}=0
		\end{split}
	\end{align}
	The vanishing $C_{0,0}=0$ holds due to the symmetry of the bare vertex  $\matr{V}^\dagger_{-\7k,0}=-\matr{V}^\dagger_{\7k,0}$. Rotational invariance   implies that in the hydrodynamic limit $C_{q \to 0,0} = \frac{\beta_0^{2,1}}{k_D^d} q^2$ and $C_{0,p \to 0} =  \frac{ \beta_0^{2,2}}{k_D^d} p^2$ holds. The same holds trivially for the high-frequency matrix $C^{(0)}_{q \to 0, 0} = \frac{\beta_0^{1,1}}{k_D^d}q^2$ and  $C^{(0)}_{0,p \to 0} =  \frac{\beta_0^{1,2}}{k_D^d} p^2$, with the Debye-wavevector  $k_D=  \sqrt[d]{2 \pi^{d-1}n}$ and the $\Gamma$-function.  	The Debye wavevector is included to render the coupling coefficients $\beta^{i,j}$   unit-less scalars.  
	Following Ref.~\cite{Ikeda_Phonon_transport}, we determine the sound damping by evaluating the dissipation at the propagation frequency $ q v_0^\perp$ taking $q,\omega \to0$:
	\begin{align} 
	\label{eq_main_text_Rayleigh_Damping_small_w_Ikeda}
	    \Gamma_q^\perp =  \left.\frac{\tilde{\Gamma}_q(\omega)}{\omega}\right|_{\omega=v_0^\perp q} \to (q v_0^\perp)^{d+1} \frac{\pi}{2 k_D^d}  \left( \frac{\beta_0^{(1)}}{(c_0^\perp)^d} + \frac{\beta_0^{(2)}}{(v_0^\perp)^d}\right)
	\end{align}
	and $\beta_0^{(1)}=\beta_0^{(1,1)} + \beta_0^{(1,2)} \frac{(v_0^\perp)^2}{(c_0^\perp)^2} $ and  $\beta_0^{(2)}=\beta_0^{(2,1)} + \beta_0^{(2,2)} $.
	Note, that $\beta_0^{(2)}$ depends inversely on density but $\beta_0^{(1)}$ is independent of the density.  Equation \eqref{eq_main_text_Rayleigh_Damping_small_w_Ikeda} proves that sound attenuation is Rayleigh-like for $d>2$.
When presenting the self-consistent Born approximation in Eq.~\eqref{Main_text_mct_planar}, we had mentioned that it predicts hydrodynamic sound damping. This can be verified from Eq.~\eqref{eq_Cancelation_Planar_Non_Planar}, where the second term would be missing in a self-consistent Born approximation, so that the necessary cancellation fails.

	\subsection{Density of states}
	
	\begin{figure}
		\centering
		\includegraphics[width=0.5\textwidth]{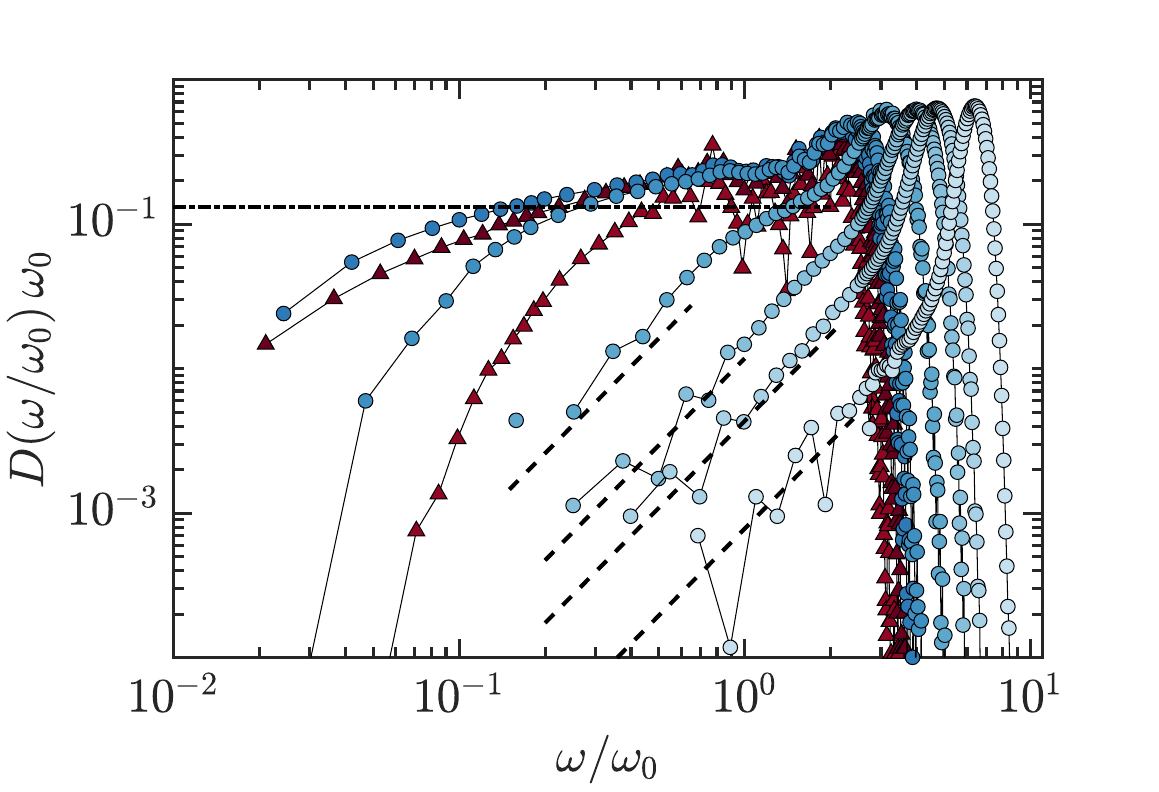}
		\caption{The vibrational density of states $D(\omega)$ obtained from the numerical diagonalization of the ERM model for the spring function $f^\Theta$ for the densities $n = 0.4,0.6, 0.85, 1.0, 2.0, 3.5, 5.0$ and $n=9.5$ with large-frequency drop ordered from left to right. The densities in the un-jammed state are shown in red and the densities in the jammed state in blue. The dashed black lines display the Debye spectrum of the stable solid as calculated from Eq.~\eqref{eq_Debye_Spectrum_DOS}, and the dashed dotted black line the constant level from Eq.~\eqref{eq:constantDOS} that develops close to un-jamming. Note that we excluded integer valued eigenvalues from the calculation of the density of states. See the App.~\ref{app:ERM} for further information.}
		\label{fig:vDOS_cutSpring}
	\end{figure}

	The susceptibility $\hat{\chi}_q(s)$ in Eq.~\eqref{susceptibility} has the form of a Green's function with the \textit{bare dispersion-relation} $(q c_q^\perp)^2$ and the self-energy $\hat{\sigma}_q(s)$, which vanishes in the limit of zero interaction.  The spectral density per particle, \textit{viz.}~the vibrational density of states (vDOS) \cite{Ashcroft76}, is now obtained by summing over all the modes of the imaginary part of the Green's function and by dividing by the number of particles $N$ \cite{Schirmacher2007}:
	\begin{align}
		\label{eq_General_equation_for_vDOS}
		D(\omega) =   \frac{2 \omega}{N \pi } \int_{|\7q|<k_D} \frac{d^d \7q}{(2 \pi)^d} \;\Im  \chi_{q}(\omega)
	\end{align}
	The Debye-wavevector $k_D$ encodes the density of the number of degrees of freedom.
  Later, we will calculate the spectra within the scalar ERM-approximation (see Eq.~\eqref{sec_Connection_to_ERM}). There is only one degree of freedom in this schematic model. Thus, the  Debye wavevector reads in this case $k_D= \sqrt[3]{2 \pi^2 n}$ in $d=3$. As our approach is anharmonic,  Eq.~\eqref{eq_General_equation_for_vDOS} is an approximation, which could be improved following    Ref.~\cite{Boss_Goetze_Zippelius_1978}.
	
	Since the shear modulus \MF{becomes a constant in the long wavelength limit}, the renormalized dispersion relation vanishes with $(v_0^\perp q)^2$. The non-analytic part of the Green's function hence leads to the Debye-Spectrum for $\omega \to 0$ 
	\begin{align}
		\label{eq_Debye_Spectrum_DOS}
		D_D(\omega) =  \frac{  \omega^{d-1} }{(k_D v_0^\perp)^{d}} \;.
	\end{align} 
	The Debye-law is re-obtained, since the jammed system is an elastic medium in the hydrodynamic limit and supports sound waves. 
	
	Figure \ref{fig:vDOS_cutSpring} displays the vDOS for the $\Theta$-spring function in the ERM model (Eq.~\ref{springf}). The data was obtained from numerically diagonalizing the Hamiltonian. The Debye-law and a peak at higher frequency are clearly visible deep in the stable jammed state. The peak frequency shifts with $\sqrt{n}$ because of the pair interactions.   Close to the transition the vDOS starts to develop a plateau for small frequencies. This is also predicted by the theory. See section \ref{sect:crit} for details. In the un-jammed state, the data appear compatible with a minimal frequency, which  shifts higher with increasing distance to the transition.

	\subsection{Mapping to the ERM-model \label{sec_Connection_to_ERM}}
	
	\subsubsection{The ERM approach}

The vibrations in low-temperature glass have been studied in the ERM model (ERM) \cite{Vogel_ERM,M_zard_1999,Martin_Mayor_2001,ciliberti2003brillouin,grigera2011high, Ganter_Schirmacher, Ganter_2011_Diagrams,schirmacher2019self}, which considers the harmonic motion of particles around fixed random positions.  
 The system is specified by the random matrix $\matr{\mathcal{H}}$ which constitutes the potential energy 
	\begin{align}\label{ermmodel}
		U= \frac 12 \sum_{i,j} \phi_i \matr{\mathcal{H}}_{ij} \phi_j  \equiv m \frac{\omega_0^2}{4} \sum_{i,j} f(\7r_i-\7r_j)(\phi_i-\phi_j)^2\;.
	\end{align}
	The $\phi_i$ denotes the small displacement of particle $i$  around the frozen position $\7r_i$. The vector character of the displacements is neglected here, but can be considered \cite{ciliberti2003brillouin}. The  dimensionless spring function $f(r)$ models the (pair) interaction strength depending on the distance of two particles. The frequency scale $\omega_0$ is set by the microscopic length scale $\sigma$, modelling the particle diameter,  and the high frequency shear modulus  $n^*\omega_0^2 = \lim_{q \to \infty} q^2( c_{q}^\perp)^2$, where $n^*=(N/V)\sigma^d$ is the dimensionless density. The (Newtonian, $\xi=0$) harmonic  motion is solved by determining eigenvalues and eigenvectors of $\matr{\mathcal{H}}$, which is a symmetric and positive semi-definite matrix. Finally, disorder averages are performed by taking uniformly and independently distributed positions  $\7r_i$.
	More details on the ERM model and the description of the employed numeric techniques to solve it can be found in appendix Sect.~\ref{app:ERM}. 
	
We have numerically analysed two different spring functions. The Gaussian spring function reads $f^G = \exp[-r^2/(2 \sigma^2)]$ (more results on the Gaussian $f^G$ can be found in \cite{Baumgaertel2023properties}) and the $\Theta$-spring function $f^\Theta = \Theta(\sigma-r)$. In both cases, the dimensionless density $n^*$ is the single control parameter. Importantly, the $\Theta$-spring function $f^\Theta$ gives (harmonic) interactions  of finite extent, where our analysis will show that an un-jamming instability happens at a critical density denoted $n^*_c$. We find from the theory $n^*_c = 0.830$ and from the numerics $n^*_c = 0.80(6)$.
	
	\subsubsection{Identifying the ERM vertex }
	
	The present theory can be applied to the ERM model via the self-consistent equations  for the susceptibility (Eq.~\ref{eq16}), which plays the role of the propagator in the field theory of the ERM model  \cite{Vogel_ERM,M_zard_1999,Martin_Mayor_2001,ciliberti2003brillouin,grigera2011high}. The link starts from neglecting the tensor structure of the vertex (introduced in the fluidity in Eq.~\ref{eq_fluidity_transverse_def}): ${V}^{ERM}_{\7q,\7k}= \frac{q (c_q^\perp)^2}{d-1} {\rm Tr}\{\matr{V}_{\7q,\7k}\}$. This scalar vertex vanishes with $a q^2 + b\7q \cdot \7k$ for $\7q \to 0$ and has all the properties of the general vertex used in ERM investigations \cite{grigera2011high, Vogel_ERM}.  Taking the trace, its explicit expression can readily be obtained from Eq.~\eqref{app_eq_Exact_Expression_Vertex} in the appendix.

Looking there at expression \eqref{eq_app_Calculation_c_q^perp}  for the  high frequency transverse velocity $c_q^\perp$, one recognises 
	\begin{align} 
		\begin{split}
			V_{\7q,\7q}^{ERM} 
			=-(qc_q^\perp)^2\;.
		\end{split}
	\end{align}
	As a consequence, the high-frequency asymptote of the susceptibility interpreted as a bare propagator, $\chi_q^{(0)}(s)=\Big[s(s+\xi) + 
	(q c_q^\perp)^2 \Big]^{-1} $,  gives the expression which was also found for the ERM-model \cite{ciliberti2003brillouin,grigera2011high, Ganter_Schirmacher}. Moreover, in an high density expansion the vertex was connected to the spatial Fourier-transform of the spring function \cite{ciliberti2003brillouin, grigera2011high, Martin_Mayor_2001, Vogel_ERM, goetschy2013euclidean}
	\begin{align}  \label{springf}
		V^{ERM}_{\7q,\7k} \longrightarrow n^*\omega_0^2 \frac{f(|\7k|)-f(|\7q-\7k|)}{f(q=0)}  \;.
	\end{align}
Note, that the spatial Fourier-transformation is simply indicated by the argument. 
Using the scalar nature of the bare ERM-vertex, we  simplify the tensorial structure of the renormalized vertex,   $ q (c_q^\perp)^2 \matr{\mathcal{V}}_{\7q,\7k}(s) \to \hat{\chi}_k^{(0)}(s)S_{|\7q-\7k|}\mathcal{V}^{ERM}_{\7q,\7k}(s) ( \mathbb{1} -\hat{\7q} \hat{\7q})$.  Lastly, the ERM model neglects excluded volume effects and chooses a uniform distribution of the particles. This results in the simplification $S_q = 1$.  With these steps we have simplified our equations to solve the ERM model; the resulting equations of motion are summarized in Eq.~\eqref{eq_SI_ERM_Equation} in the appendix Sect.~\ref{app_sec_ERM}.   Performing additionally a small frequency approximation, we effectively end up with the \textit{F1}-model (Eq.~\ref{eq_F1_model_Main_text}),  which is easily solvable with the scalar ERM-vertex $V_{\7q,\7k}^{ERM}$. The defining expression is given in Eq.~\eqref{eq_app_F1_approximation_ERM}.  Even though the \textit{F1}-approximation loses its validity below the un-jamming transition, we still use  it  to investigate the un-jammed state.

	\subsubsection{Recovering the high-density expansion of the ERM model} \label{sec_diagrams}
	
	In the jammed phase, when the system is very densely packed, $\hat{\sigma}_q(s)$ becomes small because of  the increase of the intermolecular forces. The integrated fluidity therefore is also small in the high density regime and the self-energy (Eq.~\ref{Main_text_eq_self_energy}) approximately reads  
	\beq{highdens}
	\hat{\sigma}_q(s) \approx -  q^2 (c_q^\perp)^4 \hat{w}_q^\perp(s) \;.
	\eeq
	or within in the ERM simplification 
$	\hat{\sigma}^{ERM}_q(s) \approx  - \hat{w}_q^{ERM}(s)$. 
Additionally, the bare dispersion relation $c_q^\perp$ becomes large deep in the jammed state, so that a small frequency approximation $s^2\ll (q c_q^\perp)^2$ can be performed. This simplifies Eq.~\eqref{eq_app_F1_approximation_ERM} further to a result for the self-energy which shall be summarized diagrammatically:
		\begin{align}\label{vertex_ERM_final}
		\begin{split}
			\hat{\sigma}^{ERM}_q(s) \approx&  \;	
			\vcenter{\hbox{\begin{tikzpicture}
						\draw[particle] (0,0) -- (0.89,0);
						\fill[](0,0) circle (0.1);
						\node at (1,0) [rectangle,draw] {};
						\draw[photon] (0,0) arc(180:11.9:0.5) ;
			\end{tikzpicture}}}\;  \\ 
			\hbox{\begin{tikzpicture}
					\draw[photon] (0.0,0.5) arc(90:11.9:0.5);
					\draw[] (0,0) -- (0.39,0);
					\node at (0.5,0) [rectangle,draw] {};
			\end{tikzpicture}}=&
			\hbox{\begin{tikzpicture}
					\draw[photon] (0.0,0.5) arc(90:11.9:0.5);
					\draw[] (0,0) -- (0.39,0);
					\fill[](0.5,0) circle (0.1);
			\end{tikzpicture}}+  \;	  \underset{\text{planar}}{\hbox{\begin{tikzpicture};
						\draw[propagator] (0,0) -- (1.05,0);
						\draw[particle] (1.1,0) -- (1.618,0);
						\fill[](0,0) circle (0.1);
						\fill[](1.05,0) circle (0.1);
						\draw[photon] (0,0) arc(180:0:0.525) ; 
						\fill[](0,0) circle (0.1);
						\fill[](1.74,0) circle (0.1);
						\draw[photon] (1.174,0.84) arc(60:11.9:1.1) ;
			\end{tikzpicture}}}+  \;	 
			\underset{\text{non-planar}}{\hbox{\begin{tikzpicture}
						\draw[propagator] (0,0) -- (1.05,0);
						\draw[particle] (1.1,0) -- (1.618,0);
						\fill[](0,0) circle (0.1);
						\fill[](1.05,0) circle (0.1);
						\draw[photon] (0,0) arc(180:7.5:0.86) ; 
						\fill[](0,0) circle (0.1);
						\fill[](1.74,0) circle (0.1);
						\draw[photon] (0.05,0.94) arc(90:0:1.02) ;
			\end{tikzpicture}}}
			\end{split}
			\end{align}
This equation becomes equivalent to the self-consistent model for the high density regime of the ERM-model derived in Ref.~\cite{Vogel_ERM}. A thin/ thick line stands for the bare/ dressed propagator, a circle stands for the vertex 	$V^{ERM}_{\7q,\7k}$; see section \ref{app_sec_ERM} in the appendix  and Refs.~\cite{Vogel_ERM,grigera2011high} for all details on the diagrammatic expansion and on the described steps. This calculation shows that the Leutheusser symmetry, which required to consider both contributions to $\hat{\matr{\Xi}} (\7q,\7k,\7p,s)$ in Eq.~	\eqref{Main_text_eq_Decomposition_Transverse_zweite_Sigma_memory_function}, corresponds to the equivalence of planar and nonplanar diagrams in the diagrammatic expansion in Eq.~\eqref{vertex_ERM_final}. As argued in Sect.~\ref{sec_Rayleigh_damping},  both sets are required for recovering Rayleigh dissipation \cite{Vogel_ERM}.
 If we neglected the non-planar contributions and  performed the self-consistent Born approximation (Eq.~\ref{Main_text_mct_planar}),  we would  end up with the high density theory proposed by Grigera and co-workers \cite{ciliberti2003brillouin}.

	\section{The un-jammed state \label{sec.Unjammed_state}}
	
	In the un-jammed state, fluctuating stresses (viz. forces) decorrelate with time and decay for $t \to \infty$, viz.~$G_q^\perp(t\to\infty)=0$. This causes a finite value of  
	\begin{align}\label{nonergodic}
		\kappa_q^\perp= \lim_{s \to 0} (s+\xi) \hat{K}_q^\perp(s)\;.
	\end{align}
	In the Newtonian case $(\xi=0)$, the $\kappa_q^\perp$ can be interpreted as non-ergodicity parameter \cite{Gotze}.  A part of the injected momentum does not decay because there are displacement modes which feel no restoring force. The transverse current response remains finite in the Newtonian case, even at long times, $K_q(t \to \infty)=\kappa_q^\perp$. 
	In the  Langevin case, the friction causes a relaxation of the initial transverse momentum with rate $\xi$. 
	The consequences of Eq.~\eqref{nonergodic} are that the kernel $ \hat{\mathcal{W}}_q(s)=  (s+\xi) \hat{W}_q^\perp(s) $ and the generalized vertex $ \matr{\hat{\Upsilon}}(\7q,\7k,s)) = (s+\xi)  \matr{\hat{\mathcal{V}}}(\7q,\7k,s) $ also yield finite contributions  for $s \to 0$.  \MF{The final equation for the amplitude $\kappa_q^\perp$ reads}
	\begin{align}
		\label{main_text_NEP_II}
		\frac{\kappa_q^\perp}{1-\kappa_q^\perp} = \frac{ \hat{\mathcal{W}}_q(s=0)}{q^2}\;,
	\end{align}
	with the equations giving $ \hat{\mathcal{W}}_q(s=0)$ in terms of $\kappa_q^\perp$ specified in 
	Eq.~\eqref{app_eq_unjammed_Vertex_small_s}.
	Contrary to the jammed case, all terms in Eq.~\eqref{app_eq_unjammed_Vertex_small_s}  contribute even for $s=0$. However, $\kappa_q^\perp$ becomes small at the transition, and one can neglect  all terms of higher order in $\kappa_q^\perp$ then. One  ends up with a linear integral equation for the long-time limit of the fluidity, with the coupling coefficients again given by the stability matrix $C_{q,p} $
	\begin{align}
		\label{main_text_NEP_I}
		\hat{\mathcal{W}}_q(s=0)=  \sum_{p} p^{d-1}  C_{q,p}  \kappa_p
	\end{align}		
	The  zero-frequency 	fluidity,
	$\hat{\mathcal{W}}_q(s=0) = \mu_{q}^\perp/\xi$,  sets the ($q$-dependent generalized) shear viscosity $\eta_q^\perp=(mn \xi)/\mu_q^\perp$. Note that a Newtonian ($\xi=0$) un-jammed system has a vanishing viscosity which agrees with Bagnold's considerations of granular media  \cite{Kranz2018,Ness2022}. The stationary fluidity can also be related to a length scale $\lambda_-^\perp= \sqrt{1/\mu_0^\perp},$ which determines the Lorentzian form of the quasi-static current response,  $\kappa_q^\perp \to 1 /[1+(q \lambda_-^\perp)^2]$ for $q \to 0$. In the Newtonian case, it describes the distance over which momenta remain correlated in a ballistic motion. In the Langevin case,  it describes the distance over which the externally injected momentum  is dissipated into the damped un-jammed system. Considering the un-jammed system to consist of unconnected clusters of cooperatively moving particles, the length $\lambda_-^\perp$ would be linked to the average cluster size \cite{Degiuli2015,Ness2022}. 
	
\begin{figure}
		\centering
		\includegraphics[width=0.5\textwidth]{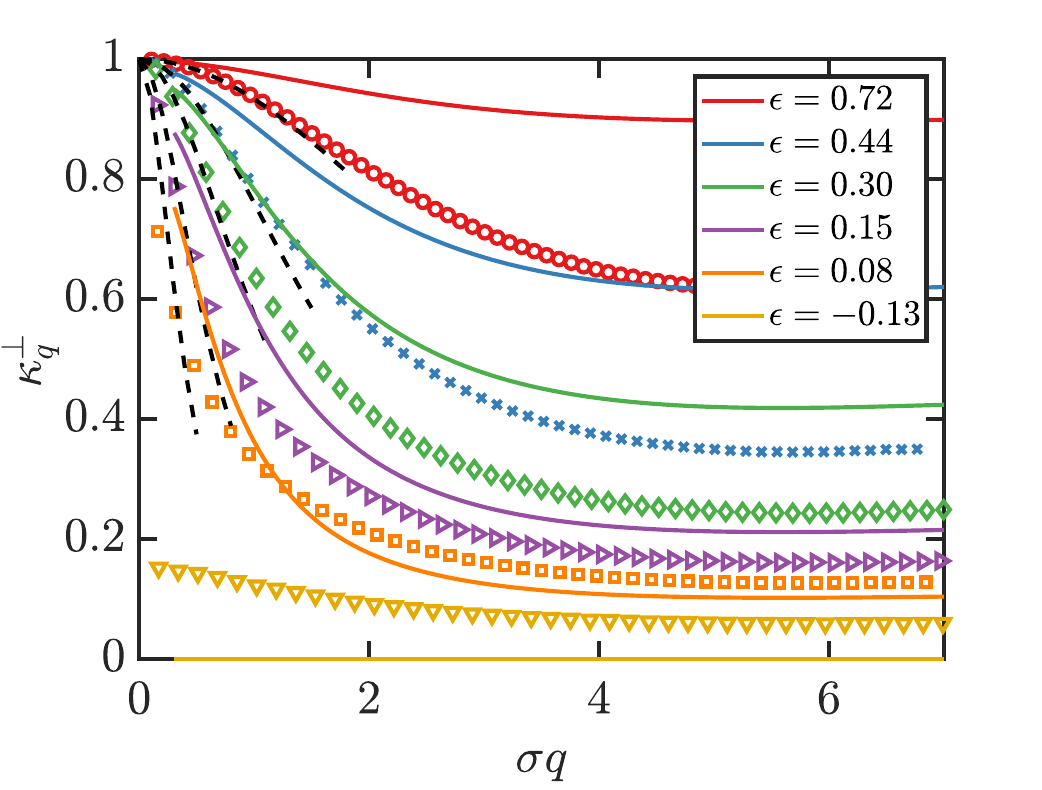}
		\caption{The non-ergodicty parameter $\kappa_q^\perp$ for the spring function $f^\Theta$ of finite extent $\sigma$ (full lines) in comparison with the results of the numerical diagonalization of the ERM model calculated with Eq.~\eqref{eq:NEPNumerics} (symbols). For the results of the numerical diagonalizazion the length scale $\lambda_-^\perp$ is obtained by Lorentzian fits indicated by the dashed black lines. The linear $\epsilon(n)$-relation   determined from the sound velocities in the jammed state gives the values in the legend.}
		\label{fig:Shear_Viscosity}
	\end{figure}

	Figure \ref{fig:Shear_Viscosity} shows the non-ergodicity parameter  $\kappa_q^\perp$ in the scalar ERM-approximation for the spring function of finite range, $f^\Theta(r)$, and compares it to numerically exact values. There, $\kappa_q^\perp$ describes the spatial structure of all of the modes with vanishing elastic eigenfrequency; see appendix \ref{app:ERM} for details. The curves  start at $\kappa_{q=0}^\perp=1$ and decrease to a $n$-dependent plateau for $q\to\infty$. For small $q$,  the curves are well described by a Lorentzian parametrized by the length $\lambda_-^\perp$.  Approaching the transition from the un-jammed side, the length increases strongly. As one can see in Fig.~\ref{fig:Shear_Viscosity}, the \textit{F1}-approximation causes quantitative differences  but captures qualitatively the shape of the $\kappa_q^\perp$ from the numeric solutions of the ERM model in the un-jammed state. Because of the finite numeric system, $\lambda_-^\perp$ can only be determined below $n^* = 0.65$. 
In the jammed state, a qualitative difference appears. The numeric solution picks up the rattlers, viz.~small particle clusters that are not elastically coupled to others and thus lead to vanishing eigenfrequencies even in the stable solid. These clusters are small, and thus $\kappa_q^\perp$ is rather $q$-independent in the jammed state. The theory does not pick up these rattlers, but identifies the divergence of the correlation length as signature of jamming from below; see Sect.~\ref{scalinglaws}.

	\section{Critical dynamics\label{sect:crit} }
	
	\MF{The transition between stable  and unstable states was identified in Eq.~\eqref{eq:pt}.
		The elastic constants $(v_q^\perp)^2$, determined from Eq.~\eqref{Main_text_eq_dispersion_relation},  and the stationary-current amplitude $\kappa_q^\perp$, determined  from Eqs.~(\ref{main_text_NEP_II},\ref{main_text_NEP_I}), both vanish when approaching the un-jamming transition; see the Figs.~\ref{fig_Dis_Vergleich} and \ref{fig:Shear_Viscosity}.  Guided by these observations of small quantities,  we follow Götze and perform a nonlinear stability analysis  \cite{Gotze1981,Gotze,Schnyder_2011} to determine the dynamics close to the transition.} Again, the details of this derivation are moved to the appendix, Section \ref{app_sec_beta_analysis}. We only go through the main aspects of the Götze stability analysis here in the main text. 
	
	\subsection{Derivation of the Götze stability equation \label{sect:beta}}
	
	We start by  \MF{introducing} an auxiliary quantity
	\begin{align} \label{eq:aux}
		s \varphi_q(s)= \frac{s(s+\xi)}{(c_q^\perp)^2} +(s+\xi)\; \hat{W}_q^\perp(s)\;.
	\end{align}
	\MF{It agrees with the fluidity kernel for low frequency and $\xi=0$.} 
	The conceptual idea is, that this quantity 
	becomes small for $s \to 0$ close to the transition.   Note, that the shear modulus can be expressed with this quantity as $\hat{G}_q^\perp(s)= \frac{s + \xi }{s\varphi_q(s)}$ .  In the un-jammed phase holds $\lim_{s, \7q  \to 0}s\varphi_q(s) = \frac{1}{(\lambda_-^\perp)^2}$, while    $\lim_{s  \to 0}s\varphi_q(s) = \frac{s(s+\xi) }{( v_q^\perp)^2 }$ is true in the jammed phase. We can trivially express the transverse velocity autocorrelation using $s\varphi_q(s)$, see Eq.~\eqref{eq:S82}. 
	Expanding the renormalized vertex in  Eq.~\eqref{app_eq_unjammed_Vertex} for small $s \varphi_q(s)$ leads to the final nonlinear equation to be solved close to the transition: 
	\begin{align}
		\label{Main_text_eq_First_B_equation}
		\begin{split}
			s \varphi_q(s) &- \frac{s(s+\xi)}{(\Tilde{ c}_q^\perp(s))^2} - \sum_{p} p^{d-3} C_{q,p} \; s \varphi_p(s) \\&= - \sum_p p^{d-3} C_{q,p}\; \frac{(s\varphi_p(s))^2}{ s\varphi_p(s)+ p^2}  + \mathcal{O}(s^2(s+\xi)\varphi_q(s))\;.
		\end{split}
	\end{align}

	\MF{The transition or critical point is now given when the linearized expansion, viz.~the first line in Eq.~\eqref{Main_text_eq_First_B_equation}, cannot be solved for the small quantity $s  \varphi_q(s) $. This happens when the critical eigenvalue of the stability matrix $C_{q,p}$ equals unity.}	
	Close to the transition, \MF{the stability matrix takes the form $p^{d-3}C_{q,p}=(1+\epsilon) h_q \hat{h}_p + C^\#_{q,p}$ with the left $h_q$ and the right $\hat{h}_q$ eigenvector to the maximal eigenvalue $1+\epsilon$, where $\epsilon$ is called separation parameter.  Because of the orthogonality of the eigenspaces,  $\sum_q \hat{h}_q C^\#_{q,p}=0$ holds.    For $|\epsilon|\to0$,}   $s \varphi_q(s)$ is dominated  by the critical eigen-direction, which suggests the ansatz 
	\begin{align}
		\label{Main_text_eq_AnsatzFried}
		s \varphi_q(s) = \sigma s_* g(s_*) h_q + \sigma^{d/2} s_* X_q^\#(s_*)
	\end{align}
	with $s_*=s t_*$. Here $\sigma$ and $t_* = 1/\omega_*$   are two  yet unspecified scales. $t_*$ sets the timescale and $\sigma$ rescales the amplitude.  Assuming that the largest eigenvalue is simple, one has  $\sum_q \hat{h}_q X_q^\#=0$. Inserting this  ansatz  \eqref{Main_text_eq_AnsatzFried} in Eq.~\eqref{Main_text_eq_First_B_equation}
	and contracting  with the left eigenvector together with the adopted normalizations 
	$\sum_q \hat{h}_q h_q=1$ and $\sum_{p}\hat{h}_p \frac{h_p h_p}{p^2} =1$,
	leads to  the scaling equation 
	\begin{align}
		\label{Main_text_eq_scaling_equation}
		\frac{t_0^2s (s+\xi) }{\sigma^2} + \frac{\epsilon}{\sigma}  s_*g_\pm(s_*) = (s_*g_\pm(s_*))^2\;.
	\end{align}
	This holds for $s \to 0$ with $\tilde{c}^\perp_q(s=0)=\tilde{c}^\perp_q$ from Eq.~\eqref{Main_text_eq_dispersion_relation}.
	The timescale  parameter results from the integral over the inhomogeneous part
	\begin{align}
		\begin{split}
			t_0^2 &= \sum_{q} \frac{ 	\hat{h}_q }{(\Tilde{c}_q^\perp)^2}\;. 
		\end{split}
	\end{align}
	Additionally, demanding all terms in Eq.~\eqref{Main_text_eq_scaling_equation} to be of the same order, one finds  $\sigma= |\epsilon|$.
	For the largest eigenvalue  becoming larger than 1, there is no unique positive solution for Eq.~\eqref{Main_text_eq_dispersion_relation}. Thus, $\epsilon >0$ corresponds to the un-jammed phase\MF{; it is negative in the jammed states; different subscripts in $g_\pm$ denote the scaling function  in either state. For $(s+\xi)\approx \xi$, Eq.~\eqref{Main_text_eq_scaling_equation}  is the scaling equation of Götze's F1-model  \cite{Gotze1981,Gotze}.  The un-jamming transition is thus of continuous type as had been suggested by Voigtmann \cite{Voigtmann2011YieldSA}.}
	
	\MF{There is an aspect  worth discussing.}	Close to the transition, $s  \varphi_q(s) $ vanishes in the small frequency limit. The first term on the right-hand side of Eq.~\eqref{Main_text_eq_First_B_equation} then features a potential divergence for $p \to 0$ \cite{Schnyder_2011}. At least in lower dimensions $d<5$. 
	\MF{We show in the appendix, section \ref{app_sec_critical_left_eigenvecor} that this divergence is eliminated by the vanishing of the critical left eigenvector in the long wavelength limit, $\hat{h}_p \propto p^2 $  for $p \to 0$.} The reason for this is, that a divergence would ultimately lead to a contradiction to the presence of Rayleigh-damping. Hence, the cancellation of the two contributions of $\hat{\matr{\Xi}}$  is at the heart of this observation, which requires us to go beyond the self-consistent Born approximation. We numerically confirm this  in the ERM-approximation in Fig.~\ref{fig_crit_l_eigenvector}.

	After having obtained the scaling functions $g_\pm(s_*)$, we can investigate the scaling behaviour of important observables in the jammed- and un-jammed states. \MF{We look at the Newtonian case in the main text and leave the Langevin one for the appendix, Sect.~\ref{AppFD}.}
	
	\subsection{Scaling laws in the Newtonian-case $\xi=0 $}
	\label{scalinglaws}
	
	From Eq.~\eqref{Main_text_eq_scaling_equation}, the characteristic frequency scale follows as $\omega_* =|\epsilon|/t_0$ for $\xi=0$.   The general solution for the scaling function is
	\begin{align}\label{eq:newtonlsg}
		g_\pm (s_*) = \frac{ \text{sign}(\epsilon)  \pm \sqrt{1+4 s_*^2  }}{2 s_*}
	\end{align}
	Note, that $g_\pm(s_*)$ is an uneven function in $s_*$.  Via Eqs.~(\ref{eq:aux},\ref{Main_text_eq_AnsatzFried}), the susceptibility $\hat{\chi}_q(s)$ becomes a symmetric function in Laplace space in the range of validity of the expansion, \textit{i.e.} after neglecting  $X_q^\#$. This  shows that the Götze stability equation (Eq.~\ref{Main_text_eq_scaling_equation}) considers the fate of the elasticity of the system. One has to choose the \textit{positive} sign of the square root in Eq.~\eqref{eq:newtonlsg} in order to get a positive spectrum. Recall, that the solution for the jammed phase is denoted $g_-$ and the solution in the un-jammed phase as $g_+$, depending on the sign of $\epsilon$. In the limits of small and large frequencies compared to $\omega_*$, one finds the asymptotic expansions:
	\begin{align}\label{asymptotes}
		\begin{split}
			g_\pm (s \gg \omega_*) & \to 1 \\
			g_-(s \ll \omega_*) & \to s_*-s_*^3 \\
			g_+(s_* \ll \omega_*)& \to \frac{1}{s_*}+s_*
		\end{split}
	\end{align}

	\paragraph{Stationary quantities}	
	Using Eq.~\eqref{eq:aux}, this gives the zero-frequency fluidity in the un-jammed state as $\mu_q= | \epsilon| h_q$ and the	length scale $ \lambda_-^\perp = 1/\sqrt{h_0 |\epsilon|}$. The dispersion relation can also be expressed with the critical eigenvector. We find
	\begin{align}
		\label{Main_text_eq_dispersopn_relation_critical_eigenvector}
		( v_q^\perp )^2= \frac{|\epsilon|}{t_0^2 h_q}\:.
	\end{align}
	This result is in accordance with the simulations in Ref.~\cite{Ikeda_Phonon_transport} where the authors found  $v_0^\perp \propto \sqrt{\hat{G}} \propto \sqrt{ \omega_*}$, with $\hat{G}$ being their shear modulus. 
	
	\begin{figure}
		\centering
		\includegraphics[width=0.5\textwidth]{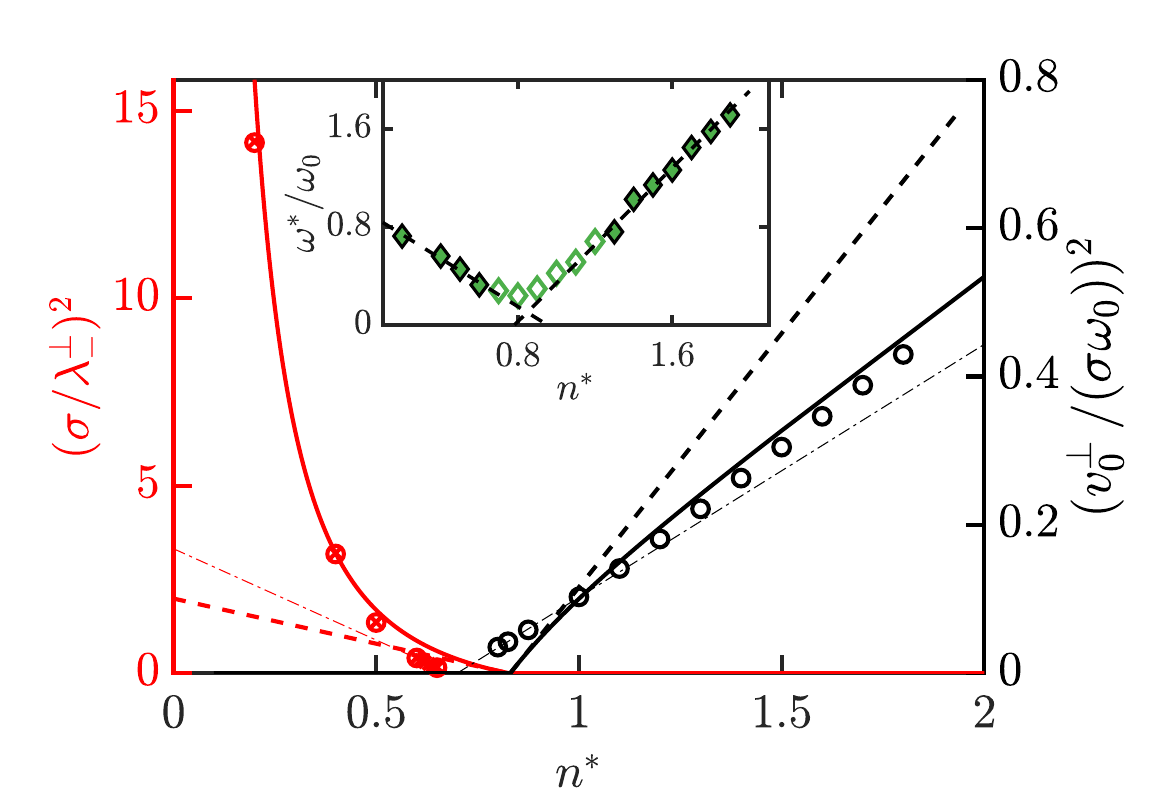}
		\caption{The length scale $\lambda_-^\perp$ and the speed of sound $v_0^\perp$ computed from Eq. \eqref{main_text_NEP_I} and Eq. \eqref{Main_text_eq_dispersion_relation} (full lines) compared to the results of the numerical diagonalization of the ERM model (symbols). The circles depict the results of calculations with $N = 4\times10^4$ and the crosses with $N = 10^4$. The dashed lines depict a linear fit of the theory results with critical densities coinciding at $n^*_c = 0.830$. The dashed dotted lines show a linear fit of the data from numerical diagonalization leading to $n^*_c = 0.70$ based on $v_0^\perp$ and to $n^*_c = 0.82$ based on $\lambda_-^\perp$. The inset shows the transition in $\omega^*$ (see text for details on the calculation). The dashed lines indicate a linear regression of the jammed and the un-jammed side, where the filled symbols depict the data used for the fits. From these fits we find $n^*_c =0.78$ and $n^*_c =0.95$.}
		\label{fig_tansition_from_left_and_right}
	\end{figure} In the ERM-approximation holds  $\epsilon \propto \frac{n_c-n}{n_c}$. Figure \ref{fig_tansition_from_left_and_right} shows the results for the length-scale $\lambda_-^\perp$ and the speed of sound  $v_0^\perp$ within the ERM-approximation for the $f^\Theta$-spring function. 
The result obtained from numerical diagonalization of the Hessian displayed in Fig.~\ref{fig_tansition_from_left_and_right} is  compared to the prediction of the theory. As one can see, the prediction for the speed of sound in the jammed phase lies close to the numerical solution of the ERM-model.
The agreement of the data  in the un-jammed phase is also good. This is somewhat surprising, since we solved only the F1-model instead of the full model in this phase.

The numerically observed un-jamming transition  of the scalar  ERM-model resembles the percolation problem.  Percolation was discussed with an exponential spring function \cite{Amir2013}, and we found a similar $n_c^*$ for a linear spring function as for $f^\Theta(r)$ (data not shown).
As discussed in the appendix, Section \ref{app_sec_geometric_multiplicity},  the system can only support extended modes in the hydrodynamic limit if a cluster of overlapping particles percolates the system.  Thus, instead of asking if a finite speed of sound exists (see equation \eqref{Main_text_eq_dispersion_relation}) in this simplified model, we can also ask if overlapping spheres of  diameter $\sigma$ build a system-spanning cluster. This network of bonds is captured by the $\Theta$-spring function.   The value for the critical density predicted by the theory $n_c^* \approx 0.830 $ then lies close to the value   ${n}^*_p =0.838$ found for the percolation in e.g.~the three-dimensional Lorentz-gas \cite{PhysRevLett.96.165901}. The mapping onto the ERM model in Sect.~\ref{sec_Connection_to_ERM} predicts a dependence of  $n_c^*$ on the functional form of the spring function. To reconcile the numerical finding  presumably requires a more accurate calculation of the vertex.

	\paragraph{Vibrational density of states}
	To consider the vDOS close to the un-jamming transition, we start from Eqs.~\eqref{eq16} and \eqref{eq:aux} to obtain the expression 
	\begin{align}
		\label{eq_Susceptibilty_via_ciritcal_m}
		\begin{split}
			\hat{\chi}_q(s)&=  \frac{\hat{K}_q^\perp(s)}{s}=  \frac{1}{s^2} \frac{s \varphi_q(s)}{s \varphi_q(s)+q^2}\; ,
		\end{split}
	\end{align}
	and investigate the two cases $\omega \ll \omega_*$ and $\omega \gg \omega_*$ with $s=-i \omega+0^+$ separately: 
	\begin{itemize}
		\item[1)] \textbf{$\omega \ll \omega_*$}. In this case, 
		the system is an elastic medium. Thus, the vDOS in this frequency regime follows Debye's law \eqref{eq_Debye_Spectrum_DOS} and is determined by the speed of sound given in Eq.~\eqref{Main_text_eq_dispersopn_relation_critical_eigenvector}.  The Debye frequency $\omega_D =k_D v_0^\perp \propto \sqrt{\epsilon}$ vanishes at the critical point. This is again the same scaling which has been found in simulations \cite{Ikeda_Phonon_transport}.

		\item[2) ] \textbf{$\omega \gg \omega_*$}. In this case, the Eq.~\eqref{Main_text_eq_AnsatzFried} and the first line of Eq.~\eqref{asymptotes} lead  in the leading order of $\sigma$ to
		\begin{align}
			s \varphi_q(s) = \sigma s_*h_q= - i \sigma \frac{\omega}{\omega_*}  h_q
		\end{align}
		The general relation for the vDOS, Eq.~\eqref{eq_General_equation_for_vDOS}, and Eq.~\eqref{eq_Susceptibilty_via_ciritcal_m} for the susceptibility give the asymptotic result:
		\begin{align}\label{eq:constantDOS}
			\begin{split}
				D(\omega \gg \omega_*)
				\approx  \frac{t_*  }{n 2^{d-2} \pi^{\frac{d}{2}+1} \Gamma\left(\frac{d}{2}\right)}  \int_0^{k_D} dq q^{d-3} h_q
			\end{split}
		\end{align} 
		Thus, the vDOS is constant for $\omega_* \ll \omega \ll \ \omega_{BP}$, where $\omega_{BP}$ denotes an upper frequency limit where the asymptotic analysis based on Eq.~\eqref{Main_text_eq_scaling_equation} breaks down. The value of the plateau is included in Fig.~\ref{fig_tansition_from_left_and_right} and does not depend on the distance to the critical point $\epsilon$. 
	\end{itemize}

The inset of Fig.~\ref{fig_tansition_from_left_and_right} additionally shows $\omega_*$ as obtained from the vDOS $D(\omega)$ calculated by numerical diagonalization. Here we estimated $\omega_*$ as $D(\omega_*/\omega_0)\omega_0>0.17$; see App.~\ref{app:ERM} for details. As predicted by Eq.~\eqref{eq:constantDOS}, we can linearly extrapolate $\omega_*$ approaching the transition from the jammed and the un-jammed state. However, due to the finite size of our system $\omega_*$ never vanishes completely. We thus exclude the values $\omega_*$ close to the transition from the linear fits and consider $\omega_*$ as the most imprecise method to calculate $n^*_c$. The final estimate for the critical density $n^*_c = 0.80(6)$ is the average over the four linear regressions extrapolating $\lambda_-^\perp$, $v_0^\perp$ and $\omega_*$, where the two results of $\omega_*$ are counted as one weight.

\paragraph{Rayleigh damping} 
	Lastly, we discuss the divergence of the Rayleigh-damping in the jammed phase (we set $\epsilon<0$).  Looking at equation \eqref{eq_main_text_Rayleigh_Damping_small_w_Ikeda},  one expects a divergence of the  sound attenuation $ \propto 1/\omega_*^{d/2}$. The damping results from sound modes losing their momentum to the disorder as captured in the non-analytic self-energy (Eq.~\ref{eq_Main_text_Rayleigh_damping_general_equation}). We checked that this singularity is compatible with the un-jamming singularity, and recovered the amplitude. 
 While this agrees with simulation  \cite{Ikeda_Phonon_transport}, the divergence observed there is stronger, $\propto 1/\omega_*^d$. It is unclear whether this is connected to the quasi-localized modes present in the particle based simulation.
	
	\section{Conclusion}

	Relying on the Zwanzig-Mori projection-operator formalism, a microscopic self-consistent model for the emergence of enthalpic rigidity was constructed. We started with the definition of macroscopic rigidity as the non-decaying ability to sustain long-wavelength shear forces \cite{ALEXANDER199865}. Hence, we looked at the correlation function of transverse displacements or rather their time derivatives \textit{i.e.}~the transverse velocities. We found that the macroscopic stability of the system depends of the maximal eigenvalue of a stability matrix and is hence determined by the energy landscape. This is in accordance with earlier works \cite{ALEXANDER199865,  DeGiuli2014, Mean_field_Paris_2018}. We confirmed this idea by testing our equations  by solving the scalar ERM-model, which exhibits a percolation-like transition if one chooses a function of finite extent to model the particle interactions. Our theory shows  agreement with the numerical solution of the $\Theta$-ERM model and predicts a critical density close to the know percolation value. Furthermore, our theory   predicts salient features of the jammed and of the athermal glassy phase. In the un-jammed state cooperative velocity fluctuations are correlated of a diverging length  \cite{Olsson2007,Heussinger2009,Saitoh2020}.
	Importantly, we constructed our theory based on a Langevin approach. But the emergence of rigidity is independent of the friction parameter $\xi$. To capture transport processes in the un-jammed state requires structural rearrangements and thus to go beyond our approximation of a frozen-in structure.  
	
We argue that one has to go beyond the  standard self-consistent Born approximation by including non-planar contributions. Otherwise, the theory would not capture the Rayleigh damping of sound or the correct critical dynamics.  This important finding is in accordance with the discussions of the Lorentz-gas \cite{Leutheusser,PhysRevLett.96.165901} and of the ERM-model \cite{Vogel_ERM, grigera2011high}. We solved the issue by introducing a retarded renormalized vertex $\matr{\mathcal{V}}$. We had to perform a total of four Zwanzig-Mori steps to obtain its constituting equation. The end of this sequence of projections was indicated by discovering the bare propagator of the jammed phase and by uncovering, where the self-consistent Born-approximation went  wrong.

	The theory  predicts the same critical dynamics close to the un-jamming transition as found in \cite{Ikeda_Phonon_transport} except for the divergence of the Rayleigh-Damping. A reason for this could be, that we look at the attenuation of a sound mode, but the stronger damping found in \cite{Ikeda_Phonon_transport} resulted from other vibrational modes, namely quasi-localized modes \cite{NonpnononicSpectrum, SCHOBER2011}. But these modes are conceptualized to have a vortex-like spatial structure \cite{ Lerner_2021, schirmacher2023nature}.  Such vortex-like structures couple different degrees of freedom and are hence likely to be missed by our purely transverse approach. We will generalize our theory to incorporate longitudinal modes shortly.

	Since we included non-planar contributions into the expression for the memory kernel, the positivity of the elastic constants and the fact that the solution for $K_q(t)$ is a  bounded monotonically decaying function are not necessarily conserved by our approximations. Nevertheless, first numerical results look indeed promising. We postpone a analytical investigation to a future work. 
	
	\section*{Acknowledgments} We thank Grzegorz Szamel, Annette Zippelius,  and Walter Schirmacher for fruitful discussions.   The work was supported by the Deutsche Forschungsgemeinschaft (DFG) via SFB 1432 project CO6.

	\setcounter{equation}{0}
	\setcounter{table}{0}
	\setcounter{section}{0}
    \appendix
	\makeatletter

	\section{Technical aspects in the derivation of the self-consistent model \label{app_Derivation_Model}}
	We consider $N$ point-particles with positions $\{\7r\}_{i}^N$ and momenta
	$\{\7p\}_{i}^N$ (with identical mass $m$)  in a volume $V$ in the thermodynamic limit $N,V \to \infty$ (so that the density $n=N/V$ stays constant). The state of the system is specified by the N-particle phase space distribution
	$\Psi(\Gamma,t)=\Psi(\{\pmb{r}\}_i^N,\{\pmb{p}\}_i^N,t)$, 
	whose time evolution is
	given by the Klein-Kramers equation, $\partial_t\Psi=\Omega\Psi$, with:
	\begin{align}\label{ap1}
	\begin{split}
		\Omega(\Gamma) = & \sum_i \left( \frac{\partial H}{\partial \7r_i} \cdot \frac{\partial}{\partial \7p_i}- \frac{\partial H}{\partial \7p_i}  \cdot \frac{\partial}{\partial \7r_i}   \right)
		\\&+ \zeta_0 \sum_{i}   \frac{\partial}{\partial \pmb{p}_i} \cdot\left( k_B T \frac{\partial}{\partial \pmb{p}_i}+ \frac{\partial H}{\partial \7p_i} \right) \;.
			\end{split}
	\end{align}
	It is a Fokker-Planck equation generalizing Liouville's equation by including dissipation and random forces. Thermal energy $k_BT$ sets the noise scale. 
	For simplicity, the friction coefficient $\zeta_0$ is local, independent on particle positions \cite{Dhont}.
	Further, $H(\{\pmb{r}\}_i^N,\{\pmb{p}\}_i^N)$ denotes the Hamiltonian energy function.
	A radial pair potential is considered so that the force on particle $i$ equals: $\7F_i=-\frac{\partial}{\partial \7r_i} H = -\sum_{j\ne i}\frac{\partial}{\partial \7r_i} U(|\7r_i-\7r_j|)$.
	
	An externally imposed inhomogeneous (solvent) velocity field $\7v^{ex}(t)$   (taken to be a plane wave with wavevector $\7q$) couples to the particle momenta and perturbs the systems.   The energy changes by a linear coupling to the momentum field $m \7v(\7q)= \sum_{j=1}^N \7p_j e^{- i \7q \cdot \7r_j}$
	\begin{align}
		\label{app_eq_Change_Energy}
		\delta H = - \frac{m}{V} \7v(\7q )^* \cdot  \7v^{ex}(t)\;.
	\end{align}
	Here, we relied on a spatial Fourier transformation with the wavevector $\7q$. Later, we will also employ temporal Laplace and Fourier transformation, with the Laplace frequency $s$ and the (Fourier) frequency $\omega$. We use on the conventions 
	\begin{align}
		\label{app_eq_Conventions_Troafos}
		\begin{split}
			\text{FT}[f(\7r)](\7q)&= \int_{V} d \7r e^{-i \7q \cdot \7r} f(\7r) \;, \\
			\text{L} [g(t)](s ) &= \int_{0}^{ \infty}  d t  e^{-st  } g(t)  \equiv \hat{g}(s)\;, \hspace{0.5cm}    \text{Re}   \{s\}>0\;,
		\end{split}
	\end{align} 
	Since we exclusively work in the wavevector domain, the spatially transformed function is simply indicated by the argument $\7q$. 
	Let $\delta \Omega(t)$ denote  the change in the time-evolution operator which follows from Eqs. \eqref{ap1} and \eqref{app_eq_Change_Energy}. An adiabatic perturbation, $\7v^{ex}(t) = \7v^{ex}_0 e^{\epsilon t }$ for $ t < 0$ (with 
	$\epsilon > 0$ an infinitesimal constant) and $v^{ex}(t > 0) = 0$, shall be applied assuming that the system initially was
	in a stationary distribution $\psi_0.$  It is taken in a way that $\ln \Psi_0 \to - H/(k_BT)$ holds $T\to0$ in the spirit of Graham's discussion of the weak noise limit of Fokker-Planck equations \cite{Graham1984}. Then detailed balance holds \cite{Risken}. The shift in the distribution function can be obtained, and from it averages linear in $\7v^{\rm ex}_0$.  To simplify the notation already now, only transverse modes will be considered. This requires a transverse external field, $\7q\cdot\7v^{\rm ex}(t)=0$, and introduces the transverse velocity field $ \7v^\perp(\7q,t)$, with
	$ \7v^\perp(\7q)= \hat{\7q}\times(\7v(\7q)\times\hat{\7q})$ (here $\hat{\7q}=\7q/q$). The  relaxation of a transverse momentum fluctuation is then given by the normalized momentum correlation tensor
	\beq{ap3}
	\begin{split}
	\boldsymbol{K}^\perp(\7q,t)&=
	\frac{ \int  d\Gamma \; \7v^\perp(\7q)\;  e^{\Omega t}\; \7v^\perp(-\7q)\; \Psi_{0}}{\frac{1}{(d-1)} \langle |\7v^\perp(\7q)|^2\rangle} \\&= \frac{ \braket{\7v^\perp(\7q,t) \7v^\perp(-\7q)}}{\frac{1}{(d-1)} \langle |\7v^\perp(\7q)|^2\rangle},
		\end{split}
	\eeq
	where the integral extends over the whole phase space and defines the average $\langle \rangle$. The factor  $(d-1)$ counts the transverse directions in $d$ dimensions.  From $\Psi_0$, one obtains 
	\beq{apn1}
	\langle{\rm v}_\alpha(\7q) {\rm v}_\beta(-\7q)\rangle = (Nk_BT/m)\;  \delta_{\alpha\beta}\;.
	\eeq
	Isotropy simplifies the matrix form to: $\boldsymbol{ K}^\perp(\7q,t)={\rm K}^\perp_q(t)  ( {\bf 1} - \hat{\7q}\hat{\7q})$, where the scalar transverse momentum response function is given by: ${\rm K}^\perp_q(t)= \frac{1}{(d-1)} {\rm Tr}\{\boldsymbol{ K}^\perp(\7q,t)\}$; here $\rm Tr$ denotes the trace.  At solidification, momentum correlations change because of the conversion of dissipative stresses into elastic ones. This can be exposed employing the Zwanzig-Mori (ZM) formalism as it allows to identify slow dynamics. We utilize the operator identity 
	\begin{align}
		\label{eq_ZM_EQM:Operator_Identity}
		\Big[s-P\Omega P - P\Omega Q\hat{R}_Q(s) Q \Omega P \Big] P\hat{R}(s)P=P\;,
	\end{align}
	for the Laplace transformed  resolvent $\hat{R}(s)= [s-\Omega]^{-1}$ to express the system's dynamics in terms of the reduced resolvent $\hat{R}_Q(s)= Q [s-Q \Omega Q]^{-1} Q$. Here, $P$ and $Q=1-P$ are two projection operators.   We start by projecting  onto the transverse velocity field,
	\beq{ap4}
	P_v= \ket{\7v^\perp(-\7q) } \cdot  \bra{\7v^\perp(\7q)}\frac{m}{Nk_BT }\,.
	\eeq
	This introduces the transverse force fluctuations via Newton's equation. It can be formulated with the adjoint of the operator \eqref{ap1}, $\dot{\7v}=\Omega^\dagger \7v$ \cite{Risken} :
	\beq{ap5}
	m  \dot{\7v}(\7q)  = \sum_j \Big[\frac{-i \7q\cdot \7p_j}{m} \7p_j+ 
	{\7F_j} - \frac{\zeta_0}{m}  {\7p_j} \Big] e^{-i\7q\cdot \7r_j}\;.
	\eeq
	The first two terms  can be identified with the kinetic and potential contribution to the divergence of the stress tensor, respectively.  It is a force density    $\7F(\7q)=-i\7q\cdot\matr{\sigma}(\7q)$ \protect\cite{HansenMcDonald,Wajnryb1995}. Here and throughout this work a tensor is represented by an underbar. Generally, spatial indices are displayed by greek letters.   The kinetic contribution, formally included, in $\7F(\7q)$ plays no role at $T=0$. The third term $-\zeta_0\7v(\7q)$ is due to the local frictional forces  and breaks momentum conservation. 
	The  transverse velocity
	auto-correlation tensor has been computed by Hess and Klein~\cite{HessKlein}.  The result given as Laplace transform can be found in Eq.~\eqref{eq_Transverse_velocity_Autocorrelation}, with the microscopic definition of the shear modulus  given by  
	\beq{ap6}
	\begin{split}
	\boldsymbol{\rm G}^\perp_q(t) &=   {\rm G}^\perp_q(t) \;   ( {\bf 1} - \hat{\7q}\hat{\7q}) \\&=    \frac{1}{Nmk_BT q^2 }\,    
	\langle \7F^\perp(\7q)\;  e^{Q_1\Omega Q_1 t} \; \7F^\perp(-\7q) \rangle \; .
		\end{split}
	\eeq
	It is determined by the ZM reduced resolvent, $R_1(t) = e^{Q_1\Omega Q_1 t}$, with $Q_1=1-P_1$. Note that because of time-reversal symmetry $P_1 \7F(\7q)=0$ holds. 
	Using the explicit form of $\Psi_0$ (which gives $\7F_j A\Psi_0=-k_BT\Psi_0 \nabla_j A$ for arbitrary variable $A$), the expression of the instantaneous  modulus (viz.~${\rm G}^\perp_q(t=0)=(c^\perp_q)^2$) in terms of the pair correlation function $g(r)$ established by Zwanzig and Mountain \cite{Zwanzig1961} is recovered; only the potential part survives at $T=0$ and gives the high frequency  transverse velocity: 
	\begin{align}
		\label{ap10}
		(c^\perp_q)^2 &=   \frac{n}{m}\int d\7r\; \frac{1-e^{-iqy}}{q^2} g(r) \bigg[  \frac{r^2-x^2}{r^3} U'(r)+\frac{x^2}{r^2} U''(r)\bigg]\;,
	\end{align}

	Note, that $(c_q^\perp)^2$ scales with $n$ as it arises from pair interactions.  The time-derivative of the  stress enters in an additional ZM projection step using the projector: 
	\beq{ap8}
	P_2 =  \frac{\ket{\7F^\perp(-\7q)} \cdot \bra{ \7F^{\perp}(\7q)}}{ \frac{1}{d-1} \braket{|\7F^\perp(\7q)|^2}}\;, 
	\eeq
	where the normalisation is given by: 
	\beq{ap9}
	\frac{1}{d-1} \braket{|\7F^\perp(\7q)|^2}  =  N m k_BT\;  (q\,c^\perp_q)^2\;.
	\eeq
	
	The ZM-step introduces an instantaneous damping rate 
	\beq{ap11}\gamma^\perp_q = \frac{\frac{-1}{(d-1)} {\rm Tr}\{
		\langle \7F^{\perp}(\7q) \Omega  \7F^\perp(-\7q) \rangle \}}{\frac{1}{(d-1)}  \langle |\7F^{\perp}(\7q)|^2 \rangle} = \frac{\xi k_BT}{m( c^\perp_q)^2}\hspace{0.4cm}
	\eeq
    with  $\xi = \frac{\zeta_0}{m}$. 
    and  leads to a new transverse memory-kernel $\7W^\perp(\7q,t)={\rm W}^\perp_q(t)  (  \mathbb{1} - \hat{\7q}\hat{\7q})$ with reduced dynamics $R_2(t) = Q_2 e^{\Omega_2 t} Q_2$ containing $\Omega_2=Q_2\Omega Q_2$ and  $ Q_2=1-P_1-P_2$. Note, that this projector takes out the linear fluctuations of velocity,  and stress at wavevector $\7q$.  We could additionally project out states which are linear in the density. But this does not make a difference in the following. One has
	\begin{align}
	\label{ap12}
	\begin{split}
	\hat{\rm G}^\perp_q(s) &= \frac{(c^\perp_q)^2}{s+ \gamma^\perp_q+ (c^\perp_q)^2\, \hat{\rm W}^\perp_q(s)} \\ 
	{\bf W}^\perp_q(t) &= {\rm W}^\perp_q(t) \;   ( {\bf 1} - \hat{\7q}\hat{\7q}) \\&= \frac{-1}{ Nmk_BT q^2 (c^\perp_q)^4} \,  \langle \7F^{\perp}(\7q) \Omega \, R_2(t) \, \Omega \7F^\perp(-\7q) \rangle 
	\end{split}
	\end{align}
	
	Since the kinetic part of the stress tensor gives only rise to contributions vanishing with the temperature, we only consider the potential  stresses $\sigma_{\alpha \beta}^{(p)}$ in the  present athermal case.   The potential stresses and its time derivative can be written with a pair  of density fields or a density and a velocity field respectively  \cite{Schmid_PHD_Theses}
	\begin{align}
		\begin{split}
			\ket{\sigma_{\alpha \beta}^{(p)}(\7q)}  &= \frac{1}{V} \sum_{\7\kappa} X_{\alpha\beta} (\7\kappa,\7q) \ket{ \rho(\7q-\7\kappa) \rho(\7\kappa)}\;, \\
			\ket{\dot{ \sigma}_{\alpha \beta}^{(p)}(\7q)}  &= \frac{-2i }{V} \sum_{\7\kappa} \Re\{ X_{\alpha\beta} (\7\kappa,\7q)\} \ket{ \rho(\7q-\7\kappa) ( k_\eta v_\eta (\7\kappa))}
		\end{split}
	\end{align}
	with the weights $$X_{\alpha \beta}(\7\kappa,\7q)= - \frac{1}{2}
	\int d\7r e^{i\7\kappa\cdot \7r} \frac{r_\alpha r_\beta}{r} U'(r) \frac{1-e^{-i\7q\cdot \7r}}{ i\7q\cdot \7r}$$
	 obeying the symmetry $\matr{X}(\7k,\7q)=\matr{X}^*(\7q-\7k,\7q)$ and importantly staying finite at the origin $X_{\alpha \beta}(0,0)=  \frac{1}{2} \int d\7r  \frac{r_\alpha r_\beta}{r} U'(r) \propto \delta_{\alpha \beta}$. Generally, we use Einstein' sum convention for the spatial indices.  The super-scripted ${}^*$ denotes the complex conjugate. We abbreviated the spatial derivative in the definition of the weights $U'(r)= \partial_r U(r)$. It is important to notice, that the weights do not depend on any phase space variable and can hence be pulled out of any average.  Due to the coupling  to  the density, 	$\ket{\dot{ \sigma}_{\alpha \beta}^{(p)}(\7q)}$   has a finite overlap with the transverse and longitudinal velocity field or respectively with  the product state of velocity and density fluctuation:
	 \begin{widetext}
	\begin{align}\label{eq:translong}
		\begin{split}
			&\braket{\dot{ \sigma}_{\alpha \beta}^{(p)}(\7q), \delta \rho(\7k-\7q) v_\gamma(-\7k) }\\& =   -  \frac{2i}{V} \sum_{\7\kappa}  \kappa_\eta \Re  \{X_{\alpha \beta} (\7\kappa,\7q) \} \Big( \braket{ \rho(\7q-\7\kappa) v_\eta(\7\kappa) v_\gamma^\perp(-\7k)\delta \rho(\7k-\7q)} +\braket{ \rho(\7q-\7\kappa) v_\eta(\7\kappa) v^\parallel_\gamma(-\7k)\delta \rho(\7k-\7q)} \Big) \\ &=  -2i \frac{NV k_B T }{m} \sum_{\7\kappa}  \kappa_\zeta \Re \{X_{ \alpha \beta } (\7\kappa,\7q) \} \Big[  (\delta_{\gamma  \zeta}-\hat{k}_\gamma \hat{k}_\zeta)  \braket{ \rho(\7\kappa-\7k)  \rho(\7q-\7\kappa) \delta \rho(\7k-\7q)} + \hat{k}_\gamma \hat{k}_\zeta   \braket{ \rho(\7\kappa-\7k)  \rho(\7q-\7\kappa) \delta \rho(\7k-\7q)} \Big]\;.
		\end{split}
	\end{align}
		 \end{widetext}
	The density fluctuation couples the longitudinal and transverse component of the velocities.  In the following, we neglect  the second term in the square bracket, which represents the coupling to the longitudinal fields. The reason is, that we aim for a self-consistent theory for the transverse fields. We build an enthalpic theory for the stability at the bottom of the potential energy landscape.   Thus, we assume the structure  to be time independent and hence do we only take transverse modes into consideration. The generalisation with incorporated longitudinal modes is straight forward and work in progress.
	
	All of this  suggests the next projector to be  built with density  and transverse momentum
	\beq{ap13}
	P_3 =   \sum_{\7k}  \frac{ \ket{ \7v^\perp(-\7k) \delta\varrho(\7k-\7q)}  \cdot   \bra{ \7v^{\perp}(\7k) \delta\varrho(\7q-\7k)} }{NS_{|\7q-\7k|} \frac{1}{(d-1)}  \braket{|\7v^\perp(\7k)|^2}    }  \; .
	\eeq
 Here $S_q=\langle| \delta\rho(\7q)|^2\rangle/N$ is the equilibrium structure factor.  It is connected to the pair correlation function via: $S_q=1+n \int d\7r e^{-i\7q\cdot\7r}(g(r)-1)$ \cite{HansenMcDonald}. Note, that the structure factor is calculated with the density fluctuations $\delta \rho(\7q)= \rho(\7q) - \braket{\rho(\7q)}= \rho(\7q) - N \delta_{\7q,0}$.  We use the projector $P_2$ to express the fluctuating $ \ket{\dot{\7\sigma}(\7q)}$ by its overlap with structure and time-dependent displacements. This introduces the vertex 
	\begin{align}
		\matr{V}_{\7q,\7k}= \frac{-\braket{\7F^\perp(\7q) \Omega Q_2 \delta \rho(\7k-\7q) \7v^\perp(-\7k) }}{ N  k_B T q (c^\perp_q)^2\, S_{|\7q-\7k|}}\;.
	\end{align}
	The adjoint of this vertex is given by $\matr{V}^\dagger_{\7k,\7q}=\matr{V}_{-\7q,-\7k}=\matr{V}_{\7q,\7k}$ 
	The vertex is further analysed and approximated in section \ref{app_sec_Vertex_Evaluation} of this appendix.  It turns out to be convenient to express also the high frequency transverse velocity via the static structure factor. Only considering the potential part of the stresses, one finds
		\begin{align}
	\label{eq_app_Calculation_c_q^perp}
			\begin{split}
				q^2 (c_q^\perp)^2=2 \frac{(\delta_{\nu \gamma } -\hat{q}_\gamma \hat{q}_\nu)}{(d-1)Vm} q_\mu  \sum_{\7\kappa}  \kappa_\nu \Re  \{X_{\mu \gamma} (\7\kappa,\7q) \}  S_{|\7q-\7\kappa|}
			\end{split}
	\end{align}

	We   express the fluidity 
	with a frequency-dependent retarded renormalized vertex

	\begin{align}
		\label{app_eq_Fluidity_mit_RM}
		\begin{split}
		\hat{W}_q^\perp(s) &= \frac{1}{ (d-1)} \, \text{Tr}\left\{\;  \frac 1N\, \sum_{\7k}\;  \matr{V}_{\7q,\7k}  \cdot \hat{\matr{\mathcal{V}}}_{\7q,\7k}(s) \;\right\} \\&= \frac 1{N}\, \sum_{\7k}\;  \matr{V}_{\7q,\7k}  \, : \, \hat{\matr{\mathcal{V}}}_{\7q,\7k}(s)\;.   
				\end{split}
	\end{align}
	Its constituting equation reads
	\begin{align}
		\begin{split}\label{app_eq_Def_Rernormalisef_Vertex} 
		\hat{\matr{\mathcal{V}}}_{\7q,\7k}&(s) =   \sum_{\7p} \frac{m}{N^2 k_B T} \\ & \times \braket{ \7v^\perp(\7k)  \delta \rho(\7q-\7k) \hat{R}_2(s)\, \delta \rho(\7p-\7q) \7v^\perp(-\7p) }\cdot  \matr{V}^\dagger_{\7p,\7q}\;.
		\end{split}
	\end{align}
	with the  resolvent $\hat{R}_2(s) = Q_2 ( s - Q_2 \Omega Q_2 )^{-1} Q_2$ obtained by Laplace-transformation. Also note the symbol $:$ which connecting two tensors renders a scalar.

	The Zwanzig-Mori equation for the occurring 4-point function can be re-written to a constituting equation for the renormalized vertex 
	\begin{align}
	\begin{split}
		(s + \xi )& \; \hat{\matr{\mathcal{V}}}_{\7q,\7k }(s) +	\sum_{\7p}  \hat{\matr{\Sigma}} (\7q,\7k,\7p,s) \cdot \hat{\matr{\mathcal{V}}}_{\7q,\7p }  (s)\\&= S_{|\7q-\7k|}  \cdot \matr{V}^\dagger_{\7k,\7q}\;.
		\label{eq:S17}
			\end{split}
	\end{align}
	The new memory function is a $d \times d $ matrix 

	\begin{align}
		\begin{split}
		&\hat{\matr{\Sigma}} (\7q,\7k,\7p,s)\\&\hspace{0.5cm}= -m \frac{\braket{ \7v^\perp(\7k) \delta \rho(\7q-\7k)  \Omega_2 \, \hat{R}_{3}(s)  \, \Omega_2  \delta \rho(\7p-\7q) \7v^\perp(-\7p) } }{N^2 k_B T S_{|\7q-\7p|}}  \\&\hspace{0.5cm}
			\approx    \frac{\braket{ \7F^\perp(\7k) \delta \rho(\7q-\7k)  \, \hat{R}_{3}(s)   \,  \delta \rho(\7p-\7q) \7F^\perp(-\7p) }}{N^2 m  k_B T S_{|\7q-\7p|}}  \;.
		\end{split}
	\end{align}

	In the last line, we again neglected any contributions arising from the longitudinal current. The reduced dynamics in the memory function is specified  by $R_3(t) = Q_3 Q_2  e^{\Omega_3 t} Q_2 Q_3$ with $Q_3=1-P_3$ and  $\Omega_3= Q_3 Q_2 \Omega Q_2 Q_3$.  Note ,that $\hat{\matr{\Sigma}} (\7q,\7k,\7p,s)$ is a complicated functional of the six-point density correlation
	function,  since  we continue to  consider potential stresses only. \\

Again,  considering $T = 0$, we assume the structure factor to be static,
	\textit{i.e.} $S_q( t) \approx S_q( t = 0)$. Consequently, the basic time dependent observable is the
	velocity-autocorrelation. Thus, we need to go to the next Zwanzig-Mori level, where the occurring  memory kernel again depends on the velocities.  Hence, we have to perform one last Zwanzig-Mori step with the projector
	\begin{align}
		P_4=  \sum_{\7k}  \frac{\ket{ \7F^\perp(-\7k) \delta \rho ( \7k-\7q)} \cdot   \bra{ \delta \rho(\7q-\7k)\7F^\perp(\7k)} }{ N^2 m k_B T (k c_k^\perp)^2 S_{|\7q-\7k|}} \; .
	\end{align}
	where we approximated the overlap by its diagonal contribution 
\begin{align}
    \begin{split}
       & \frac{1}{d-1}\braket{ |\7F^\perp(\7k) \delta \rho(\7q-\7k)  |^2 } \approx  N^2 m k_B T (k c_k^\perp)^2 S_{|\7q-\7k|} \;.
    \end{split}
\end{align}
	This gives the Zwanzig-Mori equation for $\hat{\matr{\Sigma}}$ in terms of a new memory kernel $\hat{\matr{\Xi}}$ 
	\begin{align}
		\label{eq_ZM_EQM_last_transverse_meory_function}
		\begin{split}
		s &\, \hat{\matr{\Sigma}}(\7q,\7k,\7p,s )  +
		\sum_{\7b} \;   \hat{\matr{\Xi}} (\7q,\7k,\7b,s) \cdot \hat{\matr{\Sigma}}(\7q,\7b,\7p,s ) \\& =(k c_k^\perp)^2 (\matr{1}-\hat{\7k} \hat{\7k} )\; \delta _{\7k,\7p} \;.
				\end{split}
	\end{align}
	Here, we have neglected   a non-zero
	frequency matrix, which  enters the Zwanzig-Mori equation for the $\matr{\Sigma}$-memory function and  which
	is proportional to the Langevin-Damping coefficient $\xi$. This contribution is
	also a hydrodynamic kind of damping $\matr{\Tilde{\gamma}}(\7q,\7k,\7p)$, which   vanishes with $k_B T$ and is hence indeed negligible for $T \to 0$.  The frequency matrix reads
	\begin{align*}
		\begin{split}
			\Tilde{\matr{\gamma}}(\7q,\7k,\7p) &= - \frac{\braket{\7F^\perp(\7k) \delta \rho(\7q-\7k)  \Omega_3 \delta \rho(\7p-\7q) \7F^\perp (-\7p)} }{N^2 m k_B T (k c_k^\perp)^2 S_{|\7q-\7k|} } \\=& \frac{\xi k_BT}{m(c_k^\perp)^2} \Big(  \matr{\delta}^\perp_{\7k,\7p}  S_{|\7q-\7k|}  -   \frac{ \braket{\matr{Y}(\7k,\7q)}\braket{ \matr{Y}(-\7p,-\7q) }}{m  k^2}  \Big)\;,
		\end{split}
	\end{align*} 
	with $\matr{Y}(\7k,\7q) =   \frac{(\mathbb{1}-\hat{\7k}  \hat{\7k})}{N c_q^\perp } \cdot \sum_i \nabla_i e^{-i\7k\cdot \7r_i} \delta  \rho(\7q-\7k) \7F^\perp(-\7q) $.
	
	Abbreviating the reduced resolvent by $\hat{R}_4(s)= Q_4 Q_3 Q_2 ( s - \Omega_4 )^{-1} Q_2 Q_3 Q_4$ with $\Omega_4 = Q_4\Omega_3Q_4$ and $Q_4=1-P_4$  gives
	\begin{align}
		\begin{split}
			 &\hat{\matr{\Xi}} (\7q,\7k,\7p,s) \\& = - \frac{\braket{ \7F^\perp(\7k) \delta \rho(\7q-\7k) \Omega_3 \, \hat{R}_{4}(s) \,  \Omega_3 \delta \rho(\7p-\7q) \7F^\perp(-\7p)} }{N^2m  k_B T (pc_p^\perp)^2 S_{|\7q-\7p|}} 
		\end{split}
	\end{align}
	This is our last exact result.  Now, we aim to get closure and to express $\hat{\matr{\Xi}}$ in terms of the transverse  velocity auto-correlation $K^\perp_q$, the static structure factor $S_q$ and the bare vertex $\matr{V}_{\7q,\7k}$.
		
	We start  again by only considering the potential part of the stress tensor. Additionally, we approximate $Q_4 \Omega_3  \ket{   F_\alpha } \approx 0$. This approximation is justified since  the time derivative of the fluctuating forces can be expressed via    the current density-fluctuations. After neglecting the longitudinal part of the current, one arrives   at
		\begin{align}\begin{split}
			Q_3 Q_2 \Omega Q_2 & \ket{\delta \rho (\7p-\7q) \7F^\perp(-\7p)} \\& \approx \ket{ Q_3 \delta \rho (\7p-\7q) (\Omega \7F^\perp(-\7p)) } \;.
				\end{split}
		\end{align} This gives 
		\begin{widetext}
		\begin{align}
			\label{eq_Decomposition_Transverse_zweite_Sigma_memory_function}
			\begin{split}
				\hat{\matr{\Xi}} (\7q,\7k,\7p,s) \approx &  \frac{- \bra{ \delta \rho(\7q-\7k) [\Omega^\dagger \7F^\perp (\7k)  } Q_3 R_4(s) Q_3 \ket{  \delta \rho (\7p-\7q) [\Omega \7F^\perp (-\7p)]   }}{	 N^2m  k_B T (pc_p^\perp)^2 S_{|\7q-\7p| }} \\
				\approx &  -  \frac{ \bra{  (\Omega^\dagger \7F^\perp (\7k) )  }  \otimes \bra{\delta \rho(\7q-\7k) }Q_3 R_4(s) Q_3 \ket{  \delta \rho (\7p-\7q)  } \otimes \ket{ (\Omega \7F^\perp (-\7p))} }{	 N^2m  k_B T (pc_p^\perp)^2 S_{|\7q-\7p| }} \\
				=&- \frac{ \bra{  (\Omega^\dagger \7F^\perp (\7k) )  } (Q_3+P_3) \otimes \bra{\delta \rho(\7q-\7k) }Q_3 R_4(s) Q_3 \ket{  \delta \rho (\7p-\7q)  } \otimes(Q_3+P_3)  \ket{ (\Omega \7F^\perp (-\7p))}}{ N^2m  k_B T (pc_p^\perp)^2 S_{|\7q-\7p| }} \\  \underset{\matr{\sigma}=\matr{\sigma}^{(p)}}{=}&  -  \frac{m k (c_k^\perp)^2 }{N^2  k_B T  p  S_{|\7q-\7p|}} \sum_{\7k',\7p'}    \matr{V}_{\7k,\7k'}  \braket{   \7v(\7k')      \delta \rho(\7k-\7k')   \delta  \rho(\7q-\7k) Q_3 R_{4}(s) Q_3  \delta \rho(\7p-\7q) \delta \rho(\7p'-\7p) \7v(-\7p') }  \cdot  \matr{V}^\dagger_{\7p',\7p}   \\
				\approx &  -
				\frac{(c_k^\perp)^2}{ N } \sum_{\7p'}    \matr{V}_{\7k,\7p'} \cdot  K_{p'}^\perp(s)   \Big[ S_{|\7k-\7p'|}\delta_{\7k,\7p}+ \frac{k}{p} S_{|\7q-\7k|} \delta_{\7k-\7p',\7q-\7p}\Big] 
				\cdot  \matr{V}^\dagger_{\7p',\7p}   
			\end{split}
		\end{align}
			\end{widetext}
	We  assume in the second line, that the three point state is a product state. This  is indicated by the sign $\otimes$. A pair state $\ket{AB}$ is a called a product state $\ket{A} \otimes \ket{B}$ if and only if $\braket{AB,(AB)^*}=\braket{A,A^*}\braket{B,B^*}$ holds.  The six-point correlation function was  decomposed  into the product of pair-correlators and they were  dressed in the last line. In this step, we also relied on translation
		invariance to get rid of one integral: $ \braket{\7v^\perp(\7k')| R(s)|  \7v^\perp(-\7p')} \propto \delta_{\7k',\7p'}$. Additionally, translation	invariance leads to the occurrence of the Kronecker-deltas in the square bracket.

		If we neglect the second term in the square bracket in \eqref{eq_Decomposition_Transverse_zweite_Sigma_memory_function}, we would perform the diagonalization step in the standard mode-coupling procedure \cite{pihlajamaa2023unveiling, Gotze} and reproduce a qualitatively wrong approximation, since it leads to the incorrect sound attenuation and wrong predictions for the critical dynamics \cite{Leutheusser1983, ciliberti2003brillouin, goetschy2013euclidean, grigera2011high, Vogel_ERM}. Such an approximation is equivalent to setting  $\hat{\matr{\Sigma}}(\7q,\7k,\7p,s) \approx \matr{\delta}^\perp_{\7k,\7p} k^2G_q^\perp(s)$ and $\hat{\matr{\Xi}}(\7q,\7k,\7p,s) \approx W_q^\perp \matr{\delta}_{\7k,\7p}^\perp$.  We call this a  self-consistent Born approximation \cite{Altland_Simons_2010, Kamenev_2011}.  
		
	For readability, we write   
	\begin{align}
		\label{eq_Decomposition_Transverse_zweite_Sigma_memory_function_Result}
		\begin{split}
	&\hat{	\matr{\Xi}} (\7q,\7k,\7p,t) \\&\approx (c_k^\perp)^2 \left\{  \hat{ {M}}_k(s)\,  \matr{\delta}_{\7k,\7p}^\perp +   \frac 1N  \Tilde{\matr{V}}(\7q,\7k, \7p)     K^\perp_{|\7k+\7p-\7q|}(t) \right\}\;,
	\end{split}
	\end{align}
	with  
	\beq{mct2}
	\begin{split}
		M_k(t) &=    \frac 1N \sum_{\7b}  \; \matr{V}_{\7k,\7b} : \matr{V}_{\7b,\7k}^\dagger  \;   S_{|\7k-\7b|} \; K^\perp_b(t) \;. \\
	\Tilde{\matr{V}}(\7q,\7k, \7p)  &  = \frac{k}{p} \matr{V}_{\7k,\7p+\7k-\7q} \cdot  \matr{V}_{\7k+\7p-\7q,\7p}^\dagger S_{|\7q-\7k|}\;.
	\end{split}
	\eeq
	After multiplying the  equation \eqref{eq_ZM_EQM_last_transverse_meory_function} with the renormalized vertex $\Tilde{\matr{\mathcal{V}}}_{\7q,\7p}(s)$ and integrating over  $\7p$, one finds with Eq.~\eqref{eq_ZM_EQM_last_transverse_meory_function}  (identical to Eq.~\eqref{eq:S17})
		\begin{widetext}
		\begin{align}
		\label{app_eq_RV_First_expression}
		\begin{split}
			\hat{W}_q^\perp(s) &=   \frac 1{N}\, \sum_{\7k}\;  \matr{V}_{\7q,\7k}  \, : \, \hat{\matr{\mathcal{V}}}_{\7q,\7k}(s)  \\
			\sum_{\7p} \left[	\Big(s(s+\xi) +  p^2(c_p^\perp)^2  \Big)  \delta_{\7k,\7p} + (s+\xi) \;	\hat{\matr{\Xi}} (\7q,\7k,\7p,s) \right] \cdot \hat{\matr{\mathcal{V}}}_{\7q,\7p}(s)    &= \sum_{\7p} \left[	s  \delta_{\7k,\7p} + 	\hat{\matr{\Xi}} (\7q,\7k,\7p,s) \right] \cdot \matr{V}^\dagger_{\7p,\7q}    S_{|\7q-\7p|}  
			\;.
		\end{split}
	\end{align} 
		\end{widetext}

	The equations \eqref{eq_Transverse_velocity_Autocorrelation}, \eqref{ap12}, \eqref{eq_fluidity_transverse_def}, 
	and \eqref{app_eq_RV_First_expression} constitute our self-consistent model.   Since we did not perform the standard MCT diagonalization step  when inserting the two-point projection operator $P_3$ in the equation for the fluidity  \eqref{app_eq_Fluidity_mit_RM}, a few explanatory comments are in order: 
	\begin{itemize}
		\item[a)] The derived model goes beyond standard MCT, by explicitly not diagonalizing the matrix element occurring in the first equation for the fluidity \eqref{app_eq_Fluidity_mit_RM}. Our new approach may be considered a \textit{non-planar self-consistent current response theory}. The notion \textit{non-planar} has its origin in a diagrammatic interpretation of our equations, which becomes applicable in an high density approximation deep in the jammed state. We comment on this is the sections  \ref{sec_Connection_to_ERM} and \ref{app_sec_ERM}.
		\item[b)] The sequence of Zwanzig-mori steps was stopped, when it became clear, where the standard MCT goes wrong. \textit{I.e}, the correct level is indicated by where the standard MCT-approximation scheme becomes equivalent to neglecting a (crucial) term. Also, that we have found the bare propagator $[s(s+\xi)+q^2 (c_q^\perp)^2]^{-1}$ of a short time or high density expansion at this level, suggests that we have included a sufficient  sequence of consecutive ZM-steps. 
		
		\item[d)] The essential inclusion of the essential non-diagonal correction terms, leads to the weights of the memory function $W_q$ not being necessarily positive anymore \cite{Gotze}. Hence, the positivity of the spectra of the occurring correlation function is also not guaranteed.  Nevertheless, the numerical results suggest that this is indeed the case, but a rigorous proof is still missing.
		\item[e)] The symmetry of the bare vertex  (see section  \ref{app_sec_Vertex_Evaluation})  $\matr{V}_{\7k,\7q}^\dagger=\matr{V}_{-\7q,-\7k}=\matr{V}_{\7q,\7k} $ implies $\underset{\7q \to 0}{ \text{lim}} \hat{\matr{\Xi}} (\7q,-\7k,-\7p,s) = \underset{\7q \to 0}{ \text{lim}} \hat{\matr{\Xi}} (\7q,\7k,\7p,s) $.  Together with the property $ \matr{V}_{-\7k,0}^\perp = - \matr{V}_{\7k,\70}^\perp $ (again, see section  \ref{app_sec_Vertex_Evaluation}), this gives   the symmetry of the renormalized vertex $ \hat{\matr{\mathcal{V}}}_{0,-\7k}(s) = - \hat{\matr{\mathcal{V}}}_{0,\7k}(s) $ \cite{Vogel_ERM}. This useful property ensures the correct prediction of Rayleigh-damping. 
	\end{itemize}

	\section{Jammed phase \label{app_sec_Jammed_Phase}}
Transverse stresses	can not decay  in jammed phase. This gives a finite  long time limit of the shear modulus $G_q^\perp(t \to \infty )= \lim_{s \to 0} s \hat{G}^\perp_q(s)$. Hence,  a susceptibility  $\hat{\chi}_q(s)=  \frac 1s \hat{K}^\perp_q(s)$ arises along with a self-energy (Eq.~ \eqref{susceptibility}).  $\hat{\chi}_q(s)$ and $\hat{\sigma}_q(s)$ are  evaluated with the time-integrated vertex $\matr{\hat{\mathcal{U}}}_{\7q,\7k}(s) = \frac 1s \hat{\matr{\mathcal{V}}}_{\7q,\7k}(s)   $. 
	Writing $\hat{\matr{m}}_k(s)= \frac{\hat{M}_k(s)(\mathbb{1}-\hat{\7k}\hat{\7k})}{s}$, equation  \eqref{app_eq_RV_First_expression} becomes
	\begin{widetext}
	\begin{align}
		\label{jammed}
		\begin{split}
			& \left[\Big(s(s+\xi) + k^2 (c_k^\perp)^2 \Big) + s(s+\xi)  \;	(c_k^\perp)^2 \hat{\matr{m}}_k(s) \right]  \cdot  \matr{\hat{\mathcal{U}}}_{\7q,\7k}(s) +  \;	\frac{ s(s+\xi) (c_k^\perp)^2}{N}  \sum_{\7p} \matr{\tilde{V}}(\7q,\7k,\7p)\cdot   \matr{\hat{\mathcal{U}}}_{\7q,\7p}(s)\;\hat{\chi}_{|\7q-\7k-\7p|}(s) =  \\
			=& \left[1 + (c_k^\perp)^2 \hat{\matr{m}}_k(s) \right]\cdot \matr{V}^\dagger_{\7k,\7q}S_{|\7q-\7k|} +	(c_k^\perp)^2
			\frac 1N  \sum_{\7p} \matr{\tilde{V}}(\7q,\7k,\7p) \cdot  \mV_{\7p,\7q}^\dagger \chi_{|\7q-\7k-\7p|}(s)  S_{|\7q-\7p|}
			\;.
		\end{split}
	\end{align}
		\end{widetext}
	\subsection{Dispersion-relation}  The residual transverse stresses lead to finite restoring forces in the long time limit, whose small wavevector limit defines the speed of sound. We introduce the transverse renormalized dispersion relation as $(v_q^\perp)^2 =G_q^\perp(t \to  \infty) = \lim_{s \to 0} s \hat{G}_q^ \perp(s)$. The speed of sound is then given by $v_0^\perp$. We solve our self-consistent model for $s=0$ in this paragraph and hence derive an equation for the transverse speed of sound.  
	
	Since the susceptibility stays finite for  vanishing frequency in the jammed state, the equation for the renormalized vertex can be exactly solved 
	\begin{align}
		\label{app_eq_time_integrated_renormalized_vertex}
		\begin{split}
		k^2 \matr{\hat{\mathcal{U}}}_{\7q,\7k}(s=0)&=  \matr{V}^\dagger_{\7k,\7q}S_{|\7q-\7p|} +  \hat{\matr{m}}_k(s) \cdot  \matr{V}_{\7k,\7q}^\dagger S_{|\7q-\7k|} \\ +
		\frac 1N  \sum_{\7p}&  \matr{\tilde{V}}(\7q,\7k,\7p) \cdot  \matr{V}^\dagger_{\7p,\7q} \chi_{|\7q-\7k-\7p|}(s) S_{|\7q-\7p|}
		\end{split}
	\end{align}
	Thus, the time integrated fluidity at $s=0$ can be expressed as a linear integral equation
	\begin{align}
		\label{app_eq_integrated_fluidity_linear_integral}
		\begin{split}
		\hat{w}_q(s=0)& =  \frac{1}{N} \sum_{\7k} \matr{V}_{\7q,\7k} : \matr{\hat{\mathcal{U}}}_{\7q,\7k}(s=0)\\&= \hat{\Delta}_q(s=0) +  \sum_{p} p^{d-3}  C_{q,p}  \frac{1}{ (v_p^\perp)^2}\;. 
			\end{split}
	\end{align}
	with the first order or high frequency contribution  $\hat{\Delta}_q(s) $ and the stability matrix $p^{d-3}C_{q,p}$
	\begin{subequations}
		\begin{align} 
			\label{eq_app_one_loop_cotribution}
		& \hat{\Delta}_q (s)   = \frac{1}{N}  \Bigg\{\sum_{\7k  }    \matr{V}_{\7q,\7k}  : \matr{V}_{\7k,\7q}^\dagger  \frac{S_{|\7q-\7k|}}{s^2+  k^2 (c_k^\perp)^2}  \cdot   \Bigg\}  \\
			\label{eq_app_Def_stability_matrix} 
			\begin{split}
				& C_{q,p}=  \frac{1}{N^2(d-1)}  \int d^{d-1} \hat{p}   \text{Tr} \Bigg\{\sum_{\7k  } \frac{1}{k^2} \matr{V}_{\7q,\7k}   \cdot  \matr{V}_{\7k,\7p} S_{|\7q-\7k|}  \\ &   \cdot  \Big[ \matr{V}_{\7p,\7k}^\dagger    \cdot   	S_{|\7k-\7p|}  \matr{V}_{\7k,\7q}^\dagger  +   \matr{V}_{\7p,\7p+\7q-\7k,}^\dagger\cdot 	 \frac{kS_{|\7k-\7p|}  }{|\7p+\7q-\7k| }    \matr{V}_{\7p+\7q-\7k,\7q}^\dagger  \Big] \bigg\}
			\end{split}
		\end{align}
	\end{subequations}
	We define the coupling coefficients  occurring in the equations \eqref{app_eq_integrated_fluidity_linear_integral},\eqref{eq_app_Def_stability_matrix} and later in equation \eqref{app_eq_long_time_fluidity_linear_integral} by integrating over the angular degrees of freedom $\hat{\7p}$.  This leads to 
	\begin{align}
		\label{app_eq_dispersion_relation}
		\begin{split}
			&(v_q^\perp)^2=   \frac{(c_q^\perp)^2}{1+(c_q^\perp)^2 \hat{w}_q(0)} \\
			\Longleftrightarrow & \sum_p  \Big( \delta_{q,p}- p^{d-3} C_{q,p} \Big) \frac{1}{(v_p^\perp)^2} =  \frac{1}{(\Tilde{c}_q^\perp)^2}\;.
		\end{split}
	\end{align}
	with the first order renormalised bare sound velocity $\hat{\Delta}_q(s)+ \frac{1}{(c_p^\perp)^2} =\frac{1}{(\Tilde{c}_q^\perp(s))^2} $  and  $\frac{1}{(\Tilde{c}_q^\perp(0))^2}\equiv\frac{1}{(\Tilde{c}_q^\perp)^2}$. 
	Equation \eqref{app_eq_dispersion_relation} is only guaranteed to have  a unique, non zero, solution, if the maximal eigenvalue of the stability matrix $ p^{d-3} C_{q,p}$ is smaller than unity. There are two main reasons why the small eigenvalue scenario corresponds to the jammed case: First of all, the jammed state corresponds to very high densities. The integrated fluidity vanishes with $1/n$ in the infinitely densely packed system. Secondly,  the speed of sound would be negative if the jammed state corresponded to all eigenvalues being larger than one. The eigenvalues are controlled by the density correlation functions. The solution of our model close to the transition is analysed in chapter \ref{sect:crit}.  It is important to notice, that the stability of the system is independent of $\xi$. \\

	Rotational invariance implies  $ C_{0,p \to 0}  \propto p^2 $ and $ C_{q\to 0,0 }  \propto q^2 $, since $C_{0,0}=0$ holds, as we will show in the next subsection. It is noteworthy, that the equation for the fluidity in the jammed state basically reduces to the F1-model, which is extensively studied in the literature \cite{Gotze}.\\
	
	\subsection{Rayleigh-damping \label{App_Paragraph_Proof_Rayleigh}}  
	We set $\xi=0$ in this section, in order to study the sound attenuation induced by the disorder. 
	Evaluating the self-energy on the imaginary axis $\sigma_q(s=-i \omega + 0^+)$ leads to an imaginary part in the inverse susceptibility, which, in the time domain, gives rise to an exponential decay of the correlation function. Hence, this imaginary part is associated with the damping. This non-trivial sound attenuation arises from a non-analyticity  of the propagator at the sound pole $\omega= v_q^\perp q$ for $\omega \to 0$. We prove in this section, that the damping is Rayleigh-like, \textit{i.e} $\Im\{ \sigma_q(s=-i \omega+0^+)\} \propto \omega q^4$ in $d=3$ for $\omega \to 0$, by expanding the time integrated fluidity in powers in $s$ for $s \to 0.$ The zeroth order gives the dispersion relation. As we will see, the next leading order is  in accordance with Rayleigh-damping. \\  
	
	For $s =-i \omega +0^+ \to 0$, the imaginary part of he self-energy can be written as
	\begin{align}
		\Im\{\hat{\sigma}_q(s \to 0)\}\Big|_{ s=-i \omega +i0^+} =q^2 (v_q^\perp)^4  \Im\{ \hat{w}_q(s=-i \omega +0^+)\}\;.
	\end{align}
	One can identify three different contributions to the residual self-energy $\hat{\sigma}_q(s)-\hat{\sigma}_q(0)$: 
	\begin{itemize}
		\item[i)] The first contribution arises from the Born-term of the self-energy, which dominates deep in the jammed phase. Writing \begin{align}
		\label{app_eq_First_order_Matrix_C0}
		   C^{(0)}_{q,k}=  \frac{1}{N}  \int d^{d-1} \hat{k}    \matr{V}_{\7q,\7k} : \matr{V}_{\7k,\7q}^\dagger   S_{|\7q-\7k|}\;,  \end{align} the residual contribution reads in the thermodynamic limit 
		\begin{align}
			\begin{split}
				& \hat{w}_q^{(1)}(s \to 0 )-\hat{w}_q^{(1)}(0) \\& = \lim_{s \to 0 } \int_0^\infty dk k^{d-1} C^{(0)}_{q,k} \left[ \frac{1}{s^2+k^2 (c_k^\perp)^2} - \frac{1}{k^2 (c_k^\perp)^2} \right]\;.\\ 
				&  \approx  \lim_{s \to 0 } \int_0^\infty dk k^{d-1} C^{(0)}_{q,k} \left[ \frac{1}{s^2+k^2 (c_0^\perp)^2} - \frac{1}{k^2 (c_0^\perp)^2}\right] \\
				& \overset{x=\frac{k {c_0}^{\perp}}{s}}{ = }   \frac{1}{s^2} \left( \frac{s}{c_0^\perp}\right)^d \int_0^\infty dx x^{d-1} C^{(0)}_{q,s \frac{x}{v_0^\perp}}  \left[ \frac{1}{1+x^2}- \frac{1}{x^2}\right] \\& \propto \frac{\beta^{1,1}}{(k_D c_0^\perp)^d}s^{d-2}q^2 + \frac{\beta^{1,2}}{(k_D c_0^\perp)^d} \frac{s^d}{(c_0^\perp)^2} \;.
			\end{split}
		\end{align}
		In the second line did we use, that the small $k$ regime mainly contributes to the residual part of the self-energy. In the last line did we rely on the behaviour of the vertex $C^{(0)}_{q \to 0, 0} \propto q^2$ and $C^{(0)}_{q =0, k \to 0} \propto k^2$. Hence, the Born term indeed gives rise to Rayleigh-damping $ \sigma_q^{(0)}(s \to 0) \propto q^4 s^{d-2}$ around the sound pole $\omega= q v_q^\perp$.  We added the Debye-wavevector to render the coefficients unitless scalars. One has $[C_{q,p}^{(0)}]=[C_{q,p}]=[q]^{2-d}=q^2/k_D^d]$. This gives $[\hat{w}_q(s)]= [\frac{q^2}{w^2}]$ and this is consistent with the requirement $[(c_q^\perp)^2 \hat{w}_q(s)]=1.$

		\item[ii)] The second contribution arises from the the self-consistent term with the stability matrix determining the weights. For $q,s \to 0$ and one finds 
		\begin{align}
			\label{eq_app_Damping_Coeff}
			\begin{split}
			 &	\lim_{q \to 0 } \Big[	\hat{w}_q^{(2)}(s \to 0 )-\hat{w}_q^{(2)}(0) \Big]  \\& = \lim_{q,s \to 0 } \int_0^\infty dp p^{d-1} C_{q,p} \left[ \frac{1}{s^2+p^2 (v_p^\perp)^2} - \frac{1}{p^2 (v_p^\perp)^2} \right]\\ 
				&  \lim_{q,s \to 0 } \int_0^\infty dp p^{d-1} C_{q,p} \left[ \frac{1}{s^2+p^2 (v_0^\perp)^2} - \frac{1}{p^2 (v_0^\perp)^2} \right]\\ 
				& \overset{x=p \frac{ {v_0}^{\perp}}{s}}{ = }   \frac{1}{s^2} \left( \frac{s}{v_0^\perp}\right)^d \int_0^\infty dx x^{d-1} C_{q,s \frac{x}{v_0^\perp}}  \left[ \frac{1}{1+x^2}- \frac{1}{x^2}\right] \\&  \propto \frac{\beta^{2,1}}{(k_D v_0^\perp)^{d}}s^{d-2}q^2 +\frac{\beta^{2,2}}{k_D^d(v_0^\perp)^{d+2}}s^{d}\;.
			\end{split}
		\end{align}
		Here, the dimensionless coefficients depend on the number density $n^*$  as $\frac{1}{n^*}$.
		The previous equation holds, since the stability matrix vanishes in the hydrodynamic limit  due to the cancellation of the planar and non-planar terms:
		\begin{align}
		\begin{split}
			C_{0,0}& =  \frac{1}{N^2(d-1)}  \int d^{d-1} \hat{p}   \text{Tr} \Bigg\{\sum_{\7k  } \frac{1}{k^2} \matr{V}_{0,\7k}   \cdot  \matr{V}_{\7k,0} S_{k}   \\&  \cdot  \Big[ \matr{V}_{0,\7k}^\dagger    \cdot   	S_{k}  \matr{V}_{\7k,0}^\dagger  +   \matr{V}_{0,-\7k}^\dagger\cdot 	   S_{k}   \matr{V}_{-\7k,0}^\dagger  \Big] \bigg\}=0\;.
					\end{split}
		\end{align} 
		This proves that the contribution is at least Rayleigh-like.  It is important to notice, how crucial the inclusion of the non-planar terms is: A standard MCT-approximation which involves a diagonalization already when determining the fluidity, and which is essentially a self-consistent Born approximation or a planar theory,  predicts  a hydrodynamic damping $\propto q^2s $ instead of $q^4s$ in $d=3.$ \\
		
		Note, that we have set  $ \frac{(c_k^\perp)^2}{s^2 +k^2 (c_k^\perp)^2} \approx \frac{1}{k^2}$. This approximation is justified since the contribution of this term to the residual self-energy is Rayleigh-like anyways. This can be proven in similar fashion to point [i)].

		\item[iii)] The last contribution arises from 
		\begin{align}
		    \hat{w}_q^{(3)}(s) = \frac{1}{N} \sum_{\7k} \matr{V}_{\7q,\7k}  : \matr{\hat{\mathcal{U}}}^{(3)}_ {\7q,\7k}(s)
		    		\end{align}
		    with  
		\begin{align}
		\begin{split}
			& \matr{\hat{\mathcal{U}}}^{(3)}_{\7q,\7k}(s) 	 = -  \frac{s^2}{k^2} \matr{m}_k(s) \cdot \matr{\hat{\mathcal{U}}}_{\7q,\7k}(s)   \\&  -    \frac{s^2}{k^2N}
			\sum_{\7p} \matr{\tilde{V}}(\7q,\7k,\7p)\cdot
			\matr{\hat{\mathcal{U}}}_{\7q,\7p}(s) \chi_{|\7q-\7k-\7p|}(s) \;.  
					\end{split}
		\end{align}
		Expanding $  \matr{\hat{\mathcal{U}}}_{\7q,\7p}(s) $ on the right hand side by inserting equation \eqref{app_eq_time_integrated_renormalized_vertex} and then performing again the steps [i)], [ii)] and [iii)] leads immediately to the conclusion, that the contribution to the self-energy is at least of order $\mathcal{O}(q^2 s^{d})$ and hence in accordance with Rayleigh-damping. 
		
	\end{itemize}
	
	The calculations (i),(ii) and (iii) prove that the sound attenuation is in any dimension in accordance with Rayleigh-damping but the displayed small $s$ expansion allows only the calculation of the coefficients for $2<d<4$.  One can rely on  the Sokhotski–Plemelj theorem in Eq.~\eqref{eq_main_text_Sokhotski_Plemelj_theorem} to calculate 
	the damping in arbitrary dimensions $d>2$ for $s=-i \omega +0^+$:  
	\begin{align}\begin{split}
	 \Gamma_q^\perp(s& =-i \omega + i 0^+) = - \Im \bigg\{ \frac{(q c_q^\perp)^2}{1+(c_q^\perp)^2 w_q^\perp(-i \omega+0^+)}\bigg\} \\& \underset{\omega \to 0}{= }   q^2 (v_q^\perp)^4 \frac{\pi}{2}   \Bigg[ \frac{C^{(0)}_{q,\omega/c_0^\perp}}{(c_0^\perp)^d} +\frac{C_{q,\omega/v_0^\perp}}{(v_0^\perp)^d} \Bigg] \omega^{d-2}\;.
	 	\end{split}
	\end{align}
	The coefficients  can now be easily obtained by expanding the weights $\matr{C}^{(0)}$ and  $\matr{C}$. 
	\begin{align}
	\begin{split}
	    \beta^{(2,1)} &= k_D^d  \lim_{q,p\to 0} \frac{1}{q^2} C_{q,p}\;,\\ \hspace{1cm}  \beta^{(2,2)} &= k_D^d \lim_{q,p\to 0} \frac{1}{p^2}C_{q,p} \;. 
	    	\end{split}
	\end{align}
	The equations for $\beta^{(1,1)}$ and $\beta^{(1,2)}$ read analogously. \\
	
	\subsection{ F1-model \label{sec_app_F1_model}} 
	
	Since the transition is indicated by the system becoming soft and hence by a vanishing speed of sound, we  can neglect the contributions of $w_q^{(3)}(s)$. As it was shown, this does neither effect the prediction of Rayleigh-damping nor affects the dispersion relation quantitatively. $\hat{w}_q^{(3)}(s)$ becomes non-negligible only for higher frequencies. Thus, we can propose a simple self-consistent model, which is valid in the low frequency regime and which is  mathematically  equivalent to the well studied F1-model \cite{Gotze, Gotze1981}: 
	\begin{align}
		\begin{split}
			\hat{w}_q(s)&=\hat{\Delta}_q(s)+\hat{w}_q^{(2)}(s) \\
			&=  \hat{\Delta}_q(s) +  \sum_{p} p^{d-1} C_{q,p}   \hat{\chi}_p(s)\;.
		\end{split}
	\end{align}
	Since  the vertices $\matr{V}_{\7q,\7k}$  can be calculated from the  pair-potential $U(r)$ and density correlation functions, this self-consistent model can potentially be solved numerically and compared to experiments.  Since it its numerically quite costly to determine the stability matrix, this is postponed to a future work.

\section{Mapping to the field theory of the Euclidean Random matrix model \label{app_sec_ERM}}
	
	The euclidean random matrix model introduced by Mezard, Parisi and Zee \cite{M_zard_1999} has gained a lot of attention in the last 25 years \cite{Vogel_ERM,M_zard_1999,Martin_Mayor_2001,ciliberti2003brillouin,grigera2011high, Ganter_Schirmacher, Ganter_2011_Diagrams,schirmacher2019self}.  The model is about investigating particles  harmonically oscillating around random and fixed positions. Thus, it is a simplistic model for stable topological disorder. The system is specified by the random matrix $\matr{\mathcal{H}}$ which constitutes the potential energy 
	\begin{align}
		U= \frac 12 \sum_{i,j} \phi_i \matr{\mathcal{H}}_{ij} \phi_j  \equiv m \frac{\omega_0^2}{4} \sum_{i,j} f(\7r_i-\7r_j)(\phi_i-\phi_j)^2\;.
	\end{align}
	Here, $\phi_i$ denote  small fluctuations  around the frozen positions $\7r_i$. Note, that the vector character of the displacements is neglected. But still these scalar fluctuations have the dimension of a displacement.  It is   requested   that the
	Fourier transformation $f(\7q)$ of the unitless spring function exists. Further, we   assume rotational invariace, which implies  that $f$ only depends on the absolute value of the wavevector $q = |\7q|$. Lastly, we assume that the spring
	function is regular. All in all,  this implies $f(0)-f(\7q) \propto q^2 $. Note, that the ERM-model is supposed to describe a Newtonian system. Thus, we set $\xi=0$ in this section.    All the proposed solutions of the ERM-model, known by the authors, assume a uniform distribution of the particles. This amounts to $S_q=1$. Performing a high density expansion, the susceptibility can be written as
	\begin{align} 
	\begin{split}
		\hat{\chi}^{ERM}_q(s) & = \overline{\sum_{i,j} e^{-i \7q \cdot (\7r_i-\7r_j)} \left[ \frac{1}{s^2 + \matr{\mathcal{H}}} \right]_{ij}} \\& = \frac{1}{s^2- V_{\7q,\7q}^{ERM} +\sigma_q^{ERM}(s)}\;.
	\end{split}
	\end{align}
	The supersrcipted \textit{ERM} indicates that the quantities refer to the ERM model. The susceptibility or Green's function $\hat{\chi}^{ERM}_q(s)$ is here the resolvent of the random matrix   $\matr{\mathcal{H}}$. The overline means averaging over the disorder, \textit{i.e.} the random positions.  A self-consistent Born approximation was proposed in \cite{ciliberti2003brillouin}. The proposed equation for the self-energy reads 
	\begin{align}
	\label{eq_app_ERM_Grigera}
		\hat{\sigma}_q(s)^{SCB,ERM} =- \frac{1}{n} \int \frac{d^d \7k}{(2 \pi)^d} V^{ERM}_{\7q,\7k} \hat{\chi}_k^{ERM}(s) V^{ERM}_{\7q,\7k}\;.
	\end{align}
Note, that we consider the thermodynamic limit $V,N \to \infty$ and $\sum_{\7k} \to \frac{V}{(2 \pi)^d} \int d^d \7k$ through out this section, being in accordance with the earlier works on the ERM model.
	The model defined in equation \eqref{eq_app_ERM_Grigera} fails to predict Rayleigh-damping. One finds in the hydrodynamic limit: $\Im\{\sigma_q(s=-i \omega +0^+)^{SCB,ERM}\} \propto \omega^{d-2} q^2 $. Another self-consistent theory was proposed in \cite{Vogel_ERM} by two of the present authors, which takes non-planar contributions into account. The self-energy reads 
	\begin{subequations}
		\label{eq_Main_text_PRL_theory_I}
		\begin{align}
			& \hat{\sigma}_q(s)^{VF,ERM}=- \frac{1}{n} \int \frac{d^d \7k}{(2 \pi)^d} V_{\7q,\7k}^{ERM} \hat{\chi}_k^{(0)}(s) \hat{\mathcal{U}}_{\7q,\7k}^{ERM}(s)\;,\\
			\begin{split}
				 & \bigg [1 - \frac{1}{n} \int \frac{d^d \7p}{(2 \pi)^d}  V^{ERM}_{\7k,\7p}   \hat{\chi}^{ERM}_{\7p}(s)  V^{ERM}_{\7k,\7p} \hat{\chi}_k^{(0)}(s)  \bigg] \hat{\mathcal{U}}^{ERM}_{\7q,\7k}(s) \\ & \approx  		 V^{ERM}_{\7q,\7k}  +    
				\frac{1}{n} \int \frac{d^d \7p}{(2 \pi)^d} 
				V^{ERM}_{\7k,\7p+\7k-\7q}  \hat{\chi}^{ERM}_{|\7q-\7p-\7k|}(s)\\& \hspace{1.4cm}\times   V^{ERM}_{\7p,\7q-\7p-\7k} \hat{\chi}_p^{(0)}(s) \hat{\mathcal{U}}^{ERM}_{\7q,\7p}(s)  \;.
			\end{split}
		\end{align}
	\end{subequations}
	with the bare propagator $\chi_q^{(0)} (s) = \frac{1}{s^2 - V_{\7q,\7q}^{ERM}}$. This model predicts the correct sound attenuation. There are different models proposed in \cite{Ganter_Schirmacher}. But here,  the authors dismantle the vertex $V_{\7q,\7k}^{ERM}$ in more elementary parts before summing up parts of them again. Since its structure is hence not easily accessible, we leave this work aside here.  We will show in the following how the two self-consistent solutions for the self-energy $ \hat{\sigma}_q(s)^{VF,ERM}$ and $ \hat{\sigma}_q(s)^{SCB,ERM}$ arise in an high density or small integrated fluidity approximation.

	Hereinafter, we look at the susceptibility $\hat{\chi}_q(s)$ in the high density limit, were $c_q^\perp$ becomes large.  The self-energy (Eq.~\ref{Main_text_eq_self_energy}) approximately reads  
	\beq{highdens_app}
	\hat{\sigma}_q(s) \approx -  q^2 (c_q^\perp)^4 \hat{w}_q^\perp(s) \;.
	\eeq
	We start by defining  the scalar ERM-vertex ${V}^{ERM}_{\7q,\7k}= \frac{q (c_q^\perp)^2}{d-1} {\rm Tr}\{\matr{V}_{\7q,\7k}\}$. The vertex vanishes with $a q^2 + b\7q \cdot \7k$ for $\7q \to 0$ and has all the properties of the general vertex used in ERM investigations \cite{grigera2011high, Vogel_ERM}. Starting from equation \eqref{app_eq_Exact_Expression_Vertex} and utilising  Einsteins sum convention one finds
	\begin{align}
		\begin{split}
			V_{\7q,\7k}^{ERM}= & 2   \frac{q_\gamma (\delta_{\alpha \tau}-\hat{q}_\alpha \hat{q}_\tau)  }{mV(d-1)  }  \sum_{\7\kappa}  \kappa_\zeta  \Re \{X_{\tau \gamma} (\7\kappa,\7q) \}   (\delta_{\alpha  \zeta}-\hat{k}_\alpha \hat{k}_\zeta) \\& \bigg[ \frac{\braket{\delta \rho(\7\kappa-\7k) \delta \rho(\7q-\7\kappa) ,\delta \rho(\7q-\7k)}}{NS_{|\7q-\7k|}}  -  S_{|\7q-\7\kappa|}  \bigg] \;.
		\end{split}
	\end{align}
	Thus, we can express the high frequency 
	transverse velocity $c_q^\perp$ via the scalar ERM-vertex
	\begin{align} 
		\begin{split}
			V_{\7q,\7q}^{ERM} &= \frac{-2(\delta_{\nu \gamma } -\hat{q}_\gamma \hat{q}_\nu)}{(d-1)Vm} q_\mu  \sum_{\7\kappa}  \kappa_\nu \Re  \{X_{\mu \gamma} (\7\kappa,\7q) \}  S_{|\7q-\7\kappa|}\\&=-(qc_q^\perp)^2\;.
		\end{split}
	\end{align}
	This holds due to the fact of 
	$\delta \rho(0)=0$. As a consequence, we recover the bare propagator of  the ERM-theory  $\chi_q^{(0)}(s)=\Big[s^2 -  V_{\7q,\7q}^{ERM} \Big]^{-1} $. To relate our theory to the Euclidean Random Matrix model, we   express the vertex as well with  a structure independent scalar and isotropic function 
	\begin{align} \label{springf_app}
		V^{ERM}_{\7q,\7k} \longrightarrow 	 n^*\omega_0^2 \frac{f(|\7k|)-f(|\7q-\7k|)}{f(q=0)}  \;.
	\end{align}

	Similarly to the bare vertex, we set  $ q (c_q^\perp)^2 \hat{ \matr{\mathcal{U}}}_{\7q,\7k} \to \hat{\chi}_k^{(0),\perp}(s)S_{|\7q-\7k|} \hat{\mathcal{U}}^{ERM}_{\7q,\7k} ( \mathbb{1} -\hat{\7q} \hat{\7q})$. Here, we also \textit{pulled} out a bare propagator and a static structure factor of the renormalised vertex. This allows the diagrammatic representation given in the main text. See  equation \eqref{vertex_ERM_final}. The associated Feynman rules read: \begin{align}
		\label{Feynmann-rules}
		\begin{split}
\begin{tikzpicture}
					\draw[propagator] (0,0) -- (0.85,0)node at (0.425,0.25) {$\7k$};
			\end{tikzpicture} \equiv  & \chi^{\perp}_q(s) \;, \;\;
			\hbox{\begin{tikzpicture}
					\draw[particle] (0,0) -- (0.85,0)node at (0.425,0.25) {$\7q$};
			\end{tikzpicture}}\equiv   \chi^{(0),\perp}_q(s) \;, \;\;
			\hbox{\begin{tikzpicture}
					\draw[photon] (0,0) -- (0.85,0) node at (0.425,0.25) {$\7k$};
			\end{tikzpicture}} \equiv  S_k \;,\\ \vspace{0.15cm}
			\vcenter{	\hbox{\begin{tikzpicture}
						\draw[photon] (0.0,0.61) 	node at (0.5,0.85) {$\7q-\7k$} arc(90:11.9:0.61);
						\draw[] (0,0) -- (0.5,0)  node at (0.25,-0.25) {$\7k$};
						\node at (0.61,0) [rectangle,draw] {};
			\end{tikzpicture}}  }&\equiv \matr{\mathcal{U}}_{\7q,\7k}(s) \;, \;\;
			\vcenter{	\hbox{\begin{tikzpicture}
						\draw[photon] (0.0,0.0) 	node at (0.4,0.85) {$\7q-\7k$} arc(178.1:90:0.61);
							\draw[] (-0.5,0) -- (0.6,0) node at (-0.25,-0.25) {$\7q$};
						\draw[] (0,0) -- (0.6,0) node at (0.25,-0.25) {$\7k$};
						\fill[](0,0) circle (0.1);
			\end{tikzpicture}} }  \equiv  \matr{V}_{\7q,\7k} \;,\; \;
			\hbox{\begin{tikzpicture}
					\draw[dashed] (0,0) -- (0.85,0);
			\end{tikzpicture}} \equiv  1 \;.
		\end{split}
	\end{align} The self-energy is constituted by all irreducible diagrams. Drawing them, one has to keep in mind that the 
	momentum is conserved at every vertex. One has to integrate over the internal momenta. \\

	Multiplying  equation \eqref{eq_ZM_EQM_last_transverse_meory_function} with the renormalized vertex   and integrating over  $\7p$  leads to
	\begin{widetext}
	\begin{align}
		\label{eq_Transverse-Equation_Sigma}
		\begin{split}
			\int \frac{d^d \7p}{(2 \pi)^d}\hat{\matr{\Sigma}} (\7q,\7k,\7p,s) & \cdot ( \mathbb{1} -\hat{\7q} \hat{\7q}) \chi_{p}^{(0)}(s)S_{|\7q-\7p|} \hat{\mathcal{U}}^{ERM}_{\7q,\7p}(s) = \frac{1}{s}k^2 (c^\perp_k)^2 \chi_{k'}^{(0)}(s)S_{|\7q-\7k'|} \;  ( \mathbb{1} -\hat{\7q} \hat{\7q}) \;  \hat{\mathcal{U}}^{ERM}_{\7q,\7k}(s)  \\& - \frac{1}{n}
			\int \frac{d^d \7p d^d\7b}{(2 \pi)^{2d}} {V}^{ERM}_{\7k,\7b}   \hat{\chi}_{\7b}(s)S_{|\7k-\7b|}  {V}^{ERM}_{\7k,\7b}   \frac{  	\hat{\matr{\Sigma}}( \7q,\7k, \7p,s) \chi_{p}^{(0)}(s)S_{|\7q-\7p|} \cdot ( \mathbb{1} -\hat{\7q} \hat{\7q})\; \hat{\mathcal{U}}^{ERM}_{\7q,\7p}(s) }{   k^2(c_k^\perp)^2}  \\ &-\frac{1}{n}
		\int \frac{d^d \7p d^d\7b}{(2 \pi)^{2d}}
			{V}^{ERM}_{\7k,\7b+\7k-\7q}  \hat{\chi}_{|\7q-\7b-\7k|}(s) S_{\7k-\7b} {V}^{ERM}_{\7p,\7q-\7b-\7k} \frac{  	\hat{\matr{\Sigma}}( \7q,\7b, \7p,s) \chi_{p}^{(0)}(s)S_{|\7q-\7p|} \cdot ( \mathbb{1} -\hat{\7q} \hat{\7q})\; \hat{\mathcal{U}}^{ERM}_{\7q,\7p}(s) }{  p^2(c_p^\perp)^2}  \;.
		\end{split}
	\end{align}
	\end{widetext}
	
	Since the numerators of the integral kernels  in the lines two and three of equation \eqref{eq_Transverse-Equation_Sigma} vanish at least  with $k^{d+1}$ and $p^{d+1}$, the small wavevector contributions to the integral are negligible. Also, the contribution of the sound pole $-s^2=p^2(c_p^\perp)^2$ in the third and forth line becomes small in the considered limit. The contribution to the damping is in accordance with Rayleigh-damping, as shown in section \ref{App_Paragraph_Proof_Rayleigh}. However, the contribution becomes small, due to the factor $1/(c_q^\perp)^2$.  Thus, the first term   dominates for $c^\perp_k \to \infty$  and  we find in this  high density approximation  
	\begin{align}
		\label{app_eq_renormalized_Transverse_Vertex_ERM}
		\begin{split}
			 	\int \frac{d^d \7p }{(2 \pi)^{d}} &  \hat{\matr{\Sigma}} (\7q,\7k,\7p,s)  \cdot   ( \mathbb{1} -\hat{\7q} \hat{\7q}) \chi_{p}^{(0)}(s)S_{|\7q-\7p|}  \hat{\mathcal{U}}^{ERM}_{\7q,\7k}(s) \\& \approx \frac{1}{s}k^2 (c^\perp_k)^2 \chi_{\7k}^{(0)}(s)S_{|\7q-\7k|}  ( \mathbb{1} -\hat{\7q} \hat{\7q})   \hat{\mathcal{U}}^{ERM}_{\7q,\7k}(s)\;.
		\end{split}
	\end{align}
	Having this, we arrive at a concise equation for the scalar renormalised ERM vertex 
		\begin{widetext}
	\begin{align}	    \begin{split}
&	\hat{\mathcal{U}}^{ERM}_{\7q,\7k} (s)  \approx   {V}_{\7q,\7k}+ \frac{1}{n}
			\int \frac{d^d\7p}{(2 \pi)^d}  {V}^{ERM}_{\7k,\7p} \hat{\chi}_{\7p}(s)S_{|\7k-\7p|}  {V}^{ERM}_{\7k,\7p}  \hat{\chi}_{\7k}^{(0)}  (s) V^{ERM}_{\7q,\7k}(s)  \\ &+ \frac{1}{n}
			\int \frac{d^d\7p}{(2 \pi)^d}
			{V}^{ERM}_{\7k,-\7q+\7p+\7k}  \hat{\chi}_{|-\7q+\7p+\7k|}(s) S_{|\7q-\7p|} {V}^{ERM}_{\7p,-\7q+\7p+\7k}\hat{\chi}_{\7p}^{(0)}(s) V^{ERM}_{\7q,\7p}(s) \;.
		\end{split}
	\end{align}
	\end{widetext}
		This equation is diagrammatically represented in the main text in equation
	\eqref{vertex_ERM_final}. 
	In the last line we can approximate $\frac{1}{k^2 (pc_p^\perp)^2} \matr{V}^{ERM}_{\7q,\7p} \approx \hat{\chi}_p^{(0),ERM} \hat{\mathcal{U}}_{\7q,\7p}^{ERM}(s)$. This becomes an exact approximation deep in the jammed state for small frequencies. Looking at the diagrammatic representation  \eqref{vertex_ERM_final}, this approximation means adding more loop-diagrams. Since every new loop comes with a factor $\frac{1}{n}$ this does neither quantitatively nor qualitatively  change the high density state. The contributions to the sound attenuation of the \textit{erroneously} introduced sound pole vanishes faster that the Rayleigh sound attenuation for $s,q \to 0.$ Meaning, the new pole  only leads to higher corrections of the sound attenuation for $s, q \to 0$.
	Thus, we find for the susceptibility of the transverse modes deep in the jammed state
	\begin{subequations}
		\label{app_Result_ERM_Approximation}
	\begin{align}
		\hat{\chi}_q(s)&= \frac{1}{s(s+\xi)+q^2(c_q^\perp)^2+ \sigma^{ERM}_q(s)}\;,\\ 
		\hat{\sigma}^{ERM}_q(s)&=-\frac{1}{n} \int \frac{d^d \7k}{(2 \pi)^d} V_{\7q,\7k}^{ERM}  \hat{\chi}_k^{(0)} S_{|\7q-\7k|} \hat{\mathcal{U}}^{ERM}_{\7q,\7k}(s)\;. 
	\end{align}
		\end{subequations}
		with the diagrammatic representation of the self energy
			\begin{align}
		\begin{split}
			\hat{\sigma}^{ERM}_q(s) \approx&  \;	
			\vcenter{\hbox{\begin{tikzpicture}
						\draw[particle] (0,0) -- (0.89,0);
						\fill[](0,0) circle (0.1);
						\node at (1,0) [rectangle,draw] {};
						\draw[photon] (0,0) arc(180:11.9:0.5) ;
			\end{tikzpicture}}}\;  \\ 
			\hbox{\begin{tikzpicture}
					\draw[photon] (0.0,0.5) arc(90:11.9:0.5);
					\draw[] (0,0) -- (0.39,0);
					\node at (0.5,0) [rectangle,draw] {};
			\end{tikzpicture}}=&
			\hbox{\begin{tikzpicture}
					\draw[photon] (0.0,0.5) arc(90:11.9:0.5);
					\draw[] (0,0) -- (0.39,0);
					\fill[](0.5,0) circle (0.1);
			\end{tikzpicture}}+  \;	  \hbox{\begin{tikzpicture};
						\draw[propagator] (0,0) -- (1.05,0);
						\draw[particle] (1.1,0) -- (1.618,0);
						\fill[](0,0) circle (0.1);
						\fill[](1.05,0) circle (0.1);
						\draw[photon] (0,0) arc(180:0:0.525) ; 
						\fill[](0,0) circle (0.1);
					\node at (1.75,0) [rectangle,draw] {};
						\draw[photon] (1.174,0.84) arc(60:11.9:1.1) ;
			\end{tikzpicture}}+  \;	 
			\hbox{\begin{tikzpicture}
						\draw[propagator] (0,0) -- (1.05,0);
						\draw[particle] (1.1,0) -- (1.618,0);
						\fill[](0,0) circle (0.1);
						\fill[](1.05,0) circle (0.1);
						\draw[photon] (0,0) arc(180:7.5:0.86) ; 
						\fill[](0,0) circle (0.1);
						\node at (1.75,0) [rectangle,draw] {};
						\draw[photon] (0.05,0.94) arc(90:0:1.02) ;
			\end{tikzpicture}}
			\end{split}
			\end{align}
	These equations are   the same as the ones derived in \cite{Vogel_ERM} for the renormalized vertex in the ERM-model: $\sigma^{ERM}_q(s)=\sigma^{VF,ERM}_q(s)$. The same set of irreducible diagrams survives in the high density regime.  One has to 
	neglect excluded volume effects and choose a uniform distribution of the particles to  fully recover the self-consistent model in \cite{Vogel_ERM}. This results in the approximation $S_q \approx 1.$

	All the numerical results shown in the main text were obtained by neglecting the tensor structure of the vertices 
	\begin{align}
	\begin{split}
		&(c_q^\perp)^2 \matr{\mathcal{V}}_{\7q,\7k}  \to \hat{\chi}_k^{(0),\perp}(s) S_{|\7q-\7k|} \mathcal{V}^{ERM}_{\7q,\7k} ( \mathbb{1} -\hat{\7q} \hat{\7q})\;,\\ &{V}^{ERM}_{\7q,\7k}= \frac{q (c_q^\perp)^2}{d-1} {\rm Tr}\{\matr{V}_{\7q,\7k}\} \equiv   \frac{n^*\omega_0^2 f(|\7k|)-f(|\7q-\7k|)}{f(q=0)-f(q=2 \pi/\sigma)}\;,
			\end{split}
	\end{align}
	but by explicitly not performing the high density expansion discussed above. 
	Neglecting excluded volume effects and setting $S_q =1$ leads to the following self-consistent set of equations
	\begin{widetext}
	\begin{subequations}
	\label{eq_SI_ERM_Equation}
		\begin{align}
	\hat{K}_q^{ERM}(s)=& \left[ s+\xi- \frac{V^{ERM}_{\7q,\7q}}{s- \frac{1}{V^{ERM}_{\7q,\7q}}	\hat{W}_q^{ERM}(s)}\right]^{-1} \;\\
			\hat{W}_q^{ERM}(s) &=	\frac{1}{n}\int \frac{d^d \7k}{(2 \pi)^d}{V}_{\7q,\7k}^{ERM} \hat{\chi}_k^{(0)}(s)  {\mathcal{V}}^{ERM}_{\7q,\7k}(s)\;,\\
	\hat{\Xi}^{ERM}(\7q,\7k,\7p,s) &=  \frac{  \delta_{\7k,\7p}}{n} \int \frac{d^d \7b}{(2 \pi)^d}	 V^{ERM}_{\7k,\7b}   \hat{K}^{ERM}_{\7b}(s)  V^{ERM}_{\7k,\7b}  + \frac{1}{n} 	V^{ERM}_{\7k,\7p+\7k-\7q}  \hat{K}^{ERM}_{|\7q-\7p-\7k|}(s)  V^{ERM}_{\7p,-\7q+\7p+\7k} \\
	\begin{split} 	\label{app_SI_renormalised_vertex_scalar_ERM}
 		\int \frac{d^d \7p}{(2 \pi)^d}	\bigg [\delta_{\7k,\7p}  - (s + \xi) &\;	\hat{\Xi}^{ERM}(\7q,\7k,\7p,s) \chi^{(0)}_p(s) \bigg] \hat{\mathcal{V}}^{ERM}_{\7q,\7p}(s) = 	\int \frac{d^d \7p}{(2 \pi)^d} \Big[ s \delta_{\7k,\7p}+  \hat{\Xi}^{ERM}(\7q,\7k,\7p,s) \hat{\chi}_p^{(0)}(0) \bigg] V^{ERM}_{\7q,\7p}\;.
			\end{split}
		\end{align}
	\end{subequations} 
		\end{widetext}

	We approximate the left hand side to be equal to the renormalised vertex for $s \to 0$. This basically amounts to an \textit{F1} approximation. Note, that this approximation is exact for $s=0$ in the jammed state. We extend its validity to the small frequency regime of the un-jammed phase, in order to test our equations in this phase as well.  One arrives at 
	\begin{align}
	\label{eq_app_F1_approximation_ERM}
		\begin{split}
	 &	\hat{\mathcal{V}}^{ERM}_{\7q,\7k}(s) \\&  \underset{s \to 0}{\approx} 	\int \frac{d^d \7p}{(2 \pi)^d} \Big[ s \delta_{\7k,\7p}+  \hat{\Xi}^{ERM}(\7q,\7k,\7p,s) \chi_p^{(0)}(s) \bigg] V^{ERM}_{\7q,\7p}\;.
				\end{split}
	\end{align}
	All numerical results presented in this work are derived from the equations \eqref{eq_SI_ERM_Equation} and \eqref{eq_app_F1_approximation_ERM}.

	\section{Un-jammed  phase}
	In the un-jammed state decorrelate stresses  with time and decay for $t \to \infty$. This causes a finite value of 
	\begin{align}
		\kappa_q^\perp= \lim_{s \to 0} (s+\xi) \hat{K}_q^\perp(s)\;.
	\end{align}
	In the Newtonian case can  	$\kappa^\perp_q$ be interpreted as a non-ergodicity parameter.  As a consequence,$   (s+\xi) \hat{W}_q(s) \equiv  \hat{\mathcal{W}}_q(s)$ and $ (s+\xi)  \hat{\matr{\mathcal{V}}}_{\7q,\7k}(s) \equiv \hat{\matr{\Upsilon}}_{\7q,\7k}(s)$ also yield finite contributions  for $s \to 0$.   Writing $\hat{\matr{\mathcal{M}}}_k(s)= (s+\xi) \hat{M}_k(s)(\mathbb{1}-\hat{\7k}\hat{\7k})$, equation  \eqref{app_eq_RV_First_expression} reads after multiplying with $(s+\xi)$  
	\begin{widetext}
	\begin{align}
		\label{app_eq_unjammed_Vertex}
		\begin{split}
			& \left[ s(s+\xi)+k^2(c_k^\perp)^2  + 	(c_k^\perp)^2 \hat{\matr{\mathcal{M}}}_k(s)  \right]  \cdot  \hat{\matr{\Upsilon}}_{\7q,\7k}(s)  +  (s+\xi) \;
			\frac{	(c_k^\perp)^2}{N}  \sum_{\7p} \matr{\tilde{V}}(\7q,\7k,\7p)\cdot    \hat{\matr{\Upsilon}}_{\7q,\7p}(s) \;\hat{K}_{|\7q-\7k-\7p|}(s) =  \\
			=& \left[s(s+\xi) + (c_k^\perp)^2 \hat{\matr{\mathcal{M}}}_k(s) \right] \cdot
			\matr{V}_{\7k,\7q}^\dagger S_{|\7q-\7k|} + (s + \xi)	
			\frac{(c_k^\perp)^2}{N}  \sum_{\7p} \matr{\tilde{V}}(\7q,\7k,\7p)\cdot    \matr{V}_{\7p,\7q}^\dagger \hat{K}_{|\7q-\7k-\7p|}(s)	S_{|\7q-\7p|}\;.
		\end{split}
	\end{align}
		\end{widetext}
	The set of equations does not simplify much in the $s \to 0$ limit. One finds
	\begin{align}
		\label{app_eq_unjammed_Vertex_small_s}
		\begin{split}
			& \sum_{\7p} \bigg\{ \left[	k^2 + 	 \hat{\matr{\mathcal{M}}}_k(s) \right]  \delta_{\7k,\7p}  +  \matr{\tilde{V}}(\7q,\7k,\7p) \frac{\kappa^\perp_{|\7q-\7k-\7p|}  }{N} \bigg\} \cdot  \hat{\matr{\Upsilon}}_{\7q,\7p}(s)\\
			=&  \sum_{\7p} \left\{ \hat{\matr{\mathcal{M}}}_p(s) \delta_{\7k,\7p}   + 	
			\frac 1N   \matr{\tilde{V}}(\7q,\7k,\7p)\;  \kappa^\perp_{|\7q-\7k-\7p|} \right\}  \cdot  \matr{V}_{\7p,\7q}^\dagger	 S_{|\7q-\7p|} \;.
		\end{split}
	\end{align}
	
	However, $\lim_{s \to 0} (s + \xi) \hat{K}_q^\perp(s)$ is suppose to become small, when approaching the jamming transition. Since we are mainly interested in this case, we can  neglect higher orders in $\kappa^\perp_q$. This leads to the  this simplifies to 
	\begin{align}
		\label{app_eq_unjammed_Approx}
		\begin{split}
		k^2 \hat{	\matr{\Upsilon}}_{\7q,\7k}(s)  
			\approx &  \hat{\matr{\mathcal{M}}}_k(s=0)\cdot \matr{V}_{\7k,\7q}^\dagger \;
			S_{|\7q-\7k|} \\& +  
			\frac 1N  \sum_{\7p} \matr{\tilde{V}}(\7q,\7k,\7p)\cdot \matr{V}_{\7p,\7q}^\dagger \; \kappa^\perp_{|\7q-\7k-\7p|} \;  	\;.
		\end{split}
	\end{align}
	Again, we end up with a simple expression for the $s =0$ limit 
	\begin{align}
		\label{app_eq_long_time_fluidity_linear_integral}
	\hat{	\mathcal{W}}_q(s=0)=  \sum_{p} p^{d-1} C_{q,p}  \kappa_p
	\end{align}
	
	Looking at the long time limit, $\hat{K}^\perp_q(t \to \infty) = \lim_{s \to 0} s \hat{K}_q^\perp(s),$ one finds in the Langevin case
	\begin{align}
		\lim_{s \to 0} s \hat{K}_q^\perp(s) = \frac{s}{\xi} \frac{1}{1+ \frac{q^2}{ \mathcal{W}_q(s=0)}}\;. 
	\end{align}

	\section{Vertex evaluation \label{app_sec_Vertex_Evaluation}}
	In this section do we analyse the  vertex
	\begin{align}
		\label{eq_Full_Vertex}
		{V}_{\alpha \beta}(\7k,\7q)= -  \frac{\braket{F^\perp_{\alpha}(\7q) \Omega Q \delta \rho(\7k-\7q) v^\perp_\beta(-\7k) }}{N  k_B  T q(c_q^\perp)^2   S_{|\7q-\7k|}}\;,
	\end{align} 
	explicitly.  We are going to express $	{V}_{\alpha \beta}(\7k,\7q)$ in terms of  density correlation functions. Hence, $	{V}_{\alpha \beta}(\7k,\7q)$ can be calculated for real systems.
	We use Einstein's sum convention in this section. 
We end up with  \begin{widetext}
		\begin{align}
			\label{app_eq_Exact_Expression_Vertex}
	\begin{split}
	- \frac{\braket{F^\perp_\alpha(\7q)\Omega Q \delta \rho(\7q-\7k) v^\perp_\beta(-\7k } }{Nq(c_q^\perp)^2k_B T S_{|\7q-\7k|}}= &
				2   \frac{q_\gamma }{mV q(c_q^\perp)^2 } (\delta_{\alpha \tau}-\hat{q}_\alpha \hat{q}_\tau)  \sum_{\7\kappa}  \kappa_\zeta (\delta_{\beta  \zeta}-\hat{k}_\beta \hat{k}_\zeta) \Re \{X_{\tau \gamma} (\7\kappa,\7q) \} \\& \times \bigg[ \frac{\braket{\delta \rho(\7\kappa-\7k) \delta \rho(\7q-\7\kappa) \delta \rho(\7q-\7k)}}{NS_{|\7k-\7q|}}  -  S_{|\7q-\7\kappa|}  \bigg] 
			\end{split}
	\end{align} 	\end{widetext} 
	 Applying the convolution approximation $\frac{1}{N} \braket{\delta \rho(\7k)\delta \rho(\7k'),\delta \rho(-\7k-\7k')} \approx S_k S_{k'}S_{|\7k+\7k'|}$ \cite{HansenMcDonald} for the three-point correlator, gives the concise formulae
	\begin{align} \begin{split}
		- &  \frac{\braket{F^\perp_\alpha(\7q)\Omega  \delta \rho(\7q-\7k) v^\perp_\beta(-\7k } }{N q(c_q^\perp)^2 k_B T S_{|\7q-\7k|}} \approx - 2 \frac{q_\gamma (\delta_{\alpha \tau}-\hat{q}_\alpha \hat{q}_\tau)  }{mV q(c_q^\perp)^2 }  \\&  \times  \sum_{\7\kappa}  \kappa_\zeta (\delta_{\beta  \zeta}-\hat{k}_\beta \hat{k}_\zeta) \Re \{X_{\tau \gamma} (\7\kappa,\7q) \}  S_{|\7q-\7\kappa|}  \Big[1-S_{|\7k-\7\kappa|}   \Big] 
			\end{split}
	\end{align}
	The weights are determined by the interaction potential
	\begin{align}
		X_{\alpha \beta}(\7\kappa,\7q)= - \frac{1}{2}
		\int d\7r e^{i\7\kappa\cdot \7r} \frac{r_\alpha r_\beta}{r} U'(r) \frac{1-e^{-i\7q\cdot \7r}}{ i\7q\cdot \7r}\;,
	\end{align}
	Note, that the vertex depends on the density  and the local structure.

	\subsection{Symmetry of the Vertex \label{app_sub_Symmetry_normal_vertex}}
	It is easy to see, that $
	\Re[X_{\alpha \beta}(-\7k,-\7q)]=\Re[X_{\alpha \beta}(\7k,\7q)]$ holds.
	Additionally, the correlation function of scalar quantities only depends on the absolute modulus of the wavevector \cite{Gotze}. 
	After a change of the integration variable, we hence find
	\begin{align}
		\begin{split}
		V_{\alpha \beta}(\7q,\7k) & =V_{\alpha \beta}(-\7q,-\7k) = 	V_{\beta \alpha}^\dagger(\7k,\7q) \;,  \\
		\matr{V}_{\7q,0} &= 	\PS{-}\matr{V}_{\7q,0}\;, \hspace{0.7cm}
				\matr{V}_{0,\7k}= 	-\matr{V}_{0,-\7k}
				\end{split}
	\end{align}

	\subsection{Hydrodynamic limit}
	The weights are a symmetric function of $\7\kappa$ in $\7q=0$, \textit{i.e} $\matr{X}(\7\kappa,0)=\matr{X}(-\7\kappa,0)$. But the whole argument $  \kappa_\zeta \Re \{X_{\tau \gamma} (\7\kappa,\70) \} \ \bigg[ \frac{\braket{\delta \rho(\7\kappa) \delta \rho(\7\kappa) ,\delta \rho(\7q-\7k)}}{NS_{|\7q-\7k|}}  -  S_{\kappa}  \bigg] $ is an odd-function of $\7\kappa$, 
	since the average over scalar quantities does only depend on the absolutest modulus of the wavevector. Thus,  the zeroth order in $\7q,\7k$  of the integral over the three-point function vanishes. Since the integrand stays non-zero for only $\7k \to 0$ or $\7q \to 0$, we find in the hydrodynamic limit.
	\begin{align}
		V_{\alpha \beta}(\7k\to 0,\7q \to 0)=V^{(1)}_{\alpha\beta \gamma \eta}(0,0) \frac{q_{\gamma } q_\eta}{q} + V^{(2)}_{\alpha \beta \gamma \eta}(0,0) \frac{q_{\gamma } k_\eta}{q} 
	\end{align}
	In the case of the ERM-approximation, (see chapter \ref{app_sec_ERM}) or when approximating the vertex to have a diagonal form, one  finds 
	\begin{align} \begin{split}
		V^{ERM}_{\7q,\7k}&= \frac{q (c_q^\perp)^2}{d-1}\Tr \matr{V}_{\7q,\7k}\;\\
		V^{ERM}_{\7k\to 0,\7q \to 
		0} &=V^{(1)}q^2 + V^{(2)}_{\alpha \beta}  q_\alpha  k_\beta \\&=
		V^{(1)}q + V^{(2)}  \7q \cdot   \7k
			\end{split}
	\end{align}
	The last equality holds to the assumed isotropy of the system, which implies $V^{(2)}_{\alpha \beta} \propto \delta_{\alpha \beta}$.

	\section{Götze-scaling analysis \label{app_sec_beta_analysis}}
	The dispersion relation $v_q^\perp$ and the non-ergodicity parameter $\kappa_q^\perp$  both vanish,when approaching the (un-) jamming transition, or respectively near the critical density $n_c$ where the maximal eigenvalue of the stability matrix becomes $1$. Utilising that the dynamics close to the transition is governed by the eigenvector of the stability matrix $C_{q,p}$  belonging to the largest eigenvalue, we derive the scaling of important quantities like $v_q^\perp, \kappa_q^\perp$ close to $n_c$. Basically, our analysis involves standard linear algebra (Fredholm's theorem ) and is based on the assumption that the largest eigenvalue is simple.   
	
	\subsection{The scaling equation}
	We start by re-writing equation \eqref{app_eq_RV_First_expression}: 
	\begin{widetext} 
	\begin{align}
		\label{app_eq_starting_point_beta_scaling}
		\begin{split}
		&\matr{\Upsilon}_{\7q,\7k}(s) = - (c_k^\perp)^2 \matr{\mathcal{M}}_k(s)\cdot \matr{\Upsilon}_{\7q,\7k}(s)  -  	(c_k^\perp)^2
			\frac 1N  \sum_{\7p} \matr{\tilde{V}}(\7q,\7k,\7p)\cdot     \matr{\Upsilon}_{\7q,\7p}(s) \;\overline{K}_{|\7q-\7k-\7p|}(s)   \\
			=& \left[s(s+\xi) + (c_k^\perp)^2 \matr{\mathcal{M}}_k(s) \right] \cdot 
			\matr{V}^\dagger_{\7k,\7q} + 	(c_k^\perp)^2
			\frac 1N  \sum_{\7p} \matr{\tilde{V}}(\7q,\7k,\7p)\cdot   \matr{V}_{\7p,\7p}^\dagger\;\overline{K}_{|\7q-\7k-\7p|}(s) \;	\;.
		\end{split}
	\end{align}
	\end{widetext}
	Here, we have introduced for the purpose of readability $(s+\xi)\hat{K}_q^\perp(s)= \overline{K}^\perp_q(s)$ and $\7b=\7q-\7p-\7k$. Further, we wrote  again $(s+\xi)\hat{M}_k (s)(\mathbb{1}-\hat{\7k}\hat{\7k})= \hat{\matr{\mathcal{M}}}_k(s)$ .
	The conceptual idea is that the quantity 
	\begin{align}
		s \varphi_q(s)= \frac{s(s+\xi)}{(c_q^\perp)^2} +(s+\xi) \hat{W}^\perp_q(s)
	\end{align}
	becomes small for $s \to 0$ when approaching the transition. Note, that the shear modulus can be expressed with this quantity as $G_q^\perp(s)= \frac{s + \xi }{s\varphi_q(s)}$.  In the un-jammed phase holds  $\lim_{s, \7q  \to 0}s\varphi_q(s) = \frac{1}{(\lambda_-)^2}$. In the jammed phase holds    $\lim_{s  \to 0}s\varphi_q(s) = \frac{s(s+\xi) }{( v_q^\perp)^2 }$.  We can express the transverse velocity autocorrelation as 
	\begin{align}\label{eq:S82}
		\overline{K}_q(s)= \frac{s \varphi_q(s)}{s \varphi_q(s)+q^2} = \frac{s \varphi_q(s)}{q^2} - \frac{(s \varphi_q(s))^2}{s \varphi_q(s)+q^2} \frac{1}{q^2}\;.
	\end{align}
	In order to deal with our set of  self-consistent equations, we expand the renormalized vertex on the right hand side of equation  \eqref{app_eq_starting_point_beta_scaling} according to 
	\begin{align}
	 \hat{\matr{\Upsilon}}_{\7q,\7p}(s)=  \frac{s(s+\xi)}{s(s+\xi)+(p c_p^\perp)^2} S_{|\7q-\7p|} \matr{V}_{\7p,\7q}^\dagger\;.
	\end{align}
	This ansatz leads to 
	\begin{widetext}
	\begin{align}
		\begin{split}
			s \varphi_q(s)-& \frac{s(s+\xi)}{(c_q^\perp)^2} =  \frac{1}{N}  \sum_{\7k} \mV_{\7q,\7k} : \hat{\matr{\Upsilon}}_{\7q,\7k}(s)\\
			=&  - \frac{s	(s+\xi)}{N} \sum_{\7k,\7p}  \mV_{\7q,\7k} : \mV_{\7k,\7q}^\dagger   \frac{(c_k^\perp)^2 s \varphi_p(s) S_{|\7q-\7k|}}{[s\varphi_p(s)+p^2][s(s+\xi)+k^2(c_k^\perp)^2]}
		\mV_{\7k,\7p}:	\mV_{\7p,\7k}^\dagger S_{|\7k-\7p|} 	   \\ &-  \frac{s	(s+\xi)}{N}  \sum_{\7k,\7p}    \frac{(c_k^\perp)^2 s \varphi_b(s)S_{|\7q-\7p|}}{[s\varphi_b(s)+b^2][s(s+\xi)+p^2(c_p^\perp)^2]} 
		\mV_{\7q,\7k}:(	\Tilde{\mV}(\7q,\7k,\7p)  \cdot  \mV_{\7p,\7q}^\dagger )   \\
			&+ \frac{1}{N} \sum_{\7k} \mV_{\7q,\7k}  :  \left\{ \Big[ s(s+\xi) +  ( c_k^\perp)^2 \matr{\mathcal{M}}_k(s)  \Big] \cdot  \mV_{\7k,\7p}^\dagger S_{|\7q-\7p|}  +   \frac{(c_k^\perp)^2 }{N} \sum_{\7p}  \Tilde{\mV}(\7q,\7k,\7p) \cdot    \mV_{\7p,\7q}^\dagger\; \overline{K}_{b}(s)   \right\} S_{|\7q-\7p|} \;,
		\end{split}
	\end{align}
	\end{widetext}
	Since the terms in the second and third  line are proportional to $s(s+\xi)$, they can be neglected for  small frequencies. The term $s(s+\xi)$ in the fourth line has to be kept, since it plays a role in the jammed phase. Relying on the decomposition of $\overline{K}$, we arrive at   
	\begin{align}
		\label{app_eq_First_B_equation}
		\begin{split}
		s \varphi_q(s) &- \frac{s(s+\xi)}{(\Tilde{ c}_q^\perp(s))^2} - \sum_{p} p^{d-3} C_{q,p}  s \varphi_p(s) \\&  = - \sum_p p^{d-3} C_{q,p}\frac{(s\varphi_p(s))^2}{ s\varphi_p(s)+ p^2}  \;.
			\end{split}
	\end{align}

In the small $s$ limit can we set $\Tilde{c}_q^\perp(s) \approx \Tilde{c}_q^\perp(0)  \equiv \Tilde{c}^\perp_q$. Its contribution of the sound pole becomes subdominant close to the transition. \\

	One has to notice a few things here
	
	\begin{itemize}
		
		\item Close to the transition vanishes  $s \varphi_p(s) $  in the small frequency limit. Hence, this term can be set to zero in the denominators on the right hand side, since it will eventually be smaller than $p^2$ for any wavenumber $p$
		\item  Note, that the  term on the right hand side of equation \eqref{app_eq_First_B_equation} features a potential divergence for $p \to 0$ close to the transition, where $s \varphi_p(s) \approx 0$ can be set to zero in the denominator \cite{Schnyder_2011}.  At least in in lower dimensions $d<5$. But it turns out, that this is not a problem, since the memory function $\hat{W}^\perp_q(s)$ close to the transition is determined by the left and right eigenvector $\hat{h}_p$ and $h_q$ of the stability matrix associated with its largest eigenvector. As it will be shown in the next subsection, the critical left eigenvector vanishes $\hat{h}_p \propto p^2 $ for $p \to 0$. This prevents the divergence.
	\end{itemize} 
	We interpret $\matr{C} \in \mathbb{R}^{m \times m}$, where $m \to \infty$ is the number of wavevectors considered. Hence, $\matr{C}$ as a square matrix can expressed via its Jordanform
	\begin{align}
		\matr{C}= \matr{E}^{-1} \matr{J} \matr{E}\;,
	\end{align}
	where $\matr{J}$ is the Jordan block-matrix.  $\matr{E}$ is constituted by the  generalised eigenvectors of $\matr{C}$.  One has to remember that, if $(\7v^{(i)})^T$ is a left eigenvector to the eigenvalue $\lambda_i$  and $\7u^{(j)}$ is a right generalised eigenvector to $\lambda_j \neq \lambda_i$ then $(\7v^{(i)})^T \cdot \7u^{(j)}=0$ holds.  \\

	We assume, that the largest eigenvalue of $\matr{C}$ is simple, \textit{i.e} its geometric multiplicity is 1. In this case can we    order the eigenvalues and set 
	\begin{align}
		J_{1,1}= \lambda_{max}=1+\epsilon\;, \hspace{1cm} \left(\matr{E}^{-1} \right)_{j,i} = \hat{h}_j\;,\; \; E_{1,j}=h_j\,.
	\end{align}
	As mentioned earlier, the critical eigenvalue of the stability matrix is $\lambda_{max}=1$. We analyse the dynamics close to the transition by setting $\lambda_{max}=1+\epsilon$ with $|\epsilon|$ small. The spectrum is determined by the density or to be more precise by density correlations functions.  We adopt the normalisation  $\sum_q \hat{h}_q h_q=1$ 
    Assuming the largest eigenvalue being simple, gives  
	\begin{align}
		C_{q,p}=(1+\epsilon) h_q \hat{h}_p + C^\#_{q,p}\;,
	\end{align} 
with $\sum_p C^\#_{q,p}h_p=0$ and $\sum_q \hat{h}_q C^\#_{q,p}=0$.
	Again, $s \varphi_q(s)$ is mainly determined by the critical eigenvector. Hence, we make the ansatz \begin{align}
		\label{eq_AnsatzFried}
		s \varphi_q(s) = \sigma s_* g(s_*) h_q + \sigma^\frac{d}{2} s_* X_q^\#(s_*)
	\end{align}
	with $s_*=s t_*$, where $\sigma$ and $t_* = 1/\omega_*$   are yet unspecified. $t_*$ sets the time scale and $\sigma$ is a measure for the distance to the critical point. The orthogonality of the generalised eigenvectors gives $\hat{\7h}^T \cdot \7X^\#=0$. Inserting this  ansatz  \eqref{eq_AnsatzFried} in equation  \eqref{app_eq_First_B_equation} gives for $s \to 0$
	\begin{align}
		\begin{split}
			- \frac{s(s+\xi)}{(\tilde{c}^\perp_q)^2} - & \sigma  \epsilon s_* g(s_*) h_q + (s_*g(s_*))^2 \sigma^2 \sum_{p}(1+\epsilon) h_q \hat{h}_p    \frac{h_p ^2}{p^2} \\  \approx &  - \sigma^\frac{d}{2} \sum_p \Big( \delta_{q,p}-C^\#{q,p} \Big)s_*X_p^\# \\& - \sum_p C^\#_{q,p}  \frac{\sigma^2 (s_* g(s_*))^2 h_p^2}{p^2+s \varphi_p(s)}+ \mathcal{O}(\sigma^{2.5})\;. 
		\end{split}
	\end{align}
	Contracting from the left side with $\hat{\7h}$ together with the adopted normalisation 
	\begin{align}
		\label{app_eq_second_normalisation_Beta_Scaling}
		\sum_{p} \hat{h}_p  \frac{h_p^2}{p^2} =1\;,
	\end{align}
	leads to  the scaling equation 
	\begin{align}
		\label{app_eq_scaling equation}
		\frac{t_*^2s (s+\xi) }{\sigma^2} + \frac{\epsilon}{\sigma}  s_*g(s_*) = (1-\epsilon)(s_*g(s_*))^2 \underset{\epsilon \to 0}{\longrightarrow } (s_*g(s_*))^2
	\end{align}
	with the time-scale  parameter resulting from the integral over the inhomogeneous part
	\begin{align}
		\begin{split}
			t_*^2 &= \int_0^\infty  dq \frac{ 	\hat{h}_q }{(\Tilde{c}_q^\perp)^2}\;. 
		\end{split}
	\end{align}
	Additionally, we find the important result \begin{align}\sigma= |\epsilon|.\end{align} The minus sign corresponds to the jammed phase.
	\subsection{The critical left eigenvector \label{app_sec_critical_left_eigenvecor}}
	As it was explained in the previous section, it is crucial in lower dimensions that the left critical eigenvector vanishes with $p^2$. We will argue in this section that this is indeed the case. First, we give a intuitive physical argument. Secondly we will provide a  mathematical proof in $d=3$ and lastly, we will show numerical results to confirm this prediction.
	
	\paragraph{The physical picture}
	The  matrix $\matr{C}$ has a finite contribution for $p=0$. However, this contribution is of higher order in $q$. More specifically, $C_{q \to 0,0} \propto q^2$ holds. The un-jamming transition can be understood as the system becoming macroscopically soft. The ability to sustain shear stress ceases and  an applied perturbation does not lead to propagating sound waves anymore in the thermodynamic limit but causes macroscopic plastic deformations even for a vanishing amplitude of the perturbation. Nevertheless, the emergence of stability in disordered materials is not a small wavevector phenomenon since the macroscopic stability supervenes on the microscopic structure, which is resolved by higher wavenumbers.   Approaching the un-jamming transition from above, more and larger subregions of the system become weakly connected. The restoring forces vanish here. If the number or size of those weakly connected parts becomes extensive, the system becomes macroscopically soft. Thus, the low wavenumber columns and not the rows of the stability matrix $C_{q,p}$ are crucial to investigate the un-jamming transition.  This is also why the critical part of the stability matrix $C_{q,p}$ close to the transition is $p$-dependent but not $q$-dependent. This is confirmed by the observation that the maximal eigenvalue of $C_{q,p}$ decreases when the maximal value of the array $p=\{p_1,p_2....p_m\}$ is chosen too small.  
	
	As a consequence, the low $q$-limit of the stability matrix should determine the rigidity of the system. Since this part vanishes with $C_{0,p \to 0} \propto p^2$, no divergence should affect the critical dynamics of the dispersion relation close to the un-jamming transition for $d > 2$. This, together with the assumed rotational invariance, suggests that the critical left eigenvector vanishes at least with $\hat{h}_p \propto p^2$ in $d=3$. \\

	\paragraph{Mathematical proof} If the entries of the left eigenvector $\hat{h}_p$ became a constant for $p \to 0$ in $d=3$, the sound attenuation would scale with $s q^2$ for $s,q \to 0$. However, this is in contradiction to our result in section \ref{App_Paragraph_Proof_Rayleigh}:  \\ 
	
	We consider the case $d=3.$
	If $\hat{h}_q=const.$ for $ q \to 0$, the normalisation \eqref{app_eq_second_normalisation_Beta_Scaling} would diverge. In this case we couldn't set $s \varphi_p(s)=0$ in the denominator. Rather, we had to follow \cite{Schnyder_2011} and adopt the normalisation
	\begin{align}
		1= \frac{\pi}{2} \sum_q \hat{h}_q C_{q,0} h_0^{3/2}\;.
	\end{align}
	The new scaling function  would read for $\xi=0$
	\begin{align}
		\label{app_eq_scaling_equation_2}
		\frac{t_*^2s^2  }{\sigma^{3/2}} + \frac{\epsilon}{\sqrt{\sigma}}  s_*g(s_*) = (s_*g(s_*))^{3/2}\;.
	\end{align}
	The solution in the jammed state for $s_* \ll \frac{\sigma^{3/4}}{t_*}$ becomes in this case $g_-\left(s_* \ll \frac{\sigma^{3/4}}{t_*} \right) \to s_* -s_*^2$. Since the right eigenvector must stay finite for $q \to 0$, in order for the speed of sound to be finite, the residual contribution to the self-energy becomes
	\begin{align}
		\hat{\sigma}_q(s)-\sigma_q(0) \propto s q^2\;,
	\end{align}
	in contradiction to the result obtained in  \ref{App_Paragraph_Proof_Rayleigh}. Thus, the left eigenvector must vanish at least with $\hat{h}_p \propto p^2$ in $d=3$.  \\
	
\begin{figure}
		\centering
		\includegraphics[width=0.5\textwidth]{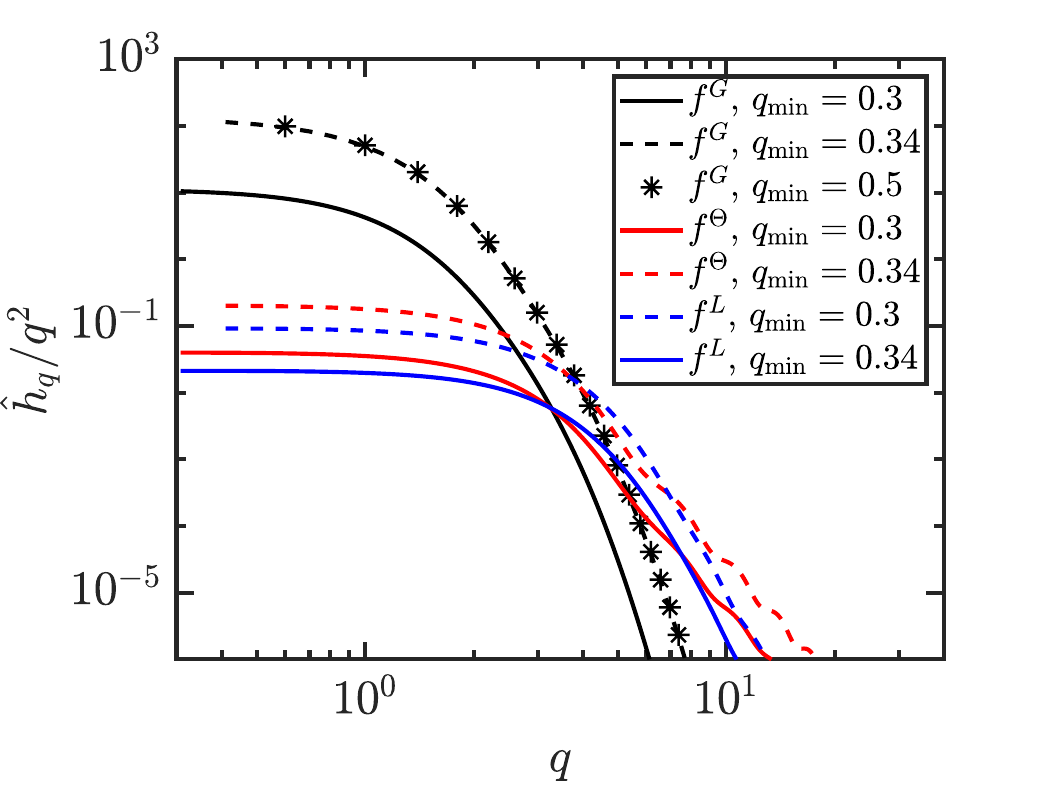}
		\caption{The critical left eigenvector divided by $q^2$ for three different spring functions. A linear  Yukawa function $f(r)=\frac{ \sigma}{r} \Theta(\sigma-r )$, the Theta-or step-spring function $f(r)=  \Theta(\sigma-r )$ and the Gaussian spring function $f(r)= e^{-r^2/(2 \sigma^2)}$.   the  left eigenvector is shown with a minimal wavevector $q_{min} \sigma =0.34$ with a linear distribution of the wavevectors and with $q_{min} \sigma =0.3$ with a logarithmic distribution. For the Gaussian-spring function we also show a linear distribution with $q_{min} \sigma =0.5$.   } 
		\label{fig_crit_l_eigenvector}
	\end{figure}

	\paragraph{Numerical confirmation} 
	Figure \ref{fig_crit_l_eigenvector} shows the critical left eigenvector divided by $q^2$. One clearly sees, that $\hat{h}_q \propto q^2$ for small wavevectors as predicted by theoretical and physical considerations.  \\

	\subsection{Scaling of the observables \label{AppFD}}
	After having obtained the scaling function $g(s_*)$, we can investigate the scaling behaviour of important observables in the jammed- and un-jammed state. We only go quickly through  the Langevin case since the Newtonian case ($\xi =0)$ is covered in the main text. \\
	
	The small frequency scenario is considered, where  $s \ll \xi$ holds. This gives $\omega_*= \frac{\epsilon^2}{\tau^2 \xi }$
	The scaling function is given by
	\begin{align}
		\begin{split}
			g_\pm(s \gg \omega_*) & \to \sqrt{s_*} \\
			g_-(s \ll \omega_*) & \to 1-s_* \\
			g_+(s \ll \omega_*) & \to \frac{1}{s_*}+1-s_*
		\end{split}
	\end{align}
	The dispersion relation is again given by 
	\begin{align}
		( v_q^\perp)^2= \frac{ |\epsilon|}{\tau^2 h_q}
	\end{align}
	and we also find $\lambda_-^\perp \propto 1/ \sqrt{\epsilon}.$  
	A density of states cannot be defined in this linear response approach to a dissipative system.

\section{Details on the Euclidean random matrix model and its numerical solutions \label{app:ERM}}

\paragraph{Numerical diagonalization }
Here, we will give a short overview of the numerical diagonalization of the ERM model. For a more comprehensive discussion we refer to Ref.~\cite{Baumgaertel2023properties}, where we studied the ERM model with a Gaussian spring function in detail.\\

\begin{figure}
		\centering
		\includegraphics[width=0.4\textwidth]{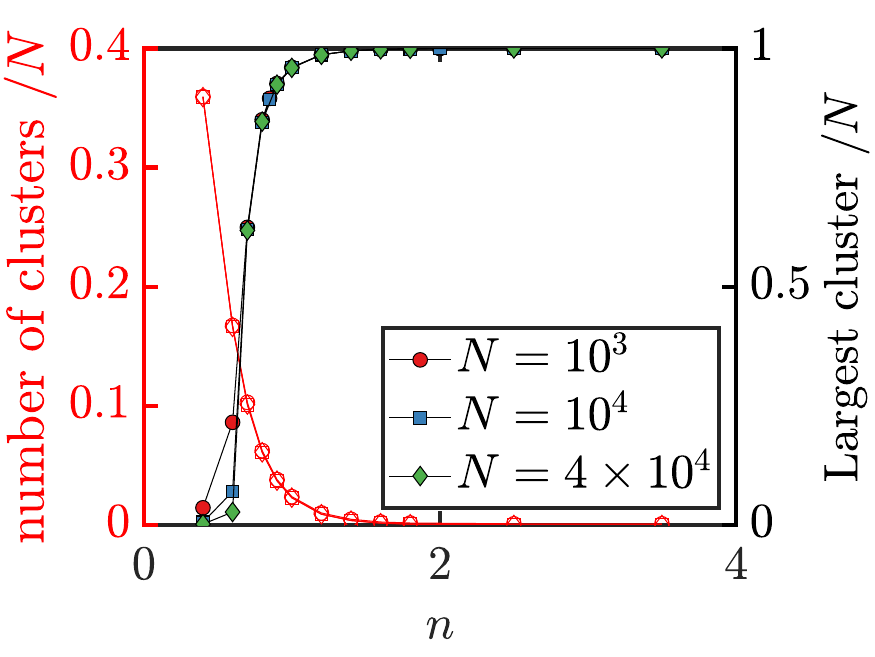}
		\caption{The number of clusters (red symbols, left axis) and the number of particles belonging to the largest cluster (black symbols, right axis) in the ERM model with spring function of finite extent, $f^\Theta(r)$, for system sizes $N = 10^3,\,10^4$ and $N = 4\times10^4$.} 
		\label{fig:NumberVsMaxCluster}
	\end{figure}

We calculate the underlying random matrix by
\begin{align}
	\matr{\mathcal{H}}_{ij} = -m \omega_0^2 f(|\7r_i-\7r_j|) +  \delta_{ij} m \omega_0^2 \sum_k f(|\7r_i-\7r_k|),
\end{align}
where $f(r)$ is the unitless spring function. The diagonalization of the matrix yields eigenvalues $\lambda_i = m \omega_i^2$  and corresponding eigenvectors $\7e_i$ from which the density of states $D(\omega)$ can be calculated straight forward.  In Fig.~\ref{fig:NumberVsMaxCluster} we show the number of clusters and the number of particles belonging to the largest cluster for different system s. At the rescaled density $n^*_p = 0.85$ we observe a distinct drop in the size of the largest cluster and strong This indicates a percolation transition at $n^*_p= 0.838$ in agreement with previous studies of overlapping spheres \cite{rintoul_precise_2000,elam_critical_1984}.  At the same time we observe an extensive number of clusters even above the transition $n^*_p$. For the $\Theta$-spring function $f^\Theta$ these unconnected clusters, or rattlers, have integer eigenvalues with a magnitude depending on the number of particles belonging to that cluster. As these eigenvalues appear as spikes in the density of states we remove them from the calculations of $D(\omega)$.\\
In Fig.~\ref{fig_tansition_from_left_and_right} we show the scaling of the frequency $\omega^*$. This is the frequency where a plateau in the density of states sets in. From the numerical diagonalization, we calculated it as $D(\omega^*/\omega_0)\omega_0>D_{\omega^*}$, where $D_{\omega^*}$ is a threshold We chose $D_{\omega^*} =0.17$ as it approximates the plateau well. 
We checked the influence of varying the threshold $D_\omega^*$ and found no qualitative differences.
\\
In Ref.~\cite{Baumgaertel2023properties} the dispersion relation $qv_q^\perp$ was obtained by fitting the dynamical structure factor and the velocity auto-correlation function with a damped harmonic oscillator. As the dispersion relation becomes very small approaching the un-jamming transition, we find that these two methods give inaccurate results. In this work we thus calculate the dispersion relation from the $s\to 0$ limit of the Laplace transformed velocity autocorrelation function
\begin{align}
	\hat{K}^\perp_q(s) = \frac{1}{N}\sum_k \frac{Q_k\,s}{s^2+\lambda_k^2},
\end{align}
where
\begin{align}
	Q_k = \left|\sum_i e_k^i\exp(i q r_{x,i})\right|^2
\end{align}
denotes the projection of the $k$-th eigenvector onto the initial conditions. Note that the normalization in the velocity autocorrelation $1/N$ comes from the initial excited plane wave $v_j(0) = \exp(i q r_{x,j})$ and holds only in the thermodynamic limit (and for uniformly distributed positions $r_{x,j}$).
With this, the dispersion relation is given by
\begin{align}\label{eq:DispersionNumerics}
	\frac{1}{(qv_q^\perp)^2} = \lim_{s\to 0}\frac{K(q,s)}{s} = \frac{1}{N}\sum_{\lambda_k >0}\frac{Q_k}{\lambda_k}.
\end{align}
This method does not introduce inaccuracies due to binning procedures as it consists of a simple sum over all eigenfrequencies $\omega_k$. It should be noted that we only sum over non-zero eigenfrequencies. Numerically, we assume an eigenfrequency to be zero if $\lambda^k < 10^{-12}$ holds.\\

In the un-jammed phase we are interested in the non-ergodicity parameter $\kappa_q$ or in other words the finite value to which the velocity autocorrelation function decays for $t\to\infty$. This value is determined by the zero frequency modes. Let $N_0$ be the number of such modes $k_0$. Then we calculate the non-ergodicity parameter by
\begin{align}\label{eq:NEPNumerics}
	\kappa^\perp_q = \frac{1}{N}\sum_{k_0}Q_{k_0}
\end{align}

\paragraph{Numerical solution of the self-consistent equations}
To solve our model (for $s=0$), we first calculated the matrix  $\matr{C}$ and the high-density correction $\hat{\Delta}_q(0)$ according to the equations    \eqref{eq_app_Def_stability_matrix} and \eqref{eq_app_one_loop_cotribution} for each of the considered spring functions. After that, we solved the linear integral equations  \eqref{Main_text_eq_dispersion_relation} and \eqref{main_text_NEP_II} with a simple linear convergence algorithm. For the dispersion relation $q v_q^\perp$ did we chose the bare dispersion $qc_q^\perp$ as the starting value and for the non-ergodicity parameter did we chose $\kappa_q=1$ as the initial value. In both cases did we let the algorithm run until the distance between two interactions was smaller then $10^{-12}$. All the other observables shown like the Debye-level  and the length scale in the un-jammed phase could be calculated with the obtained $v_q^\perp$ and $\kappa_q$.   \\

When calculating the one loop correction and the stability matrix,  we performed the angular integrals numerically.  Since a 6 dimensional integral had to be solved, a numerical error is likely. Indeed, the diagonal elements of the coupling matrix are not necessarily positive anymore for $q\sigma<0.2$. This nonphysical fact  can result from the mentioned numerical error or from taking non-diagonal contributions into account.  We took 600 wavenumbers $q\sigma =[q_{min} \sigma,60]$ into account. We chose the minimal wavenumber for all spring functions considered in  such a way that all the diagonal elements of $C_{q,p}$ were positive. We checked different distributions of the wavenumbers and considered also even higher $q_{min}$. But this did not affect our results qualitatively nor quantitatively if $q_{min} $ was not chosen too large.  \\

The critical eigenvectors $h_q$ and $\hat{h}_p$ were obtained via the standard Matlab routine. Having these two  variables, we could calculate the observables  close to the transition and compare them to the general solutions of $v_q^\perp$ and $\kappa_q$ obtained via the equations \eqref{Main_text_eq_dispersion_relation} and \eqref{main_text_NEP_II}. We also calculated the value of the plateau appearing in the density of states  with $h_q$ according to equation \eqref{eq:constantDOS}. 

\subsection{Geometric Multiplicity of the eigenvalue zero \label{app_sec_geometric_multiplicity}}
We show in this subsection that the Geometric Multiplicity of the eigenvalue $0$ equals the number of disconnected clusters if the interaction is purely repulsive $f(r)\geq 0$. This is also confirmed numerically in figure \ref{fig:NumberVsMaxCluster}. \\

Let $\mathcal{A}$ be a partition  of $\{1,....,N\}$. We consider a normalised eigenvector $\7\phi \in \mathbb{R}^N$ of  $\matr{\mathcal{H}}$ with eigenvalue $\lambda$  and ask under which condition holds $\lambda=0$. We define the sets $A_n \in \mathcal{A}$ for $1 \leq n \leq M$ in the following way: $i,j \in A_n$ if and only if $\phi_i= \phi_j$.  One has
\begin{align}
    2 U= \7\phi^T \matr{\mathcal{H}} \7\phi = \lambda\;.
\end{align}
	Thus, the energy cost of an eigenmode is zero, if and only if  its eigenvalue is zero. We get
	\begin{align}
	\begin{split}
	    \frac{4U}{m \omega_0^2}=0& = \sum_{i,j}^N \underset{\geq 0}{ f(\7r_i-\7r_j)} \underset{\geq 0}{(\phi_i-\phi_j)^2}\\
	    &= \sum_{n,m=1}^M
	        \sum_{i \in A_n} \sum_{j \in A_m} f(\7r_i-\7r_j) (\phi_i-\phi_j)^2 \\
	        &= \sum_{n \neq m}
	        \sum_{i \in A_n} \sum_{j \in A_m} f(\7r_i-\7r_j) \underset{>0}{ (\phi_i-\phi_j)^2} \;.
	    	\end{split}
	\end{align}
	Thus, $f(\7r_i-\7r_j)=0$ has to hold for $i \in A_n$ and $j \in \{1,...,N\}\subset A_n$. Hence, the sets $A_i \subset \{1,....,N\} $ denote  the indices of disconnected clusters and the number $M= \# \mathcal{A}$ of disconnected cluster equals the number of linear independent (even orthogonal) eigenvectors with eigenvalue zero.  We denote them with $\phi^{(a)}$ with $1 \leq a \leq M$. They read
	\begin{align}
	    \phi^{(a)}\propto \sum_{i \in A_a} \hat{e}_i\;.
	\end{align}
Here, $\hat{e}_i$ denotes an element of the standard basis of $\mathbb{R}^N$. Note, that the assumed translational invariance implies that $M \geq 1$ holds. \\

As a consequence, finite restoring forces in the limit $q \to 0$ in the ERM-model can only exists if a cluster percolates the system. The speed of sound $v_q^\perp$ is only non-zero if the $n>n_c$, where $n_c$ is the percolation threshold.

\bibliography{sourcesCleanedUp.bib}

\begin{thebibliography}{86}%
\makeatletter
\providecommand \@ifxundefined [1]{%
 \@ifx{#1\undefined}
}%
\providecommand \@ifnum [1]{%
 \ifnum #1\expandafter \@firstoftwo
 \else \expandafter \@secondoftwo
 \fi
}%
\providecommand \@ifx [1]{%
 \ifx #1\expandafter \@firstoftwo
 \else \expandafter \@secondoftwo
 \fi
}%
\providecommand \natexlab [1]{#1}%
\providecommand \enquote  [1]{``#1''}%
\providecommand \bibnamefont  [1]{#1}%
\providecommand \bibfnamefont [1]{#1}%
\providecommand \citenamefont [1]{#1}%
\providecommand \href@noop [0]{\@secondoftwo}%
\providecommand \href [0]{\begingroup \@sanitize@url \@href}%
\providecommand \@href[1]{\@@startlink{#1}\@@href}%
\providecommand \@@href[1]{\endgroup#1\@@endlink}%
\providecommand \@sanitize@url [0]{\catcode `\\12\catcode `\$12\catcode
  `\&12\catcode `\#12\catcode `\^12\catcode `\_12\catcode `\%12\relax}%
\providecommand \@@startlink[1]{}%
\providecommand \@@endlink[0]{}%
\providecommand \url  [0]{\begingroup\@sanitize@url \@url }%
\providecommand \@url [1]{\endgroup\@href {#1}{\urlprefix }}%
\providecommand \urlprefix  [0]{URL }%
\providecommand \Eprint [0]{\href }%
\providecommand \doibase [0]{https://doi.org/}%
\providecommand \selectlanguage [0]{\@gobble}%
\providecommand \bibinfo  [0]{\@secondoftwo}%
\providecommand \bibfield  [0]{\@secondoftwo}%
\providecommand \translation [1]{[#1]}%
\providecommand \BibitemOpen [0]{}%
\providecommand \bibitemStop [0]{}%
\providecommand \bibitemNoStop [0]{.\EOS\space}%
\providecommand \EOS [0]{\spacefactor3000\relax}%
\providecommand \BibitemShut  [1]{\csname bibitem#1\endcsname}%
\let\auto@bib@innerbib\@empty
\bibitem [{\citenamefont {Ramos}(2022)}]{Ramos2022}%
  \BibitemOpen
  \bibfield  {author} {\bibinfo {author} {\bibfnamefont {M.~A.}\ \bibnamefont
  {Ramos}},\ }\href@noop {} {\emph {\bibinfo {title} {Low-temperature Thermal
  and Vibrational Properties of Disordered Solids: A Half-century of Universal"
  anomalies" of Glasses}}}\ (\bibinfo  {publisher} {World Scientific},\
  \bibinfo {year} {2022})\BibitemShut {NoStop}%
\bibitem [{\citenamefont {Ness}\ \emph {et~al.}(2022)\citenamefont {Ness},
  \citenamefont {Seto},\ and\ \citenamefont {Mari}}]{Ness2022}%
  \BibitemOpen
  \bibfield  {author} {\bibinfo {author} {\bibfnamefont {C.}~\bibnamefont
  {Ness}}, \bibinfo {author} {\bibfnamefont {R.}~\bibnamefont {Seto}},\ and\
  \bibinfo {author} {\bibfnamefont {R.}~\bibnamefont {Mari}},\ }\bibfield
  {title} {\bibinfo {title} {The physics of dense suspensions},\ }\href
  {https://doi.org/https://doi.org/10.1146/annurev-conmatphys-031620-105938}
  {\bibfield  {journal} {\bibinfo  {journal} {Ann. Rev. Condens. Mat. Phys.}\
  }\textbf {\bibinfo {volume} {13}},\ \bibinfo {pages} {97} (\bibinfo {year}
  {2022})}\BibitemShut {NoStop}%
\bibitem [{\citenamefont {Charbonneau}\ \emph {et~al.}(2017)\citenamefont
  {Charbonneau}, \citenamefont {Kurchan}, \citenamefont {Parisi}, \citenamefont
  {Urbani},\ and\ \citenamefont {Zamponi}}]{Charbonneau_Glass_and_jamming}%
  \BibitemOpen
  \bibfield  {author} {\bibinfo {author} {\bibfnamefont {P.}~\bibnamefont
  {Charbonneau}}, \bibinfo {author} {\bibfnamefont {J.}~\bibnamefont
  {Kurchan}}, \bibinfo {author} {\bibfnamefont {G.}~\bibnamefont {Parisi}},
  \bibinfo {author} {\bibfnamefont {P.}~\bibnamefont {Urbani}},\ and\ \bibinfo
  {author} {\bibfnamefont {F.}~\bibnamefont {Zamponi}},\ }\bibfield  {title}
  {\bibinfo {title} {Glass and jamming transitions: From exact results to
  finite-dimensional descriptions},\ }\href
  {https://doi.org/10.1146/annurev-conmatphys-031016-025334} {\bibfield
  {journal} {\bibinfo  {journal} {Annu. Rev. Condens. Matter Phys.}\ }\textbf
  {\bibinfo {volume} {8}},\ \bibinfo {pages} {265} (\bibinfo {year}
  {2017})}\BibitemShut {NoStop}%
\bibitem [{\citenamefont {Ikeda}\ \emph {et~al.}(2012)\citenamefont {Ikeda},
  \citenamefont {Berthier},\ and\ \citenamefont
  {Sollich}}]{Ikeda_Berthier_Sollich}%
  \BibitemOpen
  \bibfield  {author} {\bibinfo {author} {\bibfnamefont {A.}~\bibnamefont
  {Ikeda}}, \bibinfo {author} {\bibfnamefont {L.}~\bibnamefont {Berthier}},\
  and\ \bibinfo {author} {\bibfnamefont {P.}~\bibnamefont {Sollich}},\
  }\bibfield  {title} {\bibinfo {title} {Unified study of glass and jamming
  rheology in soft particle systems},\ }\href
  {https://doi.org/10.1103/PhysRevLett.109.018301} {\bibfield  {journal}
  {\bibinfo  {journal} {Phys. Rev. Lett.}\ }\textbf {\bibinfo {volume} {109}},\
  \bibinfo {pages} {018301} (\bibinfo {year} {2012})}\BibitemShut {NoStop}%
\bibitem [{\citenamefont {Benetti}\ \emph {et~al.}(2018)\citenamefont
  {Benetti}, \citenamefont {Parisi}, \citenamefont {Pietracaprina},\ and\
  \citenamefont {Sicuro}}]{Mean_field_Paris_2018}%
  \BibitemOpen
  \bibfield  {author} {\bibinfo {author} {\bibfnamefont {F.~P.~C.}\
  \bibnamefont {Benetti}}, \bibinfo {author} {\bibfnamefont {G.}~\bibnamefont
  {Parisi}}, \bibinfo {author} {\bibfnamefont {F.}~\bibnamefont
  {Pietracaprina}},\ and\ \bibinfo {author} {\bibfnamefont {G.}~\bibnamefont
  {Sicuro}},\ }\bibfield  {title} {\bibinfo {title} {Mean-field model for the
  density of states of jammed soft spheres},\ }\href
  {https://doi.org/10.1103/PhysRevE.97.062157} {\bibfield  {journal} {\bibinfo
  {journal} {Phys. Rev. E}\ }\textbf {\bibinfo {volume} {97}},\ \bibinfo
  {pages} {062157} (\bibinfo {year} {2018})}\BibitemShut {NoStop}%
\bibitem [{\citenamefont {DeGiuli}\ \emph {et~al.}(2014)\citenamefont
  {DeGiuli}, \citenamefont {Laversanne-Finot}, \citenamefont {Düring},
  \citenamefont {Lerner},\ and\ \citenamefont {Wyart}}]{DeGiuli2014}%
  \BibitemOpen
  \bibfield  {author} {\bibinfo {author} {\bibfnamefont {E.}~\bibnamefont
  {DeGiuli}}, \bibinfo {author} {\bibfnamefont {A.}~\bibnamefont
  {Laversanne-Finot}}, \bibinfo {author} {\bibfnamefont {G.}~\bibnamefont
  {Düring}}, \bibinfo {author} {\bibfnamefont {E.}~\bibnamefont {Lerner}},\
  and\ \bibinfo {author} {\bibfnamefont {M.}~\bibnamefont {Wyart}},\ }\bibfield
   {title} {\bibinfo {title} {Effects of coordination and pressure on sound
  attenuation{,} boson peak and elasticity in amorphous solids},\ }\href
  {https://doi.org/10.1039/C4SM00561A} {\bibfield  {journal} {\bibinfo
  {journal} {Soft Matter}\ }\textbf {\bibinfo {volume} {10}},\ \bibinfo {pages}
  {5628} (\bibinfo {year} {2014})}\BibitemShut {NoStop}%
\bibitem [{\citenamefont {Kauzmann}(1948)}]{Kauzmann1948TheNO}%
  \BibitemOpen
  \bibfield  {author} {\bibinfo {author} {\bibfnamefont {W.}~\bibnamefont
  {Kauzmann}},\ }\bibfield  {title} {\bibinfo {title} {The nature of the glassy
  state and the behavior of liquids at low temperatures.},\ }\href@noop {}
  {\bibfield  {journal} {\bibinfo  {journal} {Chem. Rev.}\ }\textbf {\bibinfo
  {volume} {43}},\ \bibinfo {pages} {219} (\bibinfo {year} {1948})}\BibitemShut
  {NoStop}%
\bibitem [{\citenamefont {Debenedetti}\ and\ \citenamefont
  {Stillinger}(2001)}]{g2001supercooled}%
  \BibitemOpen
  \bibfield  {author} {\bibinfo {author} {\bibfnamefont {P.~G.}\ \bibnamefont
  {Debenedetti}}\ and\ \bibinfo {author} {\bibfnamefont {F.~H.}\ \bibnamefont
  {Stillinger}},\ }\bibfield  {title} {\bibinfo {title} {Supercooled liquids
  and the glass transition},\ }\href@noop {} {\bibfield  {journal} {\bibinfo
  {journal} {Nature}\ }\textbf {\bibinfo {volume} {410}},\ \bibinfo {pages}
  {259} (\bibinfo {year} {2001})}\BibitemShut {NoStop}%
\bibitem [{\citenamefont {Feng}\ \emph {et~al.}(1985)\citenamefont {Feng},
  \citenamefont {Thorpe},\ and\ \citenamefont {Garboczi}}]{Thorpe_1985}%
  \BibitemOpen
  \bibfield  {author} {\bibinfo {author} {\bibfnamefont {S.}~\bibnamefont
  {Feng}}, \bibinfo {author} {\bibfnamefont {M.~F.}\ \bibnamefont {Thorpe}},\
  and\ \bibinfo {author} {\bibfnamefont {E.}~\bibnamefont {Garboczi}},\
  }\bibfield  {title} {\bibinfo {title} {Effective-medium theory of percolation
  on central-force elastic networks},\ }\href
  {https://doi.org/10.1103/PhysRevB.31.276} {\bibfield  {journal} {\bibinfo
  {journal} {Phys. Rev. B}\ }\textbf {\bibinfo {volume} {31}},\ \bibinfo
  {pages} {276} (\bibinfo {year} {1985})}\BibitemShut {NoStop}%
\bibitem [{\citenamefont {Franz}\ and\ \citenamefont
  {Parisi}(2016)}]{Franz_2016}%
  \BibitemOpen
  \bibfield  {author} {\bibinfo {author} {\bibfnamefont {S.}~\bibnamefont
  {Franz}}\ and\ \bibinfo {author} {\bibfnamefont {G.}~\bibnamefont {Parisi}},\
  }\bibfield  {title} {\bibinfo {title} {The simplest model of jamming},\
  }\href {https://doi.org/10.1088/1751-8113/49/14/145001} {\bibfield  {journal}
  {\bibinfo  {journal} {J. Phys. A Math. Theor.}\ }\textbf {\bibinfo {volume}
  {49}},\ \bibinfo {pages} {145001} (\bibinfo {year} {2016})}\BibitemShut
  {NoStop}%
\bibitem [{\citenamefont {Alexander}(1998)}]{ALEXANDER199865}%
  \BibitemOpen
  \bibfield  {author} {\bibinfo {author} {\bibfnamefont {S.}~\bibnamefont
  {Alexander}},\ }\bibfield  {title} {\bibinfo {title} {Amorphous solids: their
  structure, lattice dynamics and elasticity},\ }\href
  {https://doi.org/https://doi.org/10.1016/S0370-1573(97)00069-0} {\bibfield
  {journal} {\bibinfo  {journal} {Phys. Rep.}\ }\textbf {\bibinfo {volume}
  {296}},\ \bibinfo {pages} {65} (\bibinfo {year} {1998})}\BibitemShut
  {NoStop}%
\bibitem [{\citenamefont {Mizuno}\ and\ \citenamefont
  {Ikeda}(2018)}]{Ikeda_Phonon_transport}%
  \BibitemOpen
  \bibfield  {author} {\bibinfo {author} {\bibfnamefont {H.}~\bibnamefont
  {Mizuno}}\ and\ \bibinfo {author} {\bibfnamefont {A.}~\bibnamefont {Ikeda}},\
  }\bibfield  {title} {\bibinfo {title} {Phonon transport and vibrational
  excitations in amorphous solids},\ }\href
  {https://doi.org/10.1103/PhysRevE.98.062612} {\bibfield  {journal} {\bibinfo
  {journal} {Phys. Rev. E}\ }\textbf {\bibinfo {volume} {98}},\ \bibinfo
  {pages} {062612} (\bibinfo {year} {2018})}\BibitemShut {NoStop}%
\bibitem [{\citenamefont {Ikeda}\ \emph {et~al.}(2013)\citenamefont {Ikeda},
  \citenamefont {Berthier},\ and\ \citenamefont {Biroli}}]{Ikeda_2013}%
  \BibitemOpen
  \bibfield  {author} {\bibinfo {author} {\bibfnamefont {A.}~\bibnamefont
  {Ikeda}}, \bibinfo {author} {\bibfnamefont {L.}~\bibnamefont {Berthier}},\
  and\ \bibinfo {author} {\bibfnamefont {G.}~\bibnamefont {Biroli}},\
  }\bibfield  {title} {\bibinfo {title} {Dynamic criticality at the jamming
  transition},\ }\href {https://doi.org/10.1063/1.4769251} {\bibfield
  {journal} {\bibinfo  {journal} {J. Chem. Phys.}\ }\textbf {\bibinfo {volume}
  {138}},\ \bibinfo {pages} {12A507} (\bibinfo {year} {2013})}\BibitemShut
  {NoStop}%
\bibitem [{\citenamefont {Horbach}\ \emph {et~al.}(2001)\citenamefont
  {Horbach}, \citenamefont {Kob},\ and\ \citenamefont
  {Binder}}]{Horbach2001HighFS}%
  \BibitemOpen
  \bibfield  {author} {\bibinfo {author} {\bibfnamefont {J.}~\bibnamefont
  {Horbach}}, \bibinfo {author} {\bibfnamefont {W.}~\bibnamefont {Kob}},\ and\
  \bibinfo {author} {\bibfnamefont {K.}~\bibnamefont {Binder}},\ }\bibfield
  {title} {\bibinfo {title} {High frequency sound and the boson peak in
  amorphous silica},\ }\href@noop {} {\bibfield  {journal} {\bibinfo  {journal}
  {Eur. Phys. J. B}\ }\textbf {\bibinfo {volume} {19}},\ \bibinfo {pages} {531}
  (\bibinfo {year} {2001})}\BibitemShut {NoStop}%
\bibitem [{\citenamefont {Wang}\ \emph {et~al.}(2019)\citenamefont {Wang},
  \citenamefont {Berthier}, \citenamefont {Flenner}, \citenamefont {Guan},\
  and\ \citenamefont {Szamel}}]{Wang_Stable_glasses}%
  \BibitemOpen
  \bibfield  {author} {\bibinfo {author} {\bibfnamefont {L.}~\bibnamefont
  {Wang}}, \bibinfo {author} {\bibfnamefont {L.}~\bibnamefont {Berthier}},
  \bibinfo {author} {\bibfnamefont {E.}~\bibnamefont {Flenner}}, \bibinfo
  {author} {\bibfnamefont {P.}~\bibnamefont {Guan}},\ and\ \bibinfo {author}
  {\bibfnamefont {G.}~\bibnamefont {Szamel}},\ }\bibfield  {title} {\bibinfo
  {title} {Sound attenuation in stable glasses},\ }\href
  {https://doi.org/10.1039/C9SM01092K} {\bibfield  {journal} {\bibinfo
  {journal} {Soft Matter}\ }\textbf {\bibinfo {volume} {15}},\ \bibinfo {pages}
  {7018} (\bibinfo {year} {2019})}\BibitemShut {NoStop}%
\bibitem [{\citenamefont {Schirmacher}\ and\ \citenamefont
  {Ruocco}(2020)}]{Schirmacher_Heterogeneous_Elasticity}%
  \BibitemOpen
  \bibfield  {author} {\bibinfo {author} {\bibfnamefont {W.}~\bibnamefont
  {Schirmacher}}\ and\ \bibinfo {author} {\bibfnamefont {G.}~\bibnamefont
  {Ruocco}},\ }\href@noop {} {\bibinfo {title} {Heterogeneous elasticity: The
  tale of the boson peak}} (\bibinfo {year} {2020}),\ \Eprint
  {https://arxiv.org/abs/2009.05970} {arXiv:2009.05970} \BibitemShut {NoStop}%
\bibitem [{\citenamefont {Schirmacher}\ \emph {et~al.}(2007)\citenamefont
  {Schirmacher}, \citenamefont {Ruocco},\ and\ \citenamefont
  {Scopigno}}]{Schirmacher2007}%
  \BibitemOpen
  \bibfield  {author} {\bibinfo {author} {\bibfnamefont {W.}~\bibnamefont
  {Schirmacher}}, \bibinfo {author} {\bibfnamefont {G.}~\bibnamefont
  {Ruocco}},\ and\ \bibinfo {author} {\bibfnamefont {T.}~\bibnamefont
  {Scopigno}},\ }\bibfield  {title} {\bibinfo {title} {Acoustic attenuation in
  glasses and its relation with the boson peak},\ }\href
  {https://doi.org/10.1103/PhysRevLett.98.025501} {\bibfield  {journal}
  {\bibinfo  {journal} {Phys. Rev. Lett.}\ }\textbf {\bibinfo {volume} {98}},\
  \bibinfo {pages} {025501} (\bibinfo {year} {2007})}\BibitemShut {NoStop}%
\bibitem [{\citenamefont {Franz}\ \emph {et~al.}(2015)\citenamefont {Franz},
  \citenamefont {Parisi}, \citenamefont {Urbani},\ and\ \citenamefont
  {Zamponi}}]{Franz2015}%
  \BibitemOpen
  \bibfield  {author} {\bibinfo {author} {\bibfnamefont {S.}~\bibnamefont
  {Franz}}, \bibinfo {author} {\bibfnamefont {G.}~\bibnamefont {Parisi}},
  \bibinfo {author} {\bibfnamefont {P.}~\bibnamefont {Urbani}},\ and\ \bibinfo
  {author} {\bibfnamefont {F.}~\bibnamefont {Zamponi}},\ }\bibfield  {title}
  {\bibinfo {title} {Universal spectrum of normal modes in low-temperature
  glasses},\ }\href@noop {} {\bibfield  {journal} {\bibinfo  {journal} {Proc.
  Nat. Acad. Sci.}\ }\textbf {\bibinfo {volume} {112}},\ \bibinfo {pages}
  {14539} (\bibinfo {year} {2015})}\BibitemShut {NoStop}%
\bibitem [{\citenamefont {Baldi}\ \emph {et~al.}(2013)\citenamefont {Baldi},
  \citenamefont {Zanatta}, \citenamefont {Gilioli}, \citenamefont {Milman},
  \citenamefont {Refson}, \citenamefont {Wehinger}, \citenamefont {Winkler},
  \citenamefont {Fontana},\ and\ \citenamefont {Monaco}}]{BaldiEmergence}%
  \BibitemOpen
  \bibfield  {author} {\bibinfo {author} {\bibfnamefont {G.}~\bibnamefont
  {Baldi}}, \bibinfo {author} {\bibfnamefont {M.}~\bibnamefont {Zanatta}},
  \bibinfo {author} {\bibfnamefont {E.}~\bibnamefont {Gilioli}}, \bibinfo
  {author} {\bibfnamefont {V.}~\bibnamefont {Milman}}, \bibinfo {author}
  {\bibfnamefont {K.}~\bibnamefont {Refson}}, \bibinfo {author} {\bibfnamefont
  {B.}~\bibnamefont {Wehinger}}, \bibinfo {author} {\bibfnamefont
  {B.}~\bibnamefont {Winkler}}, \bibinfo {author} {\bibfnamefont
  {A.}~\bibnamefont {Fontana}},\ and\ \bibinfo {author} {\bibfnamefont
  {G.}~\bibnamefont {Monaco}},\ }\bibfield  {title} {\bibinfo {title}
  {Emergence of crystal-like atomic dynamics in glasses at the nanometer
  scale},\ }\href {https://doi.org/10.1103/PhysRevLett.110.185503} {\bibfield
  {journal} {\bibinfo  {journal} {Phys. Rev. Lett.}\ }\textbf {\bibinfo
  {volume} {110}},\ \bibinfo {pages} {185503} (\bibinfo {year}
  {2013})}\BibitemShut {NoStop}%
\bibitem [{\citenamefont {Baldi}\ \emph {et~al.}(2014)\citenamefont {Baldi},
  \citenamefont {Giordano}, \citenamefont {Ruta}, \citenamefont {Dal~Maschio},
  \citenamefont {Fontana},\ and\ \citenamefont
  {Monaco}}]{PhysRevLett.112.125502}%
  \BibitemOpen
  \bibfield  {author} {\bibinfo {author} {\bibfnamefont {G.}~\bibnamefont
  {Baldi}}, \bibinfo {author} {\bibfnamefont {V.~M.}\ \bibnamefont {Giordano}},
  \bibinfo {author} {\bibfnamefont {B.}~\bibnamefont {Ruta}}, \bibinfo {author}
  {\bibfnamefont {R.}~\bibnamefont {Dal~Maschio}}, \bibinfo {author}
  {\bibfnamefont {A.}~\bibnamefont {Fontana}},\ and\ \bibinfo {author}
  {\bibfnamefont {G.}~\bibnamefont {Monaco}},\ }\bibfield  {title} {\bibinfo
  {title} {Anharmonic damping of terahertz acoustic waves in a network glass
  and its effect on the density of vibrational states},\ }\href
  {https://doi.org/10.1103/PhysRevLett.112.125502} {\bibfield  {journal}
  {\bibinfo  {journal} {Phys. Rev. Lett.}\ }\textbf {\bibinfo {volume} {112}},\
  \bibinfo {pages} {125502} (\bibinfo {year} {2014})}\BibitemShut {NoStop}%
\bibitem [{\citenamefont {Baldi}\ \emph {et~al.}(2010)\citenamefont {Baldi},
  \citenamefont {Giordano}, \citenamefont {Monaco},\ and\ \citenamefont
  {Ruta}}]{PhysRevLett.104.195501}%
  \BibitemOpen
  \bibfield  {author} {\bibinfo {author} {\bibfnamefont {G.}~\bibnamefont
  {Baldi}}, \bibinfo {author} {\bibfnamefont {V.~M.}\ \bibnamefont {Giordano}},
  \bibinfo {author} {\bibfnamefont {G.}~\bibnamefont {Monaco}},\ and\ \bibinfo
  {author} {\bibfnamefont {B.}~\bibnamefont {Ruta}},\ }\bibfield  {title}
  {\bibinfo {title} {Sound attenuation at terahertz frequencies and the boson
  peak of vitreous silica},\ }\href
  {https://doi.org/10.1103/PhysRevLett.104.195501} {\bibfield  {journal}
  {\bibinfo  {journal} {Phys. Rev. Lett.}\ }\textbf {\bibinfo {volume} {104}},\
  \bibinfo {pages} {195501} (\bibinfo {year} {2010})}\BibitemShut {NoStop}%
\bibitem [{\citenamefont {Ciliberti}\ \emph {et~al.}(2003)\citenamefont
  {Ciliberti}, \citenamefont {Grigera}, \citenamefont {Martin-Mayor},
  \citenamefont {Parisi},\ and\ \citenamefont
  {Verrocchio}}]{ciliberti2003brillouin}%
  \BibitemOpen
  \bibfield  {author} {\bibinfo {author} {\bibfnamefont {S.}~\bibnamefont
  {Ciliberti}}, \bibinfo {author} {\bibfnamefont {T.}~\bibnamefont {Grigera}},
  \bibinfo {author} {\bibfnamefont {V.}~\bibnamefont {Martin-Mayor}}, \bibinfo
  {author} {\bibfnamefont {G.}~\bibnamefont {Parisi}},\ and\ \bibinfo {author}
  {\bibfnamefont {P.}~\bibnamefont {Verrocchio}},\ }\bibfield  {title}
  {\bibinfo {title} {Brillouin and boson peaks in glasses from vector euclidean
  random matrix theory},\ }\href@noop {} {\bibfield  {journal} {\bibinfo
  {journal} {J. Chem. Phys.}\ }\textbf {\bibinfo {volume} {119}},\ \bibinfo
  {pages} {8577} (\bibinfo {year} {2003})}\BibitemShut {NoStop}%
\bibitem [{\citenamefont {Goetschy}\ and\ \citenamefont
  {Skipetrov}(2013)}]{goetschy2013euclidean}%
  \BibitemOpen
  \bibfield  {author} {\bibinfo {author} {\bibfnamefont {A.}~\bibnamefont
  {Goetschy}}\ and\ \bibinfo {author} {\bibfnamefont {S.~E.}\ \bibnamefont
  {Skipetrov}},\ }\href@noop {} {\bibinfo {title} {Euclidean random matrices
  and their applications in physics}} (\bibinfo {year} {2013}),\ \Eprint
  {https://arxiv.org/abs/1303.2880} {arXiv:1303.2880} \BibitemShut {NoStop}%
\bibitem [{\citenamefont {Grigera}\ \emph {et~al.}(2011)\citenamefont
  {Grigera}, \citenamefont {Martin-Mayor}, \citenamefont {Parisi},
  \citenamefont {Urbani},\ and\ \citenamefont {Verrocchio}}]{grigera2011high}%
  \BibitemOpen
  \bibfield  {author} {\bibinfo {author} {\bibfnamefont {T.~S.}\ \bibnamefont
  {Grigera}}, \bibinfo {author} {\bibfnamefont {V.}~\bibnamefont
  {Martin-Mayor}}, \bibinfo {author} {\bibfnamefont {G.}~\bibnamefont
  {Parisi}}, \bibinfo {author} {\bibfnamefont {P.}~\bibnamefont {Urbani}},\
  and\ \bibinfo {author} {\bibfnamefont {P.}~\bibnamefont {Verrocchio}},\
  }\bibfield  {title} {\bibinfo {title} {On the high-density expansion for
  euclidean random matrices},\ }\href@noop {} {\bibfield  {journal} {\bibinfo
  {journal} {J. Stat. Mech.: Theory Exp.}\ }\textbf {\bibinfo {volume}
  {2011}}\bibinfo  {number} { (02)},\ \bibinfo {pages} {P02015}}\BibitemShut
  {NoStop}%
\bibitem [{\citenamefont {Vogel}\ and\ \citenamefont
  {Fuchs}(2023)}]{Vogel_ERM}%
  \BibitemOpen
\bibfield  {number} {  }\bibfield  {author} {\bibinfo {author} {\bibfnamefont
  {F.}~\bibnamefont {Vogel}}\ and\ \bibinfo {author} {\bibfnamefont
  {M.}~\bibnamefont {Fuchs}},\ }\bibfield  {title} {\bibinfo {title}
  {Vibrational phenomena in glasses at low temperatures captured by field
  theory of disordered harmonic oscillators},\ }\href
  {https://doi.org/10.1103/PhysRevLett.130.236101} {\bibfield  {journal}
  {\bibinfo  {journal} {Phys. Rev. Lett.}\ }\textbf {\bibinfo {volume} {130}},\
  \bibinfo {pages} {236101} (\bibinfo {year} {2023})}\BibitemShut {NoStop}%
\bibitem [{\citenamefont {Ganter}\ and\ \citenamefont
  {Schirmacher}(2010)}]{Ganter_Schirmacher}%
  \BibitemOpen
  \bibfield  {author} {\bibinfo {author} {\bibfnamefont {C.}~\bibnamefont
  {Ganter}}\ and\ \bibinfo {author} {\bibfnamefont {W.}~\bibnamefont
  {Schirmacher}},\ }\bibfield  {title} {\bibinfo {title} {Rayleigh scattering,
  long-time tails, and the harmonic spectrum of topologically disordered
  systems},\ }\href {https://doi.org/10.1103/PhysRevB.82.094205} {\bibfield
  {journal} {\bibinfo  {journal} {Phys. Rev. B}\ }\textbf {\bibinfo {volume}
  {82}},\ \bibinfo {pages} {094205} (\bibinfo {year} {2010})}\BibitemShut
  {NoStop}%
\bibitem [{\citenamefont {Szamel}\ and\ \citenamefont
  {Flenner}(2022)}]{Szamel2022}%
  \BibitemOpen
  \bibfield  {author} {\bibinfo {author} {\bibfnamefont {G.}~\bibnamefont
  {Szamel}}\ and\ \bibinfo {author} {\bibfnamefont {E.}~\bibnamefont
  {Flenner}},\ }\bibfield  {title} {\bibinfo {title} {Microscopic analysis of
  sound attenuation in low-temperature amorphous solids reveals quantitative
  importance of non-affine effects},\ }\href
  {https://doi.org/10.1063/5.0085199} {\bibfield  {journal} {\bibinfo
  {journal} {J. Chem. Phys.}\ }\textbf {\bibinfo {volume} {156}},\ \bibinfo
  {pages} {144502} (\bibinfo {year} {2022})}\BibitemShut {NoStop}%
\bibitem [{\citenamefont {Schirmacher}\ \emph {et~al.}(2019)\citenamefont
  {Schirmacher}, \citenamefont {Folli}, \citenamefont {Ganter},\ and\
  \citenamefont {Ruocco}}]{schirmacher2019self}%
  \BibitemOpen
  \bibfield  {author} {\bibinfo {author} {\bibfnamefont {W.}~\bibnamefont
  {Schirmacher}}, \bibinfo {author} {\bibfnamefont {V.}~\bibnamefont {Folli}},
  \bibinfo {author} {\bibfnamefont {C.}~\bibnamefont {Ganter}},\ and\ \bibinfo
  {author} {\bibfnamefont {G.}~\bibnamefont {Ruocco}},\ }\bibfield  {title}
  {\bibinfo {title} {Self-consistent euclidean-random-matrix theory},\
  }\href@noop {} {\bibfield  {journal} {\bibinfo  {journal} {J. Phys. A Math.
  Theor.}\ }\textbf {\bibinfo {volume} {52}},\ \bibinfo {pages} {464002}
  (\bibinfo {year} {2019})}\BibitemShut {NoStop}%
\bibitem [{\citenamefont {Szamel}(2023)}]{Szamel_2023_Elastic_Constant}%
  \BibitemOpen
  \bibfield  {author} {\bibinfo {author} {\bibfnamefont {G.}~\bibnamefont
  {Szamel}},\ }\bibfield  {title} {\bibinfo {title} {Elastic constants of
  zero-temperature amorphous solids},\ }\href
  {https://doi.org/10.1103/PhysRevE.107.064608} {\bibfield  {journal} {\bibinfo
   {journal} {Phys. Rev. E}\ }\textbf {\bibinfo {volume} {107}},\ \bibinfo
  {pages} {064608} (\bibinfo {year} {2023})}\BibitemShut {NoStop}%
\bibitem [{\citenamefont {Baumg\"artel}\ \emph {et~al.}(2024)\citenamefont
  {Baumg\"artel}, \citenamefont {Vogel},\ and\ \citenamefont
  {Fuchs}}]{Baumgaertel2023properties}%
  \BibitemOpen
  \bibfield  {author} {\bibinfo {author} {\bibfnamefont {P.}~\bibnamefont
  {Baumg\"artel}}, \bibinfo {author} {\bibfnamefont {F.}~\bibnamefont
  {Vogel}},\ and\ \bibinfo {author} {\bibfnamefont {M.}~\bibnamefont {Fuchs}},\
  }\bibfield  {title} {\bibinfo {title} {Properties of stable ensembles of
  euclidean random matrices},\ }\href
  {https://doi.org/10.1103/PhysRevE.109.014120} {\bibfield  {journal} {\bibinfo
   {journal} {Phys. Rev. E}\ }\textbf {\bibinfo {volume} {109}},\ \bibinfo
  {pages} {014120} (\bibinfo {year} {2024})}\BibitemShut {NoStop}%
\bibitem [{\citenamefont {Schober}\ and\ \citenamefont
  {Ruocco}(2004)}]{Schober_Localised_modes}%
  \BibitemOpen
  \bibfield  {author} {\bibinfo {author} {\bibfnamefont {H.~R.}\ \bibnamefont
  {Schober}}\ and\ \bibinfo {author} {\bibfnamefont {G.}~\bibnamefont
  {Ruocco}},\ }\bibfield  {title} {\bibinfo {title} {Size effects and
  quasilocalized vibrations},\ }\href
  {https://doi.org/10.1080/14786430310001644107} {\bibfield  {journal}
  {\bibinfo  {journal} {Phil. Mag.}\ }\textbf {\bibinfo {volume} {84}},\
  \bibinfo {pages} {1361} (\bibinfo {year} {2004})}\BibitemShut {NoStop}%
\bibitem [{\citenamefont {Schober}(2011)}]{SCHOBER2011}%
  \BibitemOpen
  \bibfield  {author} {\bibinfo {author} {\bibfnamefont {H.}~\bibnamefont
  {Schober}},\ }\bibfield  {title} {\bibinfo {title} {Quasi-localized
  vibrations and phonon damping in glasses},\ }\href
  {https://doi.org/https://doi.org/10.1016/j.jnoncrysol.2010.07.036} {\bibfield
   {journal} {\bibinfo  {journal} {J. Non-Cryst. Solids}\ }\textbf {\bibinfo
  {volume} {357}},\ \bibinfo {pages} {501} (\bibinfo {year} {2011})},\ \bibinfo
  {note} {6th International Discussion Meeting on Relaxation in Complex
  Systems}\BibitemShut {NoStop}%
\bibitem [{\citenamefont {Lerner}\ and\ \citenamefont
  {Bouchbinder}(2021)}]{Lerner_2021}%
  \BibitemOpen
  \bibfield  {author} {\bibinfo {author} {\bibfnamefont {E.}~\bibnamefont
  {Lerner}}\ and\ \bibinfo {author} {\bibfnamefont {E.}~\bibnamefont
  {Bouchbinder}},\ }\bibfield  {title} {\bibinfo {title} {Low-energy
  quasilocalized excitations in structural glasses},\ }\href
  {https://doi.org/10.1063/5.0069477} {\bibfield  {journal} {\bibinfo
  {journal} {J. Chem. Phys.}\ }\textbf {\bibinfo {volume} {155}},\ \bibinfo
  {pages} {200901} (\bibinfo {year} {2021})}\BibitemShut {NoStop}%
\bibitem [{\citenamefont {Szamel}(2003)}]{Szamel2003}%
  \BibitemOpen
  \bibfield  {author} {\bibinfo {author} {\bibfnamefont {G.}~\bibnamefont
  {Szamel}},\ }\bibfield  {title} {\bibinfo {title} {Colloidal glass
  transition: Beyond mode-coupling theory},\ }\href
  {https://doi.org/10.1103/PhysRevLett.90.228301} {\bibfield  {journal}
  {\bibinfo  {journal} {Phys. Rev. Lett.}\ }\textbf {\bibinfo {volume} {90}},\
  \bibinfo {pages} {228301} (\bibinfo {year} {2003})}\BibitemShut {NoStop}%
\bibitem [{\citenamefont {Janssen}\ and\ \citenamefont
  {Reichman}(2015)}]{Janssen2015}%
  \BibitemOpen
  \bibfield  {author} {\bibinfo {author} {\bibfnamefont {L.~M.~C.}\
  \bibnamefont {Janssen}}\ and\ \bibinfo {author} {\bibfnamefont {D.~R.}\
  \bibnamefont {Reichman}},\ }\bibfield  {title} {\bibinfo {title} {Microscopic
  dynamics of supercooled liquids from first principles},\ }\href
  {https://doi.org/10.1103/PhysRevLett.115.205701} {\bibfield  {journal}
  {\bibinfo  {journal} {Phys. Rev. Lett.}\ }\textbf {\bibinfo {volume} {115}},\
  \bibinfo {pages} {205701} (\bibinfo {year} {2015})}\BibitemShut {NoStop}%
\bibitem [{\citenamefont {Janssen}(2018)}]{Janssen2018}%
  \BibitemOpen
  \bibfield  {author} {\bibinfo {author} {\bibfnamefont {L.~M.~C.}\
  \bibnamefont {Janssen}},\ }\bibfield  {title} {\bibinfo {title}
  {Mode-coupling theory of the glass transition: A primer},\ }\href
  {https://doi.org/10.3389/fphy.2018.00097} {\bibfield  {journal} {\bibinfo
  {journal} {Front. Phys.}\ }\textbf {\bibinfo {volume} {6}},\ \bibinfo {pages}
  {97} (\bibinfo {year} {2018})}\BibitemShut {NoStop}%
\bibitem [{\citenamefont {Franosch}\ and\ \citenamefont
  {Götze}(1994)}]{T_Franosch_1994}%
  \BibitemOpen
  \bibfield  {author} {\bibinfo {author} {\bibfnamefont {T.}~\bibnamefont
  {Franosch}}\ and\ \bibinfo {author} {\bibfnamefont {W.}~\bibnamefont
  {Götze}},\ }\bibfield  {title} {\bibinfo {title} {A theory for a certain
  crossover in relaxation phenomena in glasses},\ }\href
  {https://doi.org/10.1088/0953-8984/6/26/004} {\bibfield  {journal} {\bibinfo
  {journal} {J. Phys. Condens. Mat.}\ }\textbf {\bibinfo {volume} {6}},\
  \bibinfo {pages} {4807} (\bibinfo {year} {1994})}\BibitemShut {NoStop}%
\bibitem [{\citenamefont {Goyon}\ \emph {et~al.}(2008)\citenamefont {Goyon},
  \citenamefont {Colin}, \citenamefont {Ovarlez}, \citenamefont {Ajdari},\ and\
  \citenamefont {Bocquet}}]{Goyon_Spatial_cooperativity_2008}%
  \BibitemOpen
  \bibfield  {author} {\bibinfo {author} {\bibfnamefont {J.}~\bibnamefont
  {Goyon}}, \bibinfo {author} {\bibfnamefont {A.}~\bibnamefont {Colin}},
  \bibinfo {author} {\bibfnamefont {G.}~\bibnamefont {Ovarlez}}, \bibinfo
  {author} {\bibfnamefont {A.}~\bibnamefont {Ajdari}},\ and\ \bibinfo {author}
  {\bibfnamefont {L.}~\bibnamefont {Bocquet}},\ }\bibfield  {title} {\bibinfo
  {title} {Spatial cooperativity in soft glassy flows},\ }\href
  {https://doi.org/10.1038/nature07026} {\bibfield  {journal} {\bibinfo
  {journal} {Nature}\ }\textbf {\bibinfo {volume} {454}},\ \bibinfo {pages}
  {84} (\bibinfo {year} {2008})}\BibitemShut {NoStop}%
\bibitem [{\citenamefont {Hansen}\ and\ \citenamefont
  {McDonald}(2009)}]{HansenMcDonald}%
  \BibitemOpen
  \bibfield  {author} {\bibinfo {author} {\bibfnamefont {J.}~\bibnamefont
  {Hansen}}\ and\ \bibinfo {author} {\bibfnamefont {J.}~\bibnamefont
  {McDonald}},\ }\href@noop {} {\emph {\bibinfo {title} {Theory of simple
  liquids}}},\ Vol.~\bibinfo {volume} {3}\ (\bibinfo  {publisher} {Elsevier
  Science B.V},\ \bibinfo {year} {2009})\BibitemShut {NoStop}%
\bibitem [{\citenamefont {Voigtmann}(2011)}]{Voigtmann2011YieldSA}%
  \BibitemOpen
  \bibfield  {author} {\bibinfo {author} {\bibfnamefont {T.}~\bibnamefont
  {Voigtmann}},\ }\bibfield  {title} {\bibinfo {title} {Yield stresses and flow
  curves in metallic glass formers and granular systems},\ }\href@noop {}
  {\bibfield  {journal} {\bibinfo  {journal} {Eur. Phys. J. E}\ }\textbf
  {\bibinfo {volume} {34}},\ \bibinfo {pages} {1} (\bibinfo {year}
  {2011})}\BibitemShut {NoStop}%
\bibitem [{\citenamefont {Tanguy}(2023)}]{tanguy2023vibrations}%
  \BibitemOpen
  \bibfield  {author} {\bibinfo {author} {\bibfnamefont {A.}~\bibnamefont
  {Tanguy}},\ }\href@noop {} {\bibinfo {title} {Vibrations and heat transfer in
  glasses: the role played by disorder}} (\bibinfo {year} {2023}),\ \Eprint
  {https://arxiv.org/abs/2307.15038} {arXiv:2307.15038} \BibitemShut {NoStop}%
\bibitem [{\citenamefont {Yang}\ and\ \citenamefont
  {Schweizer}(2011)}]{Yang_Schweizer_Glassy_dynamics_2011}%
  \BibitemOpen
  \bibfield  {author} {\bibinfo {author} {\bibfnamefont {J.}~\bibnamefont
  {Yang}}\ and\ \bibinfo {author} {\bibfnamefont {K.~S.}\ \bibnamefont
  {Schweizer}},\ }\bibfield  {title} {\bibinfo {title} {{Glassy dynamics and
  mechanical response in dense fluids of soft repulsive spheres. I. Activated
  relaxation, kinetic vitrification, and fragility}},\ }\href
  {https://doi.org/10.1063/1.3592563} {\bibfield  {journal} {\bibinfo
  {journal} {J. Chem. Phys.}\ }\textbf {\bibinfo {volume} {134}},\ \bibinfo
  {pages} {204908} (\bibinfo {year} {2011})}\BibitemShut {NoStop}%
\bibitem [{\citenamefont {Jacquin}\ and\ \citenamefont
  {Berthier}(2010)}]{Jacquin_Berthier_2010}%
  \BibitemOpen
  \bibfield  {author} {\bibinfo {author} {\bibfnamefont {H.}~\bibnamefont
  {Jacquin}}\ and\ \bibinfo {author} {\bibfnamefont {L.}~\bibnamefont
  {Berthier}},\ }\bibfield  {title} {\bibinfo {title} {Anomalous structural
  evolution of soft particles: equibrium liquid state theory},\ }\href
  {https://doi.org/10.1039/b926412d} {\bibfield  {journal} {\bibinfo  {journal}
  {Soft Matter}\ }\textbf {\bibinfo {volume} {6}},\ \bibinfo {pages} {2970}
  (\bibinfo {year} {2010})}\BibitemShut {NoStop}%
\bibitem [{\citenamefont {Mohan}\ and\ \citenamefont
  {Bonnecaze}(2012)}]{Mohan_Short_ranged_pair_distribution_2012}%
  \BibitemOpen
  \bibfield  {author} {\bibinfo {author} {\bibfnamefont {L.}~\bibnamefont
  {Mohan}}\ and\ \bibinfo {author} {\bibfnamefont {R.~T.}\ \bibnamefont
  {Bonnecaze}},\ }\bibfield  {title} {\bibinfo {title} {Short-ranged pair
  distribution function for concentrated suspensions of soft particles},\
  }\href {https://doi.org/10.1039/C2SM06940G} {\bibfield  {journal} {\bibinfo
  {journal} {Soft Matter}\ }\textbf {\bibinfo {volume} {8}},\ \bibinfo {pages}
  {4216} (\bibinfo {year} {2012})}\BibitemShut {NoStop}%
\bibitem [{\citenamefont {Wyart}(2010)}]{Wyart_2010}%
  \BibitemOpen
  \bibfield  {author} {\bibinfo {author} {\bibfnamefont {M.}~\bibnamefont
  {Wyart}},\ }\bibfield  {title} {\bibinfo {title} {Scaling of phononic
  transport with connectivity in amorphous solids},\ }\href
  {https://doi.org/10.1209/0295-5075/89/64001} {\bibfield  {journal} {\bibinfo
  {journal} {Europhys. Lett.}\ }\textbf {\bibinfo {volume} {89}},\ \bibinfo
  {pages} {64001} (\bibinfo {year} {2010})}\BibitemShut {NoStop}%
\bibitem [{\citenamefont {Hess}\ and\ \citenamefont
  {Klein.}(1983)}]{HessKlein}%
  \BibitemOpen
  \bibfield  {author} {\bibinfo {author} {\bibfnamefont {W.}~\bibnamefont
  {Hess}}\ and\ \bibinfo {author} {\bibfnamefont {R.}~\bibnamefont {Klein.}},\
  }\bibfield  {title} {\bibinfo {title} {{Generalized hydrodynamics of systems
  of Brownian particles}},\ }\href@noop {} {\bibfield  {journal} {\bibinfo
  {journal} {Adv. Phys.}\ }\textbf {\bibinfo {volume} {32}},\ \bibinfo {pages}
  {173} (\bibinfo {year} {1983})}\BibitemShut {NoStop}%
\bibitem [{\citenamefont {Wajnryb}\ \emph
  {et~al.}(1995{\natexlab{a}})\citenamefont {Wajnryb}, \citenamefont
  {Altenberger},\ and\ \citenamefont {Dahler}}]{Uniqueness_Stress_Tensor_1995}%
  \BibitemOpen
  \bibfield  {author} {\bibinfo {author} {\bibfnamefont {E.}~\bibnamefont
  {Wajnryb}}, \bibinfo {author} {\bibfnamefont {A.~R.}\ \bibnamefont
  {Altenberger}},\ and\ \bibinfo {author} {\bibfnamefont {J.~S.}\ \bibnamefont
  {Dahler}},\ }\bibfield  {title} {\bibinfo {title} {{Uniqueness of the
  microscopic stress tensor}},\ }\href {https://doi.org/10.1063/1.469942}
  {\bibfield  {journal} {\bibinfo  {journal} {J. Chem. Phys.}\ }\textbf
  {\bibinfo {volume} {103}},\ \bibinfo {pages} {9782} (\bibinfo {year}
  {1995}{\natexlab{a}})}\BibitemShut {NoStop}%
\bibitem [{\citenamefont {Götze}(2009)}]{Gotze}%
  \BibitemOpen
  \bibfield  {author} {\bibinfo {author} {\bibfnamefont {W.}~\bibnamefont
  {Götze}},\ }\href@noop {} {\emph {\bibinfo {title} {Complex Dynamics of
  Glass-Forming Liquid}}}\ (\bibinfo  {publisher} {Oxford University Press},\
  \bibinfo {year} {2009})\BibitemShut {NoStop}%
\bibitem [{\citenamefont {Lemaître}(2018)}]{Lemaitre2018}%
  \BibitemOpen
  \bibfield  {author} {\bibinfo {author} {\bibfnamefont {A.}~\bibnamefont
  {Lemaître}},\ }\bibfield  {title} {\bibinfo {title} {{Stress correlations in
  glasses}},\ }\href {https://doi.org/10.1063/1.5041461} {\bibfield  {journal}
  {\bibinfo  {journal} {The Journal of Chemical Physics}\ }\textbf {\bibinfo
  {volume} {149}},\ \bibinfo {pages} {104107} (\bibinfo {year}
  {2018})}\BibitemShut {NoStop}%
\bibitem [{\citenamefont {Martin}(1965)}]{Martin}%
  \BibitemOpen
  \bibfield  {author} {\bibinfo {author} {\bibfnamefont {P.}~\bibnamefont
  {Martin}},\ }\bibfield  {title} {\bibinfo {title} {Non local transport
  coefficients correlation functions},\ }in\ \href@noop {} {\emph {\bibinfo
  {booktitle} {Statistical Mechanics of Equilibrium and Non-Equilibrium}}},\
  \bibinfo {editor} {edited by\ \bibinfo {editor} {\bibfnamefont
  {J.}~\bibnamefont {Meixner}}}\ (\bibinfo  {publisher} {North-Holland Publ.},\
  \bibinfo {address} {Amsterdam},\ \bibinfo {year} {1965})\ p.\ \bibinfo
  {pages} {100}\BibitemShut {NoStop}%
\bibitem [{\citenamefont {Maier}\ \emph {et~al.}(2018)\citenamefont {Maier},
  \citenamefont {Zippelius},\ and\ \citenamefont {Fuchs}}]{Maieretall}%
  \BibitemOpen
  \bibfield  {author} {\bibinfo {author} {\bibfnamefont {M.}~\bibnamefont
  {Maier}}, \bibinfo {author} {\bibfnamefont {A.}~\bibnamefont {Zippelius}},\
  and\ \bibinfo {author} {\bibfnamefont {M.}~\bibnamefont {Fuchs}},\ }\bibfield
   {title} {\bibinfo {title} {{Stress auto-correlation tensor in glass-forming
  isothermal fluids: From viscous to elastic response}},\ }\href@noop {}
  {\bibfield  {journal} {\bibinfo  {journal} {J. Chem. Phys.}\ } (\bibinfo
  {year} {2018})}\BibitemShut {NoStop}%
\bibitem [{\citenamefont {Vogel}\ \emph {et~al.}(2019)\citenamefont {Vogel},
  \citenamefont {Zippelius},\ and\ \citenamefont {Fuchs}}]{Vogel_2019}%
  \BibitemOpen
  \bibfield  {author} {\bibinfo {author} {\bibfnamefont {F.}~\bibnamefont
  {Vogel}}, \bibinfo {author} {\bibfnamefont {A.}~\bibnamefont {Zippelius}},\
  and\ \bibinfo {author} {\bibfnamefont {M.}~\bibnamefont {Fuchs}},\ }\bibfield
   {title} {\bibinfo {title} {Emergence of goldstone excitations in stress
  correlations of glass-forming colloidal dispersions},\ }\href
  {https://doi.org/10.1209/0295-5075/125/68003} {\bibfield  {journal} {\bibinfo
   {journal} {Europhys. Lett.}\ }\textbf {\bibinfo {volume} {125}},\ \bibinfo
  {pages} {68003} (\bibinfo {year} {2019})}\BibitemShut {NoStop}%
\bibitem [{\citenamefont {Jop}\ \emph {et~al.}(2012)\citenamefont {Jop},
  \citenamefont {Mansard}, \citenamefont {Chaudhuri}, \citenamefont {Bocquet},\
  and\ \citenamefont {Colin}}]{Microscale_Rheology_2012}%
  \BibitemOpen
  \bibfield  {author} {\bibinfo {author} {\bibfnamefont {P.}~\bibnamefont
  {Jop}}, \bibinfo {author} {\bibfnamefont {V.}~\bibnamefont {Mansard}},
  \bibinfo {author} {\bibfnamefont {P.}~\bibnamefont {Chaudhuri}}, \bibinfo
  {author} {\bibfnamefont {L.}~\bibnamefont {Bocquet}},\ and\ \bibinfo {author}
  {\bibfnamefont {A.}~\bibnamefont {Colin}},\ }\bibfield  {title} {\bibinfo
  {title} {Microscale rheology of a soft glassy material close to yielding},\
  }\href {https://doi.org/10.1103/PhysRevLett.108.148301} {\bibfield  {journal}
  {\bibinfo  {journal} {Phys. Rev. Lett.}\ }\textbf {\bibinfo {volume} {108}},\
  \bibinfo {pages} {148301} (\bibinfo {year} {2012})}\BibitemShut {NoStop}%
\bibitem [{\citenamefont {Zwanzig}\ and\ \citenamefont
  {Mountain}(2004)}]{Zwanzig_High_Frequency_elastic_constants}%
  \BibitemOpen
  \bibfield  {author} {\bibinfo {author} {\bibfnamefont {R.}~\bibnamefont
  {Zwanzig}}\ and\ \bibinfo {author} {\bibfnamefont {R.~D.}\ \bibnamefont
  {Mountain}},\ }\bibfield  {title} {\bibinfo {title} {{High‐Frequency
  Elastic Moduli of Simple Fluids}},\ }\href
  {https://doi.org/10.1063/1.1696718} {\bibfield  {journal} {\bibinfo
  {journal} {J. Chem. Phys.}\ }\textbf {\bibinfo {volume} {43}},\ \bibinfo
  {pages} {4464} (\bibinfo {year} {2004})}\BibitemShut {NoStop}%
\bibitem [{\citenamefont {Schmid}(2011)}]{Schmid_PHD_Theses}%
  \BibitemOpen
  \bibfield  {author} {\bibinfo {author} {\bibfnamefont {B.}~\bibnamefont
  {Schmid}},\ }\href@noop {} {\emph {\bibinfo {title} {Mode-Coupling Theory:
  Generalizations, High Dimensions and Microscopic Dynamics}}}\ (\bibinfo
  {publisher} {PHD-Theses University of Mainz 2011},\ \bibinfo {year}
  {2011})\BibitemShut {NoStop}%
\bibitem [{\citenamefont {Zaccarelli}\ \emph {et~al.}(2001)\citenamefont
  {Zaccarelli}, \citenamefont {Foffi}, \citenamefont {Sciortino}, \citenamefont
  {Tartaglia},\ and\ \citenamefont {Dawson}}]{Zaccarelli2001}%
  \BibitemOpen
  \bibfield  {author} {\bibinfo {author} {\bibfnamefont {E.}~\bibnamefont
  {Zaccarelli}}, \bibinfo {author} {\bibfnamefont {G.}~\bibnamefont {Foffi}},
  \bibinfo {author} {\bibfnamefont {F.}~\bibnamefont {Sciortino}}, \bibinfo
  {author} {\bibfnamefont {P.}~\bibnamefont {Tartaglia}},\ and\ \bibinfo
  {author} {\bibfnamefont {K.}~\bibnamefont {Dawson}},\ }\bibfield  {title}
  {\bibinfo {title} {Gaussian density fluctuations and mode coupling theory for
  supercooled liquids},\ }\href@noop {} {\bibfield  {journal} {\bibinfo
  {journal} {Europhys. Lett.}\ }\textbf {\bibinfo {volume} {55}},\ \bibinfo
  {pages} {157} (\bibinfo {year} {2001})}\BibitemShut {NoStop}%
\bibitem [{\citenamefont {Ohta}\ and\ \citenamefont
  {Kawasaki}(1976)}]{Ohta1976}%
  \BibitemOpen
  \bibfield  {author} {\bibinfo {author} {\bibfnamefont {T.}~\bibnamefont
  {Ohta}}\ and\ \bibinfo {author} {\bibfnamefont {K.}~\bibnamefont
  {Kawasaki}},\ }\bibfield  {title} {\bibinfo {title} {{Mode Coupling Theory of
  Dynamic Critical Phenomena for Classical Liquids. I: Dynamic Critical
  Exponents}},\ }\href {https://doi.org/10.1143/ptp.55.1384} {\bibfield
  {journal} {\bibinfo  {journal} {Prog. Theor. Phys.}\ }\textbf {\bibinfo
  {volume} {55}},\ \bibinfo {pages} {1384} (\bibinfo {year}
  {1976})}\BibitemShut {NoStop}%
\bibitem [{\citenamefont {Leutheusser}(1984)}]{Leutheusser}%
  \BibitemOpen
  \bibfield  {author} {\bibinfo {author} {\bibfnamefont {E.}~\bibnamefont
  {Leutheusser}},\ }\bibfield  {title} {\bibinfo {title} {Dynamical model of
  the liquid-glass transition},\ }\href@noop {} {\bibfield  {journal} {\bibinfo
   {journal} {Phys. Rev. A}\ }\textbf {\bibinfo {volume} {29}},\ \bibinfo
  {pages} {2765} (\bibinfo {year} {1984})}\BibitemShut {NoStop}%
\bibitem [{\citenamefont {Kamenev}(2011)}]{Kamenev_2011}%
  \BibitemOpen
  \bibfield  {author} {\bibinfo {author} {\bibfnamefont {A.}~\bibnamefont
  {Kamenev}},\ }\href@noop {} {\emph {\bibinfo {title} {Field Theory of
  Non-Equilibrium Systems}}}\ (\bibinfo  {publisher} {Cambridge University
  Press},\ \bibinfo {year} {2011})\BibitemShut {NoStop}%
\bibitem [{\citenamefont {Altland}\ and\ \citenamefont
  {Simons}(2010)}]{Altland_Simons_2010}%
  \BibitemOpen
  \bibfield  {author} {\bibinfo {author} {\bibfnamefont {A.}~\bibnamefont
  {Altland}}\ and\ \bibinfo {author} {\bibfnamefont {B.~D.}\ \bibnamefont
  {Simons}},\ }\href@noop {} {\emph {\bibinfo {title} {Condensed Matter Field
  Theory}}},\ \bibinfo {edition} {2nd}\ ed.\ (\bibinfo  {publisher} {Cambridge
  University Press},\ \bibinfo {year} {2010})\BibitemShut {NoStop}%
\bibitem [{\citenamefont {Pihlajamaa}\ \emph {et~al.}(2023)\citenamefont
  {Pihlajamaa}, \citenamefont {Debets}, \citenamefont {Laudicina},\ and\
  \citenamefont {Janssen}}]{pihlajamaa2023unveiling}%
  \BibitemOpen
  \bibfield  {author} {\bibinfo {author} {\bibfnamefont {I.}~\bibnamefont
  {Pihlajamaa}}, \bibinfo {author} {\bibfnamefont {V.~E.}\ \bibnamefont
  {Debets}}, \bibinfo {author} {\bibfnamefont {C.~C.~L.}\ \bibnamefont
  {Laudicina}},\ and\ \bibinfo {author} {\bibfnamefont {L.~M.~C.}\ \bibnamefont
  {Janssen}},\ }\href@noop {} {\bibinfo {title} {Unveiling the anatomy of
  mode-coupling theory}} (\bibinfo {year} {2023}),\ \Eprint
  {https://arxiv.org/abs/2307.03443} {arXiv:2307.03443} \BibitemShut {NoStop}%
\bibitem [{\citenamefont {Hertz}\ \emph {et~al.}(2016)\citenamefont {Hertz},
  \citenamefont {Roudi},\ and\ \citenamefont {Sollich}}]{Hertz_2016}%
  \BibitemOpen
  \bibfield  {author} {\bibinfo {author} {\bibfnamefont {J.~A.}\ \bibnamefont
  {Hertz}}, \bibinfo {author} {\bibfnamefont {Y.}~\bibnamefont {Roudi}},\ and\
  \bibinfo {author} {\bibfnamefont {P.}~\bibnamefont {Sollich}},\ }\bibfield
  {title} {\bibinfo {title} {Path integral methods for the dynamics of
  stochastic and disordered systems},\ }\href
  {https://doi.org/10.1088/1751-8121/50/3/033001} {\bibfield  {journal}
  {\bibinfo  {journal} {J. Phys. A Math. Theor.}\ }\textbf {\bibinfo {volume}
  {50}},\ \bibinfo {pages} {033001} (\bibinfo {year} {2016})}\BibitemShut
  {NoStop}%
\bibitem [{\citenamefont {Leutheusser}(1983)}]{Leutheusser1983}%
  \BibitemOpen
  \bibfield  {author} {\bibinfo {author} {\bibfnamefont {E.}~\bibnamefont
  {Leutheusser}},\ }\bibfield  {title} {\bibinfo {title} {Self-consistent
  kinetic theory for the lorentz gas},\ }\href
  {https://doi.org/10.1103/PhysRevA.28.1762} {\bibfield  {journal} {\bibinfo
  {journal} {Phys. Rev. A}\ }\textbf {\bibinfo {volume} {28}},\ \bibinfo
  {pages} {1762} (\bibinfo {year} {1983})}\BibitemShut {NoStop}%
\bibitem [{\citenamefont {Ashcroft}\ and\ \citenamefont
  {Mermin}(1976)}]{Ashcroft76}%
  \BibitemOpen
  \bibfield  {author} {\bibinfo {author} {\bibfnamefont {N.~W.}\ \bibnamefont
  {Ashcroft}}\ and\ \bibinfo {author} {\bibfnamefont {N.~D.}\ \bibnamefont
  {Mermin}},\ }\href@noop {} {\emph {\bibinfo {title} {{S}olid {S}tate
  {P}hysics}}}\ (\bibinfo  {publisher} {Holt-Saunders},\ \bibinfo {year}
  {1976})\BibitemShut {NoStop}%
\bibitem [{\citenamefont {Bosse}\ \emph {et~al.}(1978)\citenamefont {Bosse},
  \citenamefont {G\"otze},\ and\ \citenamefont
  {Zippelius}}]{Boss_Goetze_Zippelius_1978}%
  \BibitemOpen
  \bibfield  {author} {\bibinfo {author} {\bibfnamefont {J.}~\bibnamefont
  {Bosse}}, \bibinfo {author} {\bibfnamefont {W.}~\bibnamefont {G\"otze}},\
  and\ \bibinfo {author} {\bibfnamefont {A.}~\bibnamefont {Zippelius}},\
  }\bibfield  {title} {\bibinfo {title} {Velocity-autocorrelation spectrum of
  simple classical liquids},\ }\href {https://doi.org/10.1103/PhysRevA.18.1214}
  {\bibfield  {journal} {\bibinfo  {journal} {Phys. Rev. A}\ }\textbf {\bibinfo
  {volume} {18}},\ \bibinfo {pages} {1214} (\bibinfo {year}
  {1978})}\BibitemShut {NoStop}%
\bibitem [{\citenamefont {Mézard}\ \emph {et~al.}(1999)\citenamefont
  {Mézard}, \citenamefont {Parisi},\ and\ \citenamefont {Zee}}]{M_zard_1999}%
  \BibitemOpen
  \bibfield  {author} {\bibinfo {author} {\bibfnamefont {M.}~\bibnamefont
  {Mézard}}, \bibinfo {author} {\bibfnamefont {G.}~\bibnamefont {Parisi}},\
  and\ \bibinfo {author} {\bibfnamefont {A.}~\bibnamefont {Zee}},\ }\bibfield
  {title} {\bibinfo {title} {Spectra of euclidean random matrices},\ }\href
  {https://doi.org/10.1016/s0550-3213(99)00428-9} {\bibfield  {journal}
  {\bibinfo  {journal} {Nucl. Phys. B.}\ }\textbf {\bibinfo {volume} {559}},\
  \bibinfo {pages} {689–701} (\bibinfo {year} {1999})}\BibitemShut {NoStop}%
\bibitem [{\citenamefont {Martin-Mayor}\ \emph {et~al.}(2001)\citenamefont
  {Martin-Mayor}, \citenamefont {Mézard}, \citenamefont {Parisi},\ and\
  \citenamefont {Verrocchio}}]{Martin_Mayor_2001}%
  \BibitemOpen
  \bibfield  {author} {\bibinfo {author} {\bibfnamefont {V.}~\bibnamefont
  {Martin-Mayor}}, \bibinfo {author} {\bibfnamefont {M.}~\bibnamefont
  {Mézard}}, \bibinfo {author} {\bibfnamefont {G.}~\bibnamefont {Parisi}},\
  and\ \bibinfo {author} {\bibfnamefont {P.}~\bibnamefont {Verrocchio}},\
  }\bibfield  {title} {\bibinfo {title} {The dynamical structure factor in
  topologically disordered systems},\ }\href
  {https://doi.org/10.1063/1.1349709} {\bibfield  {journal} {\bibinfo
  {journal} {J. Chem. Phys.}\ }\textbf {\bibinfo {volume} {114}},\ \bibinfo
  {pages} {8068–8081} (\bibinfo {year} {2001})}\BibitemShut {NoStop}%
\bibitem [{\citenamefont {Ganter}\ and\ \citenamefont
  {Schirmacher}(2011)}]{Ganter_2011_Diagrams}%
  \BibitemOpen
  \bibfield  {author} {\bibinfo {author} {\bibfnamefont {C.}~\bibnamefont
  {Ganter}}\ and\ \bibinfo {author} {\bibfnamefont {W.}~\bibnamefont
  {Schirmacher}},\ }\bibfield  {title} {\bibinfo {title} {Euclidean random
  matrix theory: low-frequency non-analyticities and rayleigh scattering},\
  }\href {https://doi.org/10.1080/14786435.2010.530619} {\bibfield  {journal}
  {\bibinfo  {journal} {Phil. Mag.}\ }\textbf {\bibinfo {volume} {91}},\
  \bibinfo {pages} {1894} (\bibinfo {year} {2011})}\BibitemShut {NoStop}%
\bibitem [{\citenamefont {Kranz}\ \emph {et~al.}(2018)\citenamefont {Kranz},
  \citenamefont {Frahsa}, \citenamefont {Zippelius}, \citenamefont {Fuchs},\
  and\ \citenamefont {Sperl}}]{Kranz2018}%
  \BibitemOpen
  \bibfield  {author} {\bibinfo {author} {\bibfnamefont {W.~T.}\ \bibnamefont
  {Kranz}}, \bibinfo {author} {\bibfnamefont {F.}~\bibnamefont {Frahsa}},
  \bibinfo {author} {\bibfnamefont {A.}~\bibnamefont {Zippelius}}, \bibinfo
  {author} {\bibfnamefont {M.}~\bibnamefont {Fuchs}},\ and\ \bibinfo {author}
  {\bibfnamefont {M.}~\bibnamefont {Sperl}},\ }\bibfield  {title} {\bibinfo
  {title} {Rheology of inelastic hard spheres at finite density and shear
  rate},\ }\href {https://doi.org/10.1103/PhysRevLett.121.148002} {\bibfield
  {journal} {\bibinfo  {journal} {Phys. Rev. Lett.}\ }\textbf {\bibinfo
  {volume} {121}},\ \bibinfo {pages} {148002} (\bibinfo {year}
  {2018})}\BibitemShut {NoStop}%
\bibitem [{\citenamefont {DeGiuli}\ \emph {et~al.}(2015)\citenamefont
  {DeGiuli}, \citenamefont {D\"uring}, \citenamefont {Lerner},\ and\
  \citenamefont {Wyart}}]{Degiuli2015}%
  \BibitemOpen
  \bibfield  {author} {\bibinfo {author} {\bibfnamefont {E.}~\bibnamefont
  {DeGiuli}}, \bibinfo {author} {\bibfnamefont {G.}~\bibnamefont {D\"uring}},
  \bibinfo {author} {\bibfnamefont {E.}~\bibnamefont {Lerner}},\ and\ \bibinfo
  {author} {\bibfnamefont {M.}~\bibnamefont {Wyart}},\ }\bibfield  {title}
  {\bibinfo {title} {Unified theory of inertial granular flows and non-brownian
  suspensions},\ }\href {https://doi.org/10.1103/PhysRevE.91.062206} {\bibfield
   {journal} {\bibinfo  {journal} {Phys. Rev. E}\ }\textbf {\bibinfo {volume}
  {91}},\ \bibinfo {pages} {062206} (\bibinfo {year} {2015})}\BibitemShut
  {NoStop}%
\bibitem [{\citenamefont {G\"otze}\ \emph {et~al.}(1981)\citenamefont
  {G\"otze}, \citenamefont {Leutheusser},\ and\ \citenamefont
  {Yip}}]{Gotze1981}%
  \BibitemOpen
  \bibfield  {author} {\bibinfo {author} {\bibfnamefont {W.}~\bibnamefont
  {G\"otze}}, \bibinfo {author} {\bibfnamefont {E.}~\bibnamefont
  {Leutheusser}},\ and\ \bibinfo {author} {\bibfnamefont {S.}~\bibnamefont
  {Yip}},\ }\bibfield  {title} {\bibinfo {title} {Dynamical theory of diffusion
  and localization in a random, static field},\ }\href
  {https://doi.org/10.1103/PhysRevA.23.2634} {\bibfield  {journal} {\bibinfo
  {journal} {Phys. Rev. A}\ }\textbf {\bibinfo {volume} {23}},\ \bibinfo
  {pages} {2634} (\bibinfo {year} {1981})}\BibitemShut {NoStop}%
\bibitem [{\citenamefont {Schnyder}\ \emph {et~al.}(2011)\citenamefont
  {Schnyder}, \citenamefont {Höfling}, \citenamefont {Franosch},\ and\
  \citenamefont {Voigtmann}}]{Schnyder_2011}%
  \BibitemOpen
  \bibfield  {author} {\bibinfo {author} {\bibfnamefont {S.}~\bibnamefont
  {Schnyder}}, \bibinfo {author} {\bibfnamefont {F.}~\bibnamefont {Höfling}},
  \bibinfo {author} {\bibfnamefont {T.}~\bibnamefont {Franosch}},\ and\
  \bibinfo {author} {\bibfnamefont {T.}~\bibnamefont {Voigtmann}},\ }\bibfield
  {title} {\bibinfo {title} {Long-wavelength anomalies in the asymptotic
  behavior of mode-coupling theory},\ }\href
  {https://doi.org/10.1088/0953-8984/23/23/234121} {\bibfield  {journal}
  {\bibinfo  {journal} {J. Phys. Condens. Mat.}\ }\textbf {\bibinfo {volume}
  {23}},\ \bibinfo {pages} {234121} (\bibinfo {year} {2011})}\BibitemShut
  {NoStop}%
\bibitem [{\citenamefont {Amir}\ \emph {et~al.}(2013)\citenamefont {Amir},
  \citenamefont {Krich}, \citenamefont {Vitelli}, \citenamefont {Oreg},\ and\
  \citenamefont {Imry}}]{Amir2013}%
  \BibitemOpen
  \bibfield  {author} {\bibinfo {author} {\bibfnamefont {A.}~\bibnamefont
  {Amir}}, \bibinfo {author} {\bibfnamefont {J.~J.}\ \bibnamefont {Krich}},
  \bibinfo {author} {\bibfnamefont {V.}~\bibnamefont {Vitelli}}, \bibinfo
  {author} {\bibfnamefont {Y.}~\bibnamefont {Oreg}},\ and\ \bibinfo {author}
  {\bibfnamefont {Y.}~\bibnamefont {Imry}},\ }\bibfield  {title} {\bibinfo
  {title} {Emergent percolation length and localization in random elastic
  networks},\ }\href {https://doi.org/10.1103/PhysRevX.3.021017} {\bibfield
  {journal} {\bibinfo  {journal} {Phys. Rev. X}\ }\textbf {\bibinfo {volume}
  {3}},\ \bibinfo {pages} {021017} (\bibinfo {year} {2013})}\BibitemShut
  {NoStop}%
\bibitem [{\citenamefont {H\"ofling}\ \emph {et~al.}(2006)\citenamefont
  {H\"ofling}, \citenamefont {Franosch},\ and\ \citenamefont
  {Frey}}]{PhysRevLett.96.165901}%
  \BibitemOpen
  \bibfield  {author} {\bibinfo {author} {\bibfnamefont {F.}~\bibnamefont
  {H\"ofling}}, \bibinfo {author} {\bibfnamefont {T.}~\bibnamefont
  {Franosch}},\ and\ \bibinfo {author} {\bibfnamefont {E.}~\bibnamefont
  {Frey}},\ }\bibfield  {title} {\bibinfo {title} {Localization transition of
  the three-dimensional lorentz model and continuum percolation},\ }\href
  {https://doi.org/10.1103/PhysRevLett.96.165901} {\bibfield  {journal}
  {\bibinfo  {journal} {Phys. Rev. Lett.}\ }\textbf {\bibinfo {volume} {96}},\
  \bibinfo {pages} {165901} (\bibinfo {year} {2006})}\BibitemShut {NoStop}%
\bibitem [{\citenamefont {Olsson}\ and\ \citenamefont
  {Teitel}(2007)}]{Olsson2007}%
  \BibitemOpen
  \bibfield  {author} {\bibinfo {author} {\bibfnamefont {P.}~\bibnamefont
  {Olsson}}\ and\ \bibinfo {author} {\bibfnamefont {S.}~\bibnamefont
  {Teitel}},\ }\bibfield  {title} {\bibinfo {title} {Critical scaling of shear
  viscosity at the jamming transition},\ }\href@noop {} {\bibfield  {journal}
  {\bibinfo  {journal} {Phys. Rev. Lett.}\ }\textbf {\bibinfo {volume} {99}},\
  \bibinfo {pages} {178001} (\bibinfo {year} {2007})}\BibitemShut {NoStop}%
\bibitem [{\citenamefont {Heussinger}\ and\ \citenamefont
  {Barrat}(2009)}]{Heussinger2009}%
  \BibitemOpen
  \bibfield  {author} {\bibinfo {author} {\bibfnamefont {C.}~\bibnamefont
  {Heussinger}}\ and\ \bibinfo {author} {\bibfnamefont {J.-L.}\ \bibnamefont
  {Barrat}},\ }\bibfield  {title} {\bibinfo {title} {Jamming transition as
  probed by quasistatic shear flow},\ }\href@noop {} {\bibfield  {journal}
  {\bibinfo  {journal} {Phys. Rev. Lett.}\ }\textbf {\bibinfo {volume} {102}},\
  \bibinfo {pages} {218303} (\bibinfo {year} {2009})}\BibitemShut {NoStop}%
\bibitem [{\citenamefont {Saitoh}\ and\ \citenamefont
  {Kawasaki}(2020)}]{Saitoh2020}%
  \BibitemOpen
  \bibfield  {author} {\bibinfo {author} {\bibfnamefont {K.}~\bibnamefont
  {Saitoh}}\ and\ \bibinfo {author} {\bibfnamefont {T.}~\bibnamefont
  {Kawasaki}},\ }\bibfield  {title} {\bibinfo {title} {Critical scaling of
  diffusion coefficients and size of rigid clusters of soft athermal particles
  under shear},\ }\href {https://doi.org/10.3389/fphy.2020.00099} {\bibfield
  {journal} {\bibinfo  {journal} {Front. Phys.}\ }\textbf {\bibinfo {volume}
  {8}},\ \bibinfo {pages} {99} (\bibinfo {year} {2020})}\BibitemShut {NoStop}%
\bibitem [{\citenamefont {Richard}\ \emph {et~al.}(2020)\citenamefont
  {Richard}, \citenamefont {Gonz\'alez-L\'opez}, \citenamefont {Kapteijns},
  \citenamefont {Pater}, \citenamefont {Vaknin}, \citenamefont {Bouchbinder},\
  and\ \citenamefont {Lerner}}]{NonpnononicSpectrum}%
  \BibitemOpen
  \bibfield  {author} {\bibinfo {author} {\bibfnamefont {D.}~\bibnamefont
  {Richard}}, \bibinfo {author} {\bibfnamefont {K.}~\bibnamefont
  {Gonz\'alez-L\'opez}}, \bibinfo {author} {\bibfnamefont {G.}~\bibnamefont
  {Kapteijns}}, \bibinfo {author} {\bibfnamefont {R.}~\bibnamefont {Pater}},
  \bibinfo {author} {\bibfnamefont {T.}~\bibnamefont {Vaknin}}, \bibinfo
  {author} {\bibfnamefont {E.}~\bibnamefont {Bouchbinder}},\ and\ \bibinfo
  {author} {\bibfnamefont {E.}~\bibnamefont {Lerner}},\ }\bibfield  {title}
  {\bibinfo {title} {Universality of the nonphononic vibrational spectrum
  across different classes of computer glasses},\ }\href
  {https://doi.org/10.1103/PhysRevLett.125.085502} {\bibfield  {journal}
  {\bibinfo  {journal} {Phys. Rev. Lett.}\ }\textbf {\bibinfo {volume} {125}},\
  \bibinfo {pages} {085502} (\bibinfo {year} {2020})}\BibitemShut {NoStop}%
\bibitem [{\citenamefont {Schirmacher}\ \emph {et~al.}(2024)\citenamefont
  {Schirmacher}, \citenamefont {Paoluzzi}, \citenamefont {Mocanu},
  \citenamefont {Khomenko}, \citenamefont {Szamel}, \citenamefont {Zamponi},\
  and\ \citenamefont {Ruocco}}]{schirmacher2023nature}%
  \BibitemOpen
  \bibfield  {author} {\bibinfo {author} {\bibfnamefont {W.}~\bibnamefont
  {Schirmacher}}, \bibinfo {author} {\bibfnamefont {M.}~\bibnamefont
  {Paoluzzi}}, \bibinfo {author} {\bibfnamefont {F.~C.}\ \bibnamefont
  {Mocanu}}, \bibinfo {author} {\bibfnamefont {D.}~\bibnamefont {Khomenko}},
  \bibinfo {author} {\bibfnamefont {G.}~\bibnamefont {Szamel}}, \bibinfo
  {author} {\bibfnamefont {F.}~\bibnamefont {Zamponi}},\ and\ \bibinfo {author}
  {\bibfnamefont {G.}~\bibnamefont {Ruocco}},\ }\bibfield  {title} {\bibinfo
  {title} {The nature of non-phononic excitations in disordered systems},\
  }\href {https://doi.org/10.1038/s41467-024-46981-7} {\bibfield  {journal}
  {\bibinfo  {journal} {Nat. Commun.}\ }\textbf {\bibinfo {volume} {15}},\
  \bibinfo {pages} {3107} (\bibinfo {year} {2024})}\BibitemShut {NoStop}%
\bibitem [{\citenamefont {Dhont}(1996)}]{Dhont}%
  \BibitemOpen
  \bibfield  {author} {\bibinfo {author} {\bibfnamefont {J.}~\bibnamefont
  {Dhont}},\ }\href@noop {} {\emph {\bibinfo {title} {An Introduction to
  Dynamics of Colloids}}}\ (\bibinfo  {publisher} {Elsevier},\ \bibinfo
  {address} {Amsterdam},\ \bibinfo {year} {1996})\BibitemShut {NoStop}%
\bibitem [{\citenamefont {Graham}\ and\ \citenamefont
  {T{\'e}l}(1984)}]{Graham1984}%
  \BibitemOpen
  \bibfield  {author} {\bibinfo {author} {\bibfnamefont {R.}~\bibnamefont
  {Graham}}\ and\ \bibinfo {author} {\bibfnamefont {T.}~\bibnamefont
  {T{\'e}l}},\ }\bibfield  {title} {\bibinfo {title} {On the weak-noise limit
  of {F}okker-{P}lanck models},\ }\href@noop {} {\bibfield  {journal} {\bibinfo
   {journal} {J. Stat. Phys.}\ }\textbf {\bibinfo {volume} {35}},\ \bibinfo
  {pages} {729} (\bibinfo {year} {1984})}\BibitemShut {NoStop}%
\bibitem [{\citenamefont {Risken}(1996)}]{Risken}%
  \BibitemOpen
  \bibfield  {author} {\bibinfo {author} {\bibfnamefont {H.}~\bibnamefont
  {Risken}},\ }\href@noop {} {\emph {\bibinfo {title} {The Fokker-Planck
  Equation}}},\ Vol.~\bibinfo {volume} {2}\ (\bibinfo  {publisher} {Springer},\
  \bibinfo {year} {1996})\BibitemShut {NoStop}%
\bibitem [{\citenamefont {Wajnryb}\ \emph
  {et~al.}(1995{\natexlab{b}})\citenamefont {Wajnryb}, \citenamefont
  {Altenberger},\ and\ \citenamefont {Dahler}}]{Wajnryb1995}%
  \BibitemOpen
  \bibfield  {author} {\bibinfo {author} {\bibfnamefont {E.}~\bibnamefont
  {Wajnryb}}, \bibinfo {author} {\bibfnamefont {A.~R.}\ \bibnamefont
  {Altenberger}},\ and\ \bibinfo {author} {\bibfnamefont {J.~S.}\ \bibnamefont
  {Dahler}},\ }\bibfield  {title} {\bibinfo {title} {Uniqueness of the
  microscopic stress tensor},\ }\href@noop {} {\bibfield  {journal} {\bibinfo
  {journal} {J. Chem. Phys.}\ }\textbf {\bibinfo {volume} {103}},\ \bibinfo
  {pages} {9782} (\bibinfo {year} {1995}{\natexlab{b}})}\BibitemShut {NoStop}%
\bibitem [{\citenamefont {Zwanzig}(1961)}]{Zwanzig1961}%
  \BibitemOpen
  \bibfield  {author} {\bibinfo {author} {\bibfnamefont {R.}~\bibnamefont
  {Zwanzig}},\ }\bibfield  {title} {\bibinfo {title} {Memory effects in
  irreversible thermodynamics},\ }\href@noop {} {\bibfield  {journal} {\bibinfo
   {journal} {Phys. Rev.}\ }\textbf {\bibinfo {volume} {124}},\ \bibinfo
  {pages} {983} (\bibinfo {year} {1961})}\BibitemShut {NoStop}%
\bibitem [{\citenamefont {Rintoul}(2000)}]{rintoul_precise_2000}%
  \BibitemOpen
  \bibfield  {author} {\bibinfo {author} {\bibfnamefont {M.~D.}\ \bibnamefont
  {Rintoul}},\ }\bibfield  {title} {\bibinfo {title} {Precise determination of
  the void percolation threshold for two distributions of overlapping
  spheres},\ }\href@noop {} {\bibfield  {journal} {\bibinfo  {journal} {Phys.
  Rev. E}\ }\textbf {\bibinfo {volume} {62}},\ \bibinfo {pages} {68} (\bibinfo
  {year} {2000})}\BibitemShut {NoStop}%
\bibitem [{\citenamefont {Elam}\ \emph {et~al.}(1984)\citenamefont {Elam},
  \citenamefont {Kerstein},\ and\ \citenamefont {Rehr}}]{elam_critical_1984}%
  \BibitemOpen
  \bibfield  {author} {\bibinfo {author} {\bibfnamefont {W.~T.}\ \bibnamefont
  {Elam}}, \bibinfo {author} {\bibfnamefont {A.~R.}\ \bibnamefont {Kerstein}},\
  and\ \bibinfo {author} {\bibfnamefont {J.~J.}\ \bibnamefont {Rehr}},\
  }\bibfield  {title} {\bibinfo {title} {Critical properties of the void
  percolation problem for spheres},\ }\href@noop {} {\bibfield  {journal}
  {\bibinfo  {journal} {Phys. Rev. Lett.}\ }\textbf {\bibinfo {volume} {52}},\
  \bibinfo {pages} {1516} (\bibinfo {year} {1984})}\BibitemShut {NoStop}%
\end{thebibliography}%

\end{document}